\documentclass[pre,amssymb,reprint,groupaddress,superscriptaddress,showpacs,longbibliography]{revtex4-1}
\usepackage{amsmath}
\usepackage{amsfonts}
\usepackage{amssymb}
\usepackage{epsfig}
\usepackage{graphicx}

\renewcommand{\phi}{\varphi}

\newcommand{\be}{\begin{equation}}
\newcommand{\ee}{\end{equation}}
\newcommand{\bea}{\begin{equnaray}}
\newcommand{\eea}{\end{equnaray}}
\newcommand{\ba}{\begin{align}}
\newcommand{\ea}{\end{align}}

\usepackage{color}

\definecolor{green}{rgb}{0.0, 0.44, 0.0}
\definecolor{red}{rgb}{1.0, 0.13, 0.32}
\definecolor{blue}{rgb}{0.06, 0.2, 0.65}
\definecolor{magenta}{rgb}{1.0, 0.0, 1.00}
\definecolor{purple}{rgb}{0.7, 0.0, 0.7}
\definecolor{cyan}{rgb}{0.0, 1.0, 1.0}

\newcommand{\MO}[1]{{#1}}
\newcommand{\SC}{S_{\rm conf}}
\usepackage{hyperref}
\definecolor{bluelink}{rgb}{0,0.24,0.62}
\hypersetup{
bookmarksopen=true,
pdftitle="",
pdfauthor="",
pdftoolbar=false, 
pdfstartview={FitH},            
pdfmenubar=true,                        
pdfhighlight=/O,                        
colorlinks=true,                        
urlcolor=bluelink,
citecolor=bluelink,             
linkcolor=bluelink}

\begin{document}

\title{Does the Adam-Gibbs relation hold in simulated supercooled liquids?}

\author{Misaki Ozawa}

\affiliation{Laboratoire Charles Coulomb (L2C), Universit\'e de
Montpellier, CNRS, Montpellier, France}

\author{Camille Scalliet}

\affiliation{Laboratoire Charles Coulomb (L2C), Universit\'e de
Montpellier, CNRS, Montpellier, France}

\author{Andrea Ninarello}

\affiliation{CNR-ISC Uos Sapienza, Piazzale A. Moro 2, IT-00185 Roma, Italy}

\author{Ludovic Berthier}

\email{ludovic.berthier@umontpellier.fr}

\affiliation{Laboratoire Charles Coulomb (L2C), Universit\'e de
Montpellier, CNRS, Montpellier, France}

\begin{abstract}
We perform stringent tests of thermodynamic theories of the glass transition over the experimentally relevant temperature regime for several simulated glass-formers. The swap Monte Carlo algorithm is used to estimate the configurational entropy and static point-to-set lengthscale, and careful extrapolations are used for the relaxation times.
We first quantify the relation between configurational entropy and the point-to-set lengthscale in two and three dimensions. We then show that the Adam-Gibbs relation is generally violated in simulated models for the experimentally relevant time window. Collecting experimental data for several supercooled molecular liquids, we show that the same trends are observed experimentally. Deviations from the Adam-Gibbs relation remain compatible with random first order transition theory, and may account for the reported discrepancies between Kauzmann and Vogel-Fulcher-Tammann temperatures. Alternatively, they may also indicate that even near $T_g$ thermodynamics is not the only driving force for slow dynamics.  
\end{abstract}

\date{\today}

\maketitle

\section{Introduction}
\label{sec:Intro}

Since its first derivation in 1965~\cite{adam1965temperature}, the Adam-Gibbs relation has played a central role in glass transition studies~\cite{berthier2011theoretical}, since it is at the core of thermodynamic approaches to the glass problem~\cite{adam1965temperature,berthier2011theoretical,KTW89,BB04,LW07,BB09,lubchenko2015theory,CKPUZ16,dudowicz2008generalized}. The Adam-Gibbs relation captures in a simple mathematical form the physical idea that the decrease of the configurational entropy $\SC$ controls the growth of the relaxation time $\tau_\alpha$ as the experimental glass transition temperature $T_g$ is approached: 
\be 
\log(\tau_\alpha / \tau_0) \propto \frac{1}{T \SC},
\label{eq:AG}
\ee
where $\tau_0$ is a microscopic timescale.
Testing the Adam-Gibbs relation has almost become synonymous to testing the thermodynamic nature of glass formation~\cite{angell1997entropy,richert1998dynamics,sastry2001relationship,dyre2009brief}. 

Since computational methods have become available in the early 2000's to measure the configurational entropy in numerical simulations~\cite{SKT99,sastry2000evaluation,reviewsconf}, the Adam-Gibbs relation has been tested in a large number of studies using many different models of glass-forming materials~\cite{sastry2001relationship,mossa2002dynamics,sciortinoPEL,saika2001fragile,foffi2005,sengupta2012adam,starr2013relationship,parmar2017length,handle2018adam}. Importantly, these simulations are all restricted to a high temperature regime (typically above the mode-coupling crossover temperature $T_{\rm mct}$~\cite{gotze2008complex}) that barely overlaps with the corresponding experimental studies. In addition simulations typically cover a dynamic window of at most 3-4 decades, much narrower than in experimental studies. Despite  these caveats, the general consensus is that the Adam-Gibbs relation is generally valid in the regime accessed by the simulations. In experiments, which typically analyse temperatures close to $T_g$, the Adam-Gibbs relation seems again to be well obeyed for a range of materials~\cite{magill1967physical,takahara1995calorimetric,angell1997entropy,richert1998dynamics,ngai1999modification,alba2001salient,roland2004adam,cangialosi2005relationship,masiewicz2015adam}. Yet, experiments indicate as well that the Adam-Gibbs relation does not hold anymore above a temperature scale close to $T_{\rm mct}$~\cite{richert1998dynamics,ngai1999modification}, in stark contrast with the numerical results. Systematic deviations from the Adam-Gibbs relation were also reported below $T_{\rm mct}$ for some systems~\cite{ngai1999modification,roland2004adam}, but imprecise entropy measurements or inappropriate timescale determinations have been invoked to rationalise them.

In the last three decades, the random first order transition (RFOT) theory of the glass transition~\cite{KTW89,LW07} has revisited the Adam-Gibbs relation in greater depth~\cite{LW07,BB04,BB09,lubchenko2015theory} to provide an increasingly precise description of the connection between thermodynamics and dynamics in supercooled liquids. This connection can be decomposed in two steps. First, the decrease of the configurational entropy is shown, by a purely thermodynamic reasoning~\cite{BB04}, to give rise to a growing `point-to-set' static correlation lengthscale: 
\be 
\xi_{\rm pts}  \propto \SC^{-1/(d-\theta)}, \label{eq:RFOT2}
\ee   
where an interface exponent $\theta$ is introduced. In the simplest approximation, one has $\theta = d-1$ which corresponds to a (hyper-)surface  in a space of dimension $d$. The value $\theta = d/2$ was also proposed~\cite{KTW89,lubchenko2015theory}, to take into account finite dimensional surface fluctuations due to the disordered nature of the amorphous phase. More generally, the inequality $\theta \leq d-1$ is expected to hold. Second, the connection to dynamics is made via the assumption that relaxation in the liquid for $T < T_{\rm mct}$ proceeds via thermally activated events correlated over a lengthscale $\xi_{\rm pts}$, resulting in the general relation~\cite{KTW89,BB04}, 
\be 
\log( \tau_{\alpha}/\tau_0) \propto  \xi_{\rm pts}^{\psi} / T, \label{eq:RFOT1}
\ee
where $\psi$ is a dynamical exponent. Various theoretical and numerical estimates of $\psi$ have been proposed~\cite{BB04,cammarota2009numerical,karmakar2009growing,hocky2012growing,Gutierrez:2015}. In the original paper by Kirkpatrick {\it et al.}~\cite{KTW89}, $\psi=\theta=d/2$ was assumed and so only one exponent had been introduced.  
 
Using Eqs.~(\ref{eq:RFOT2}, \ref{eq:RFOT1}), one finds a generalised version of the Adam-Gibbs relation, 
\be
\log(\tau_{\alpha}/\tau_0) \propto \frac{1}{T \SC^{\alpha}}, 
\label{eq:GAG} 
\ee
with a non-trivial exponent  
\be 
\alpha = \frac{\psi}{d-\theta}.
\label{eq:alpha}
\ee
This shows that $\alpha$ may or may not be equal to unity, depending on the relative values of the two independent exponents $\psi$ and $\theta$. As a consequence, Eq.~(\ref{eq:GAG}) may or may not be equivalent to Eq.~(\ref{eq:AG}).

To our knowledge, a direct test of Eqs.~(\ref{eq:RFOT1}, \ref{eq:GAG}, \ref{eq:alpha}) in the theoretically-motivated temperature regime, employing appropriate observables, has never been performed. Most  previous simulations have considered a temperature regime $T \gtrsim T_{\rm mct}$~\cite{sastry2001relationship,mossa2002dynamics,CGV12,sengupta2012adam} where the physics is expected to be non-activated and the configurational entropy and point-to-set lengthscales are not well-defined.
This is of course valuable work, but theory itself suggests that the tested scaling relations have no reason to hold in this temperature regime.
Experiments instead access the correct temperature regime, but cannot easily measure the point-to-set correlation lengthscale. As a proxy, Refs.~\cite{capaccioli2008dynamically,brun2012evidence} replaced $\xi_{\rm pts}$ by the lengthscale of dynamic heterogeneities that can be more easily estimated experimentally~\cite{BBBCEHLP05}. Many other experimental studies study Eq.~(\ref{eq:AG}) directly near $T_g$~\cite{richert1998dynamics,roland2004adam}.

In this work, we take advantage of the progress allowed by the swap Monte Carlo algorithm~\cite{berthier2016equilibrium,NBC17} to measure directly in several numerical models the temperature dependence of the configurational entropy and point-to-set lengthscale down to $T_g$. For the dynamics, we build on previous work~\cite{NBC17} and provide additional experimental support showing that one can safely estimate the temperature dependence of the relaxation time also down to $T_g$, using a careful fitting procedure. We collect data from earlier works~\cite{berthier2018zero,ceiling17,ozawa2018configurational} that we extend where needed, and perform new simulations for one additional model.  

As a result, we are in a position to provide for the first time stringent tests of the Adam-Gibbs relation and of RFOT theory for computer models simulated in the same regime as in experiments. Our results suggest that the Adam-Gibbs relation is generally not valid in computer models in the experimental regime $T_g < T < T_{\rm mct}$. To test our findings against experiments, we collect high-quality thermodynamic and dynamic data for several supercooled liquids (most of which are obtained by state-of-the-art thermodynamic measurements~\cite{tatsumi12}), and reach similar conclusions. Overall, we find that Eq.~(\ref{eq:AG}) is not obeyed for most systems, while Eq.~(\ref{eq:GAG}) is obeyed with an exponent $\alpha$ that fluctuates weakly from system to system, with typically $\alpha < 1$. Our findings can be taken either as a confirmation that RFOT theory works well, with a non-trivial set of critical exponents, or that a small $\alpha<1$ exponent indicates that thermodynamics is not the only driving force for the dynamic slowdown near $T_g$. 

This paper is organised as follows. 
In Sec.~\ref{sec:Methods} we present the numerical methods used to obtain the configurational entropy, the point-to-set lengthscale, and the relaxation time. We also describe our choice of experimental data to reliably test the Adam-Gibbs relation over a broad range of temperatures.
In Sec.~\ref{sec:Results} we present the results of our analysis of the exponents $\theta$ and $\alpha$ in simulations, then in experiments. 
We discuss the physical meaning of our results in Sec.~\ref{sec:discussion}.

\section{Description of the data}
\label{sec:Methods}

In order to analyse quantitatively the connection between dynamic and thermodynamic properties, we collect and extend data from previous numerical works. We also collect data from selected published experimental works, and motivate our selection.

\subsection{Numerical models}

The recent development of the swap Monte Carlo algorithm  allows us to access very low-temperature equilibrium configurations in computer simulations. In particular, the temperature regime  $T_g < T < T_{\rm mct}$ can be comfortably accessed. This temperature regime is the correct one to test thermodynamic theories, as it is precisely where they should apply, and it corresponds to the regime explored experimentally. 

We gather simulation data for polydisperse systems using a continuous size distribution~\cite{NBC17}. The particle diameters $\sigma$ are distributed between $\sigma_{\rm min}$ and $\sigma_{\rm max}$ from $f(\sigma) = c\sigma^{-3}$, where $c$ is a normalization constant and $\sigma_{\rm min} / \sigma_{\rm max} = 0.45$.
 We use the average diameter $\overline{\sigma}$ as the unit length. 

We study four numerical models: three-dimensional additive hard spheres (HS3D)~\cite{berthier2016equilibrium}, two and three dimensional non-additive soft disks (SSV2D)~\cite{berthier2018zero} and spheres (SSV3D)~\cite{NBC17} under an isochoric path. We also perform new simulations of three-dimensional non-additive soft spheres (SSP3D), under an isobaric path. To thermalize the last model, we use an hybrid molecular dynamics/swap Monte Carlo scheme~\cite{berthier2018efficient}.

We use the following pairwise potential for the polydisperse soft sphere/disk models~\cite{NBC17},
\begin{eqnarray}
v_{ij}(r) &=& v_0 \left( \frac{\sigma_{ij}}{r} \right)^{12} + c_0 + c_1 \left( \frac{r}{\sigma_{ij}} \right)^2 + c_2 \left( \frac{r}{\sigma_{ij}} \right)^4, \label{eq:soft_v} \quad \\
\sigma_{ij} &=& \frac{(\sigma_i + \sigma_j)}{2} (1-\epsilon |\sigma_i - \sigma_j|), \label{eq:non_additive}
\end{eqnarray}
where $v_0$ is the energy unit, and $\epsilon$  
quantifies the degree of non-additivity of the system. We set $\epsilon=0.2$ for SSV3D and SSV2D, and $\epsilon=0.1$ for SSP3D. The constants, $c_0$, $c_1$ and $c_2$, are chosen to smooth $v_{ij}(r)$ up to its second derivative at the cut-off distance $r_{\rm cut}=1.25 \sigma_{ij}$. We set the number density $\rho=N/L^3=1.02$ with $N=1500$ for SSV3D, and $\rho=N/L^2 = 1.01$ with $N=1000$ for SSV2D. For SSP3D, the pressure on the isobaric path is $P=30.0$. 
For HS3D~\cite{berthier2016equilibrium}, the pair interaction is zero for non-overlapping particles and infinite otherwise.
The relevant control parameter for hard spheres is the reduced pressure $p=P/(\rho T)$. For hard spheres, $1/p$ plays precisely the same role as temperature $T$ for a dense liquid~\cite{BW09}, and there is no distinction between isochoric and isobaric paths. 

Relaxation times for HS3D, SSV3D, and SSV2D are measured in units of MC sweeps, which comprise $N$ Monte Carlo trial moves. For SSP3D, the relaxation time is
expressed in units of $\sqrt{v_0/ m \overline{\sigma}^2}$, where $m$ is the mass of the particles. 

\subsection{Configurational entropy and point-to-set length}

The configurational entropy $S_{\rm conf}$ is measured from configurations generated with swap Monte Carlo simulations. It is 
defined as $S_{\rm conf}=S_{\rm tot}-S_{\rm glass}$, where $S_{\rm tot}$ and $S_{\rm glass}$ are the total and glass entropies, respectively~\cite{reviewsconf}.
$S_{\rm tot}$ and $S_{\rm glass}$ are computed using thermodynamic integration schemes, as explained in Ref.~\cite{ozawa2018configurational}.
In Appendix~\ref{sec:s_conf_NPT} we describe how to measure $S_{\rm conf}$ along an isobaric path using constant pressure simulations for SSP3D, as this was not documented before. 

\begin{figure}
\includegraphics[width=0.95\columnwidth]{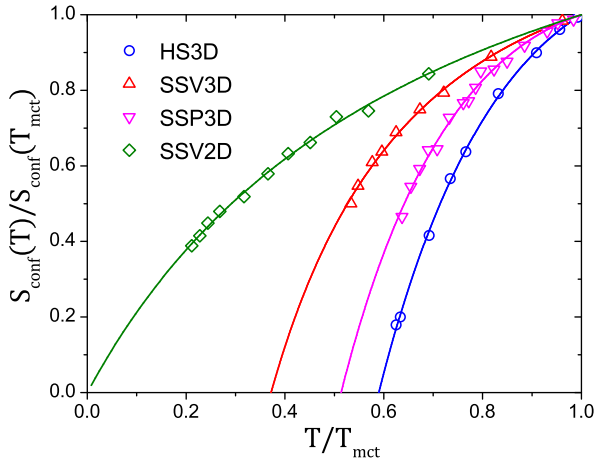}
\caption{
Configurational entropy data for the four simulated models. The data are normalized by the values at the mode coupling crossover $T_{\rm mct}$. The solid curves represent the fitting functions defined in the text and Table~\ref{tab:para_sim}.}
\label{fig:s_conf_data}
\end{figure}  

Figure~\ref{fig:s_conf_data} shows the configurational entropy that we use for latter analysis. The data for $\SC(T)$ are normalized by the values at the mode coupling crossover $T_{\rm mct}$, whose value is determined by a power law fit to the dynamic relaxation time data~\cite{gotze2008complex}. 
The actual values are $T_{\rm mct}=0.0426$, $0.104$, $0.556$, and $0.123$ for HS3D, SSV3D, SSP3D, and SSV2D, respectively.

In order to increase the accuracy of the analysis, we employ empirical fitting functions. 
For the three-dimensional models, we use a conventional fitting function plus a quadratic correction, $T S_{\rm conf}=A(T-T_{\rm K}) + B(T-T_{\rm K})^2$~\cite{richert1998dynamics,banerjee2014role}.
For the two-dimensional model, we use $1/S_{\rm conf}=A/T + B$~\cite{berthier2018zero}.
\MO{These fitting functions conveniently enable us to incorporate $S_{\rm conf}$ values in between actual data points and help us determining the exponents.} 
The fitting parameters are presented in Table~\ref{tab:para_sim}.

\begin{table}[b] 
\begin{center} 
\begin{tabular}{c|cccccccc}
\hline
\hline
\ Model  \ & \ $A$ \ & \ $B$ \ & \ $T_{\rm K}$ \ & \ $\log_{10} \tau_o$ \ & \ $T_o$ \ & \ $C$ \ & \ $m$  \\
\hline
HS3D \ & \ 3.208 \ & -37.33 \ &  0.0251 \ & 3.88 \ & 0.063 \ & 22.72 \ & 45.5 \  \\
SSV3D \  & \ 1.495 \ & -1.92 \ & 0.0386 \ & 3.02 \ & 0.266 \ & 3.15 \ & 32.0 \  \\
SSP3D \ & \ 2.082 \ & -1.74 \ &  0.2902 \ & 0.41 \ &  0.961 \ & 16.77 \ & 42.4 \  \\
SSV2D \ & \ 0.453 \ & 1.89  \ &  - \ & 2.40 \ & 1.006 \ & 0.25 \ & 31.2 \  \\
\hline \hline
\end{tabular}
\caption{
Fitting parameters for the configurational entropy ($A$, $B$, and $T_{\rm K}$), for the relaxation time ($\tau_o$, $T_o$, and $C$) and kinetic fragility index $m$ for the simulated models. Note that Monte Carlo dynamics (HS3D, SSV3D, SSV2D) and molecular dynamics (SSP3D) have different time units.}
\label{tab:para_sim}
\end{center}
\end{table}

We also collect the point-to-set lengthscale $\xi_{\rm pts}$ data for SSV2D~\cite{berthier2018zero} and HS3D~\cite{ceiling17}, obtained from recently developed computational methods~\cite{BBCGV08,berthier2016efficient}.
Together with $S_{\rm conf}$, the data for $\xi_{\rm pts}$ will allow us to estimate the exponent $\theta$ using Eq.~(\ref{eq:RFOT2}). 

\subsection{Relaxation times}
\label{sec:dynamic}

Dynamical information is obtained using either standard Monte Carlo (for HS3D, SSV3D, SSV2D) or molecular dynamics (for SSP3D). The equivalence between the two types of dynamics is well documented~\cite{berthier2007monte}. 
Both Monte Carlo and molecular dynamics simulations are run starting from initial configurations that are obtained using the swap Monte Carlo algorithm. This procedure allows us to cover about 5 orders of magnitude of relevant slow dynamics.

The relaxation time $\tau_{\alpha}$ is measured by the self-intermediate scattering function in three dimensional models.
For the two-dimensional model, we use the autocorrelation function of the bond-orientational order parameter, which is insensitive to the long-range Mermin-Wagner fluctuations that are specific to $d=2$~\cite{flenner2015fundamental}. 

\begin{figure}
\includegraphics[width=0.49\columnwidth]{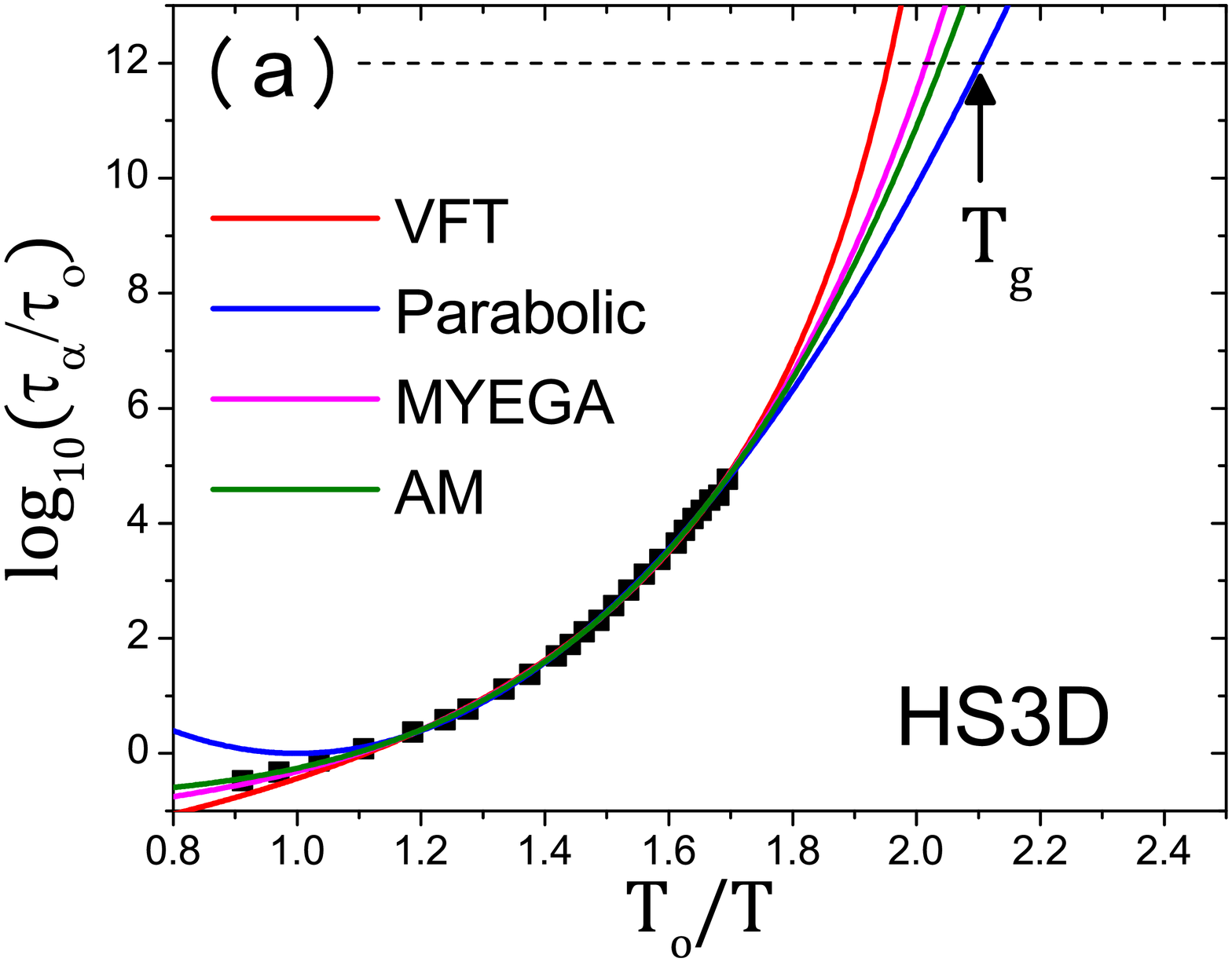}
\includegraphics[width=0.49\columnwidth]{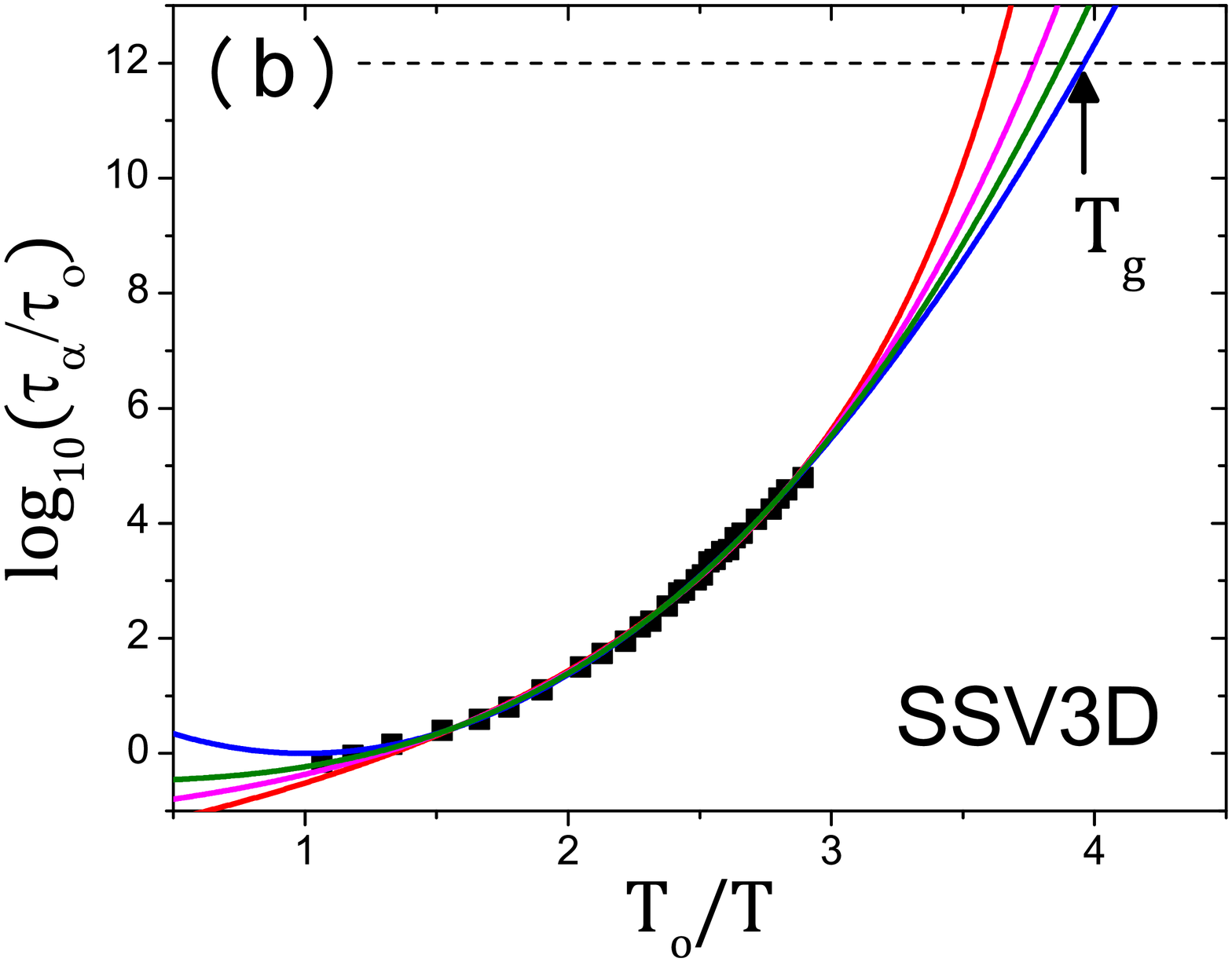}
\includegraphics[width=0.49\columnwidth]{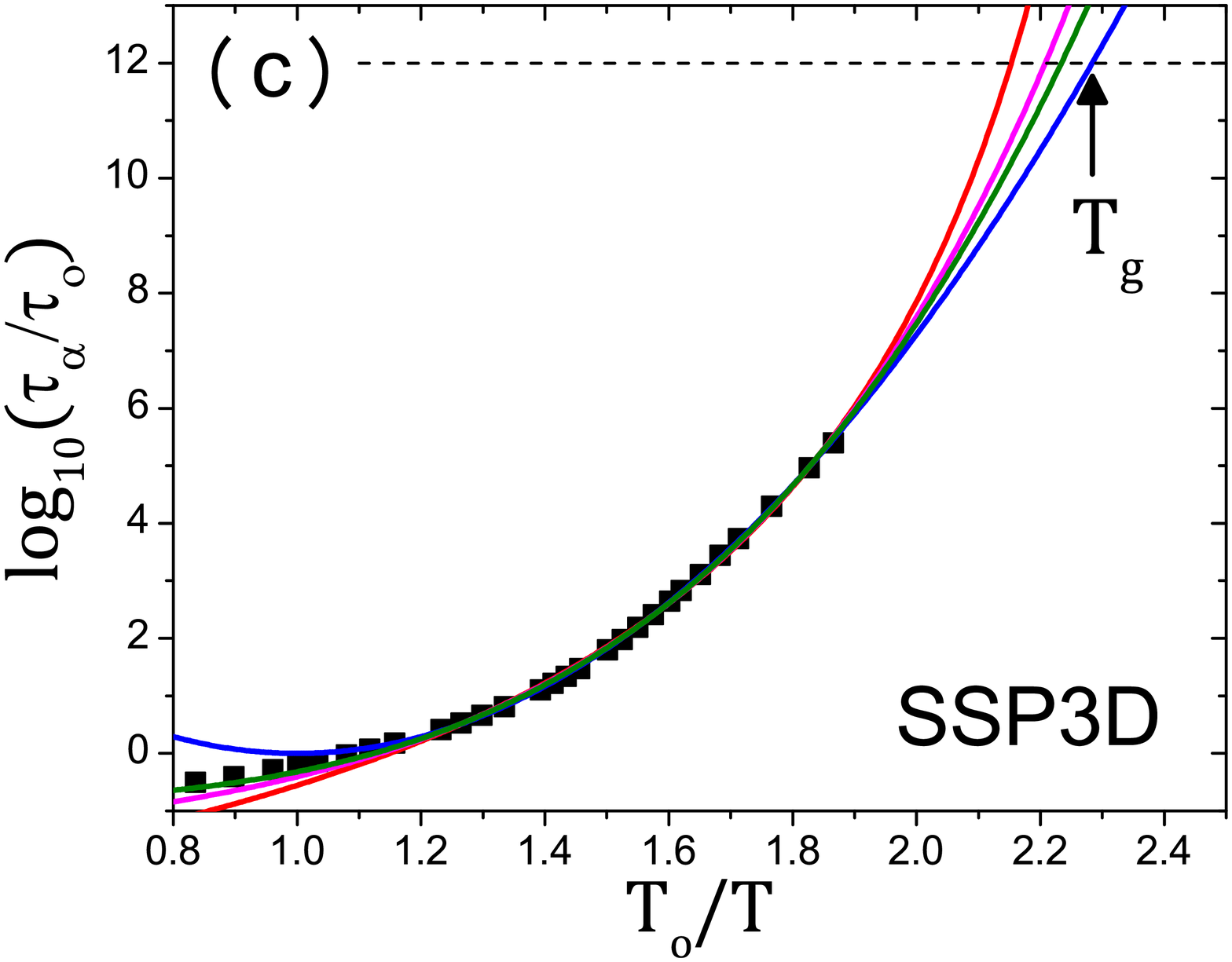}
\includegraphics[width=0.49\columnwidth]{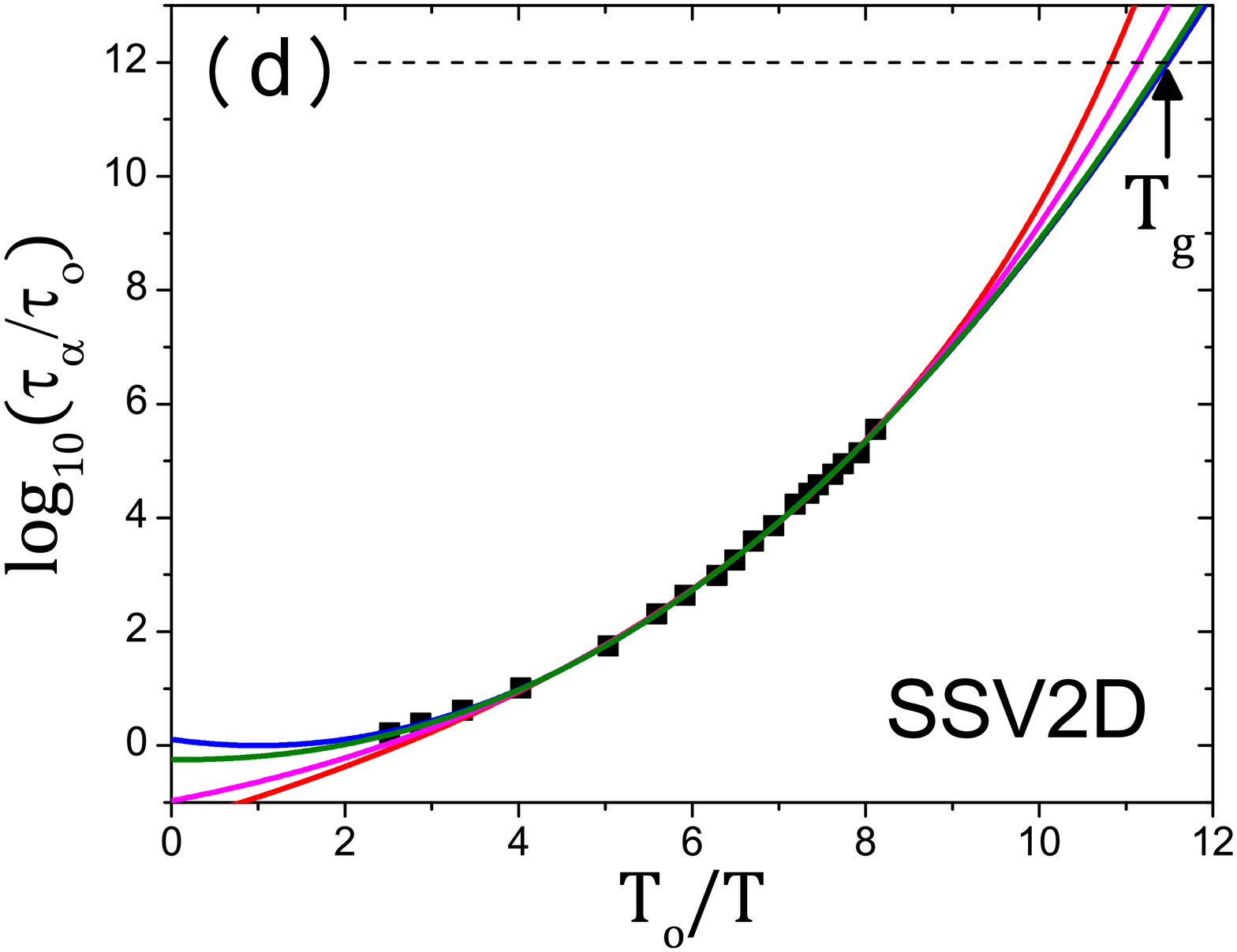}
\caption{
Relaxation time as a function of inverse temperature for the four simulated models: HS3D (a), SSV3D (b), SSP3D (c), and SSV2D (d). The data are normalized by $\tau_o$ and $T_o$, determined from a parabolic-law fitting. The horizontal dashed line indicates the timescale of the experimental glass transition, $\tau_{\alpha}/\tau_o = 10^{12}$. The vertical arrow indicates the experimental glass transition temperature $T_{\rm g}$ using the parabolic-raw fitting. Three additional fitting functions are shown.}
\label{fig:tau_data}
\end{figure}  

The relaxation time $\tau_{\alpha}$ for HS3D~\cite{ceiling17}, SSV3D~\cite{NBC17}, SSP3D (new to this work), and SSV2D~\cite{berthier2018zero} is shown in Figure~\ref{fig:tau_data}.
The data are normalized using an onset temperature $T_o$ for the emergence of slow dynamics, determined from the fitting procedure described below, and define $\tau_o = \tau_\alpha(T=T_o)$. Clearly, all simulation data show a non-Arrhenius temperature dependence of the relaxation time, which demonstrates that our models describe fragile glass-formers. 

The swap numerical schemes allow us to prepare equilibrated configurations at very low temperatures. Because they involve non-physical particle dynamics, one cannot use them to measure the relaxation time of the physical dynamics in this low-temperature regime.
Therefore, we need to extrapolate the relaxation time from the regime where $\tau_\alpha$ can be measured to the experimental regime, where this is unachievable.

We start by employing the Vogel-Fulcher-Tammann (VFT) law:
\begin{equation}
\label{eq:VFT}
\log (\tau_\alpha /\tau_0) \propto (T-T_{\rm VFT})^{-1}.
\end{equation}
where $\tau_0$ and $T_{\rm VFT}$ are fitting parameters. We fitted this function on our numerical data over the accessible time window and we concluded that it performs very badly when extrapolated at lower temperatures. We found for instance that the swap Monte Carlo algorithm easily thermalises at temperatures below the extrapolated VFT critical temperature $T_{\rm VFT}$, which invalidates directly its use to describe numerical data~\cite{NBC17}. The inability of the VFT law to describe experimental data over a wide range of temperature was discussed in detail in Refs.~\cite{stickel_dynamics_1995,blodgett2015proposal}.

It has been found in previous experimental studies that the parabolic law \begin{equation}
\tau_{\alpha}^{\rm para} = \tau_o \exp[ C (T_o/T-1)^2 ]
\label{eq:parabolic}
\end{equation}
fits accurately the data over a very large temperature range~\cite{elmatad2010corresponding,mauro2009viscosity}. Its fitting parameters are $\tau_o$, $C$, and $T_o$.

In addition to the VFT and parabolic laws, we consider two other functional forms, shown in Fig.~\ref{fig:tau_data}. One is a double exponential equation (MEYGEA) discussed in Refs.~\cite{mauro2009viscosity,elmatad2010corresponding}:
\begin{equation}
\tau_{\alpha} = \tau_0 \exp \left[ \frac{K}{T} \exp[C/T] \right],
\end{equation}
where $\tau_0$, $K$, and $C$ are the fitting parameters.
The other one is the Avramov and Milchev (AM) equation~\cite{avramov1988effect} given by
\begin{equation}
\tau_{\alpha} = \tau_0 \exp[A/T^n],
\end{equation}
where $\tau_0$, $A$, and $n$ (real exponent) are the fitting parameters.
All the fitting functions considered in this paper have three free-fitting parameters which is the minimal number to mathematically characterize non-Arrhenius behavior. 
Given the small variation of the apparent activation energy over the dynamic range studied experimentally, it is not surprising that several smooth functions of temperature can describe the evolution of $\log(\tau_\alpha)$. 
Figure~\ref{fig:tau_data} shows that different fitting functions produce slight variations in the extrapolated value for $T_g$. The key issue is therefore to choose the best fitting function, \textit{i.e.}, the one from which the low temperature data can be inferred accurately from the high temperature one.


To find the best fitting procedure, we train on experimental data (see Appendix~\ref{sec:extrapolation}). We  fit the above four equations to the data, restricting ourselves to a modest dynamic range, comparable to numerical timescales. We then extrapolate to temperatures close to $T_g$, and compare the extrapolation to the actual data. We find excellent agreement when using the parabolic law for the experimental data with kinetic fragility indexes similar to our numerical models, which validates further our procedure. Thus, we empirically find that fitting the parabolic law to the numerical time window provides an excellent description of the data close to $T_g$, as reported previously~\cite{elmatad2010corresponding}. This is a purely practical choice, and we make no assumption about the physical mechanism which could lead to such a law. 

By using the fitting parameter $\tau_o$ obtained from the parabolic law, we define two time windows. First we define the {\it simulation window} by $\tau_{\alpha}/\tau_o \in [10^0, 10^5]$. The upper bound of this timescale corresponds to recent simulation studies with very long timescales~\cite{ceiling17,coslovich2018dynamic}. The {\it experimental window} is defined by $\tau_{\alpha}/\tau_o \in [10^3, 10^{12}]$. The lower bound corresponds to a timescale around the mode-coupling crossover $T_{\rm mct}$ ($\tau_{\alpha} \simeq 10^{-7}$ s~\cite{novikov_universality_2003}), and the upper bound corresponds to the timescale at the experimental glass transition $T_{\rm g}$ ($\tau_{\alpha} \simeq 100$ s). The experimental window is therefore the appropriate regime to test the predictions made by the RFOT theory. 
Notice that in this paper, we neither try to go below $T_g$, nor to examine the fate of supercooled liquids at even lower temperature~\cite{royall2018race}. 


For numerical models, we determine the experimental glass transition temperature $T_{\rm g}$ as $\tau_{\alpha}^{\rm para}(T_{\rm g})/\tau_o = 10^{12}$.
The kinetic fragility index $m$ is determined by $m= \partial \log_{10} \tau_{\alpha}^{\rm para}/\partial (T_{\rm g}/T) |_{T= T_{\rm g}}$.
The fitting parameters and fragility indexes are given in Table~\ref{tab:para_sim}.

\subsection{Experimental data}

We select materials for which high-quality data for the configurational entropy and relaxation time over a broad temperature range is available in the literature. This allows for a comparison with computer simulations and an accurate determination of the exponent $\alpha$ in Eq.~(\ref{eq:GAG}).

We select 2-methyl tetrahydrofuran (2MTHF), ethylbenzene (ETB), ethanol, glycerol, {\it o}-terphenyl (OTP), 1-propanol, propylene carbonate (PC), salol, toluene and 3-bromopentane.
The configurational entropy data for 2MTHF, ETB, OTP, PC, salol, and toluene were recently obtained from accurate experiments by Tatsumi, Aso and Yamamuro. Some of the data is presented in Ref.~\cite{tatsumi12}. The data for 1-propanol is taken from Ref.~\cite{takahara1994heat}.
In these data for all the above materials, $S_{\rm conf}$ is measured by thermodynamic integration of the heat capacity difference between supercooled liquids and non-equilibrium glasses. This treatment should be conceptually better than using the crystal entropy~\cite{richert1998dynamics}, but this is still a rather crude approximation~\cite{yoshimori2011configurational}, whose accuracy is expected to be material-dependent~\cite{smith2017separating}.
For ethanol~\cite{takeda1999calorimetric,haida1977calorimetric}, glycerol~\cite{takeda1999calorimetric,gibson1923third}, and 3-bromopentane~\cite{takahara1995calorimetric}, $S_{\rm conf}$ is obtained using the crystal entropy $S_{\rm cry}$, {\it i.e.}, $S_{\rm conf}=S_{\rm liq}-S_{\rm cry}$. Notice that we do not seek to present thermodynamic data for ultrastable glasses prepared below $T_g$, even though we believe that these materials can be instrumental to test more precisely glass transition theories~\cite{tobemark}. 

The relaxation time data are mainly obtained from dielectric measurements, but some data are combined with other methods, such as viscosity measurements. The corresponding references are:
2MTHF~\cite{richert1998dynamics}, ETB~\cite{chen2011dynamics,barlow1966viscous,rossini1953selected}, ethanol~\cite{tatsumi12}, glycerol~\cite{schneider1998dielectric,lunkenheimer2000glassy,lunkenheimer2005glassy}, OTP~\cite{schmidtke2012boiling}, 1-propanol~\cite{hansen1997dynamics,richert1998dynamics,sillren2014liquid}, PC~\cite{schneider1999broadband,lunkenheimer2000glassy,lunkenheimer2005glassy}, salol~\cite{stickel1996dynamics}, toluene~\cite{schmidtke2012boiling}, and 3-bromopentane~\cite{berberian1986approach}.

For the experimental data, we set $\tau_o=10^{-10}$ s. Therefore the simulation and experimental time windows correspond to $\tau_{\alpha} \in [10^{-10}{~\rm s}, 10^{-5}{~\rm s}]$ and $\tau_{\alpha} \in [10^{-7}{~\rm s}, 10^{2}{~\rm s}]$, respectively. In particular, $T_g$ corresponds to the standard relaxation time $\tau_\alpha =100{~\rm s}$.

\begin{figure}
\includegraphics[width=0.85\columnwidth]{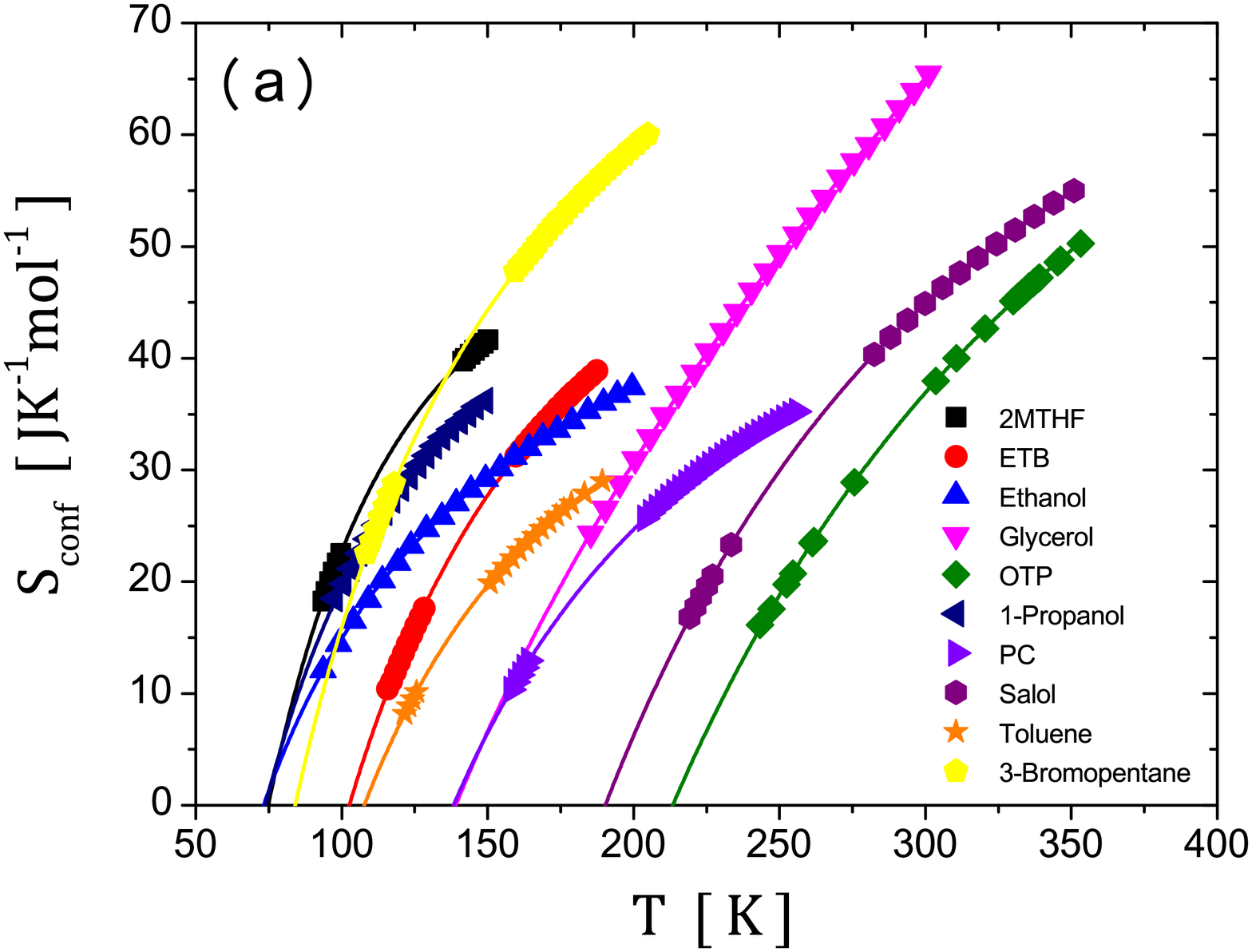}
\includegraphics[width=0.85\columnwidth]{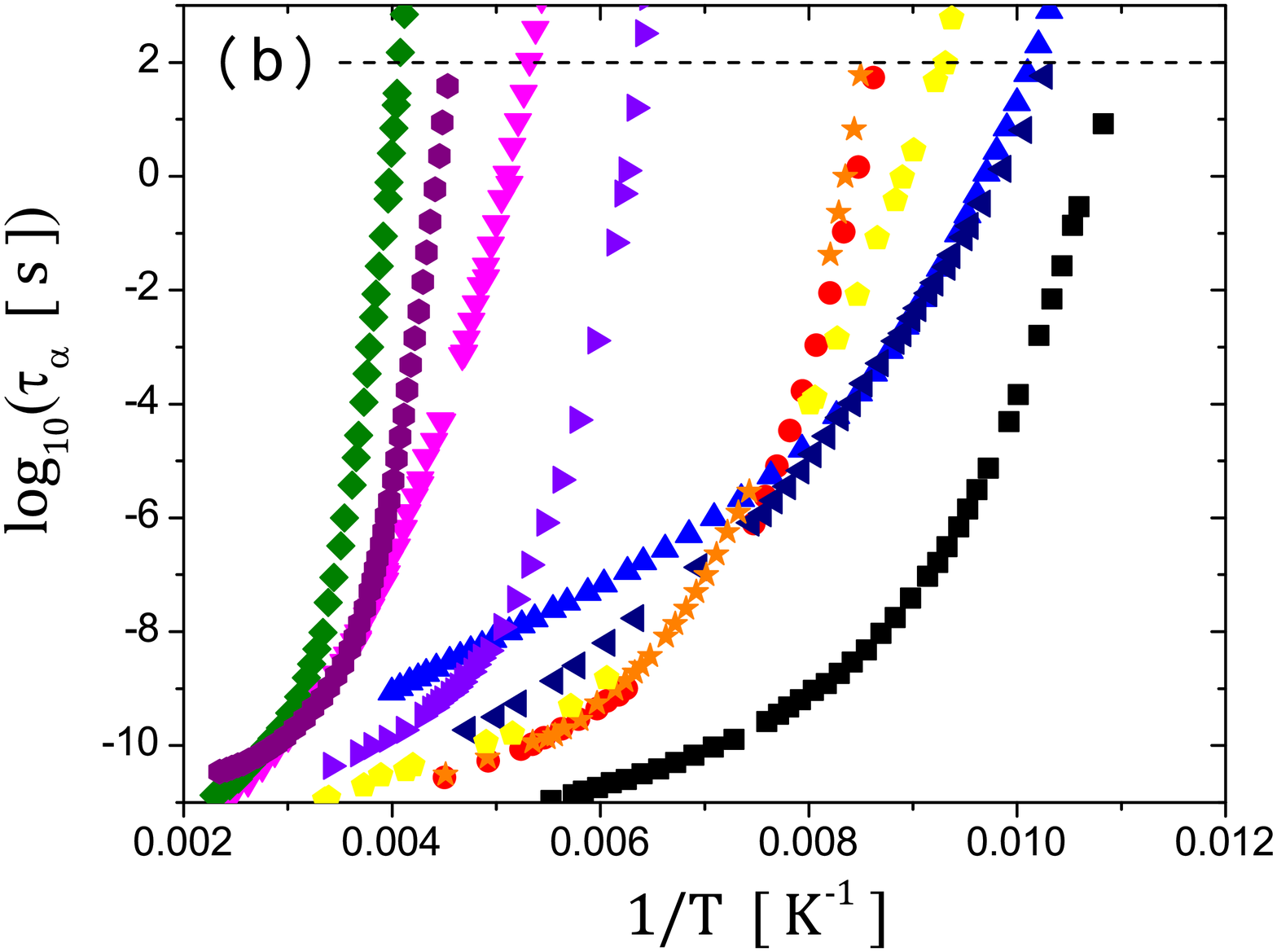}
\caption{
(a) Configurational entropy data for 2MTHF~\cite{tatsumi12}, ETB~\cite{tatsumi12}, ethanol~\cite{haida1977calorimetric}, glycerol~\cite{gibson1923third}, OTP~\cite{tatsumi_unpublished}, 1-propanol~\cite{takahara1994heat}, PC~\cite{tatsumi12}, salol~\cite{tatsumi_unpublished}, toluene~\cite{tatsumi12}, and 3-bromopentane~\cite{takahara1995calorimetric}. 
The solid curves are quadratic fitting functions, as used for the $d=3$ numerical models.
(b) Relaxation time data for 2MTHF~\cite{richert1998dynamics}, ETB~\cite{chen2011dynamics,barlow1966viscous,rossini1953selected}, ethanol~\cite{brand2000excess}, glycerol~\cite{schneider1998dielectric,lunkenheimer2000glassy,lunkenheimer2005glassy}, OTP~\cite{schmidtke2012boiling}, 1-propanol~\cite{hansen1997dynamics,richert1998dynamics,sillren2014liquid}, PC~\cite{schneider1999broadband,lunkenheimer2000glassy,lunkenheimer2005glassy}, salol~\cite{stickel1996dynamics}, toluene~\cite{schmidtke2012boiling}, and 3-bromopentane~\cite{berberian1986approach}.
The horizontal dashed line indicates the timescale of the experimental glass transition, $\tau_{\alpha}=100$ s.
}
\label{fig:experiments}
\end{figure}

The configurational entropy and relaxation time data for the materials presented above are gathered in Fig.~\ref{fig:experiments}, together with empirical quadratic fits to the configurational entropy.  

\section{Results}
\label{sec:Results}

In this section, we perform a test of Eqs.~(\ref{eq:AG}, \ref{eq:RFOT2}, \ref{eq:RFOT1}, \ref{eq:GAG}, \ref{eq:alpha}) using the experimental and numerical data presented in Sec.~\ref{sec:Methods}. 
We first study Eq.~(\ref{eq:RFOT2}) using numerical data for
$\xi_{\rm pts}$ and $S_{\rm conf}$ to estimate $\theta$. Then, 
we estimate $\alpha$ in Eq.~({\ref{eq:GAG}) by comparing $\tau_{\alpha}$ and $S_{\rm conf}$ using both computer simulations and experiments to investigate the validity of the Adam-Gibbs relation in Eq.~(\ref{eq:AG}). Finally, the values of $\theta$ and $\alpha$ allow us to discuss that taken by $\psi=(d-\theta) \alpha$, deduced from Eq.~(\ref{eq:alpha}).

\subsection{The static exponent $\theta$}

First we estimate the exponent $\theta$ in Eq.~(\ref{eq:RFOT2}) combining independent data obtained for $S_{\rm conf}$ and $\xi_{\rm pts}$.

\begin{figure}
\includegraphics[width=0.85\columnwidth]{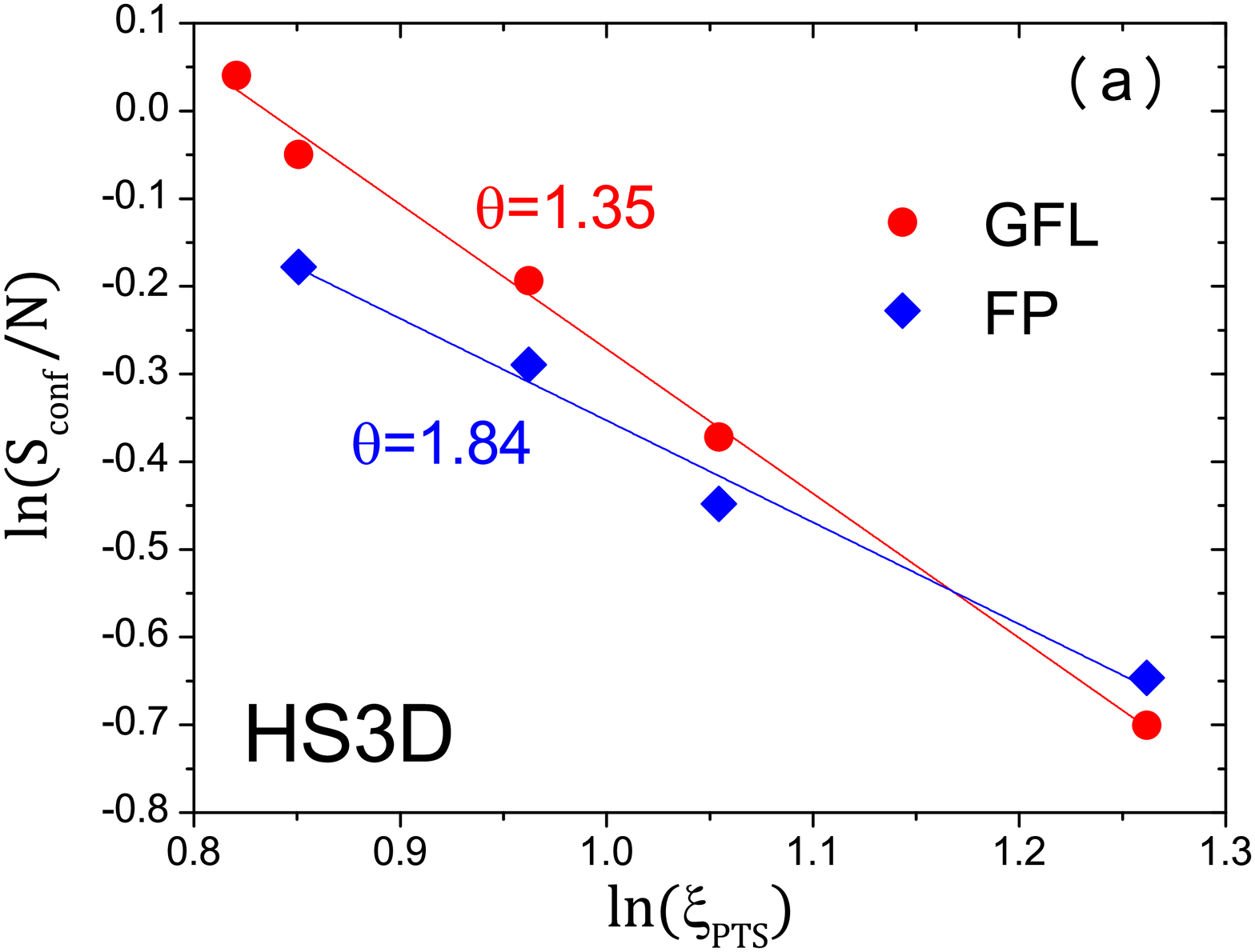}
\includegraphics[width=0.85\columnwidth]{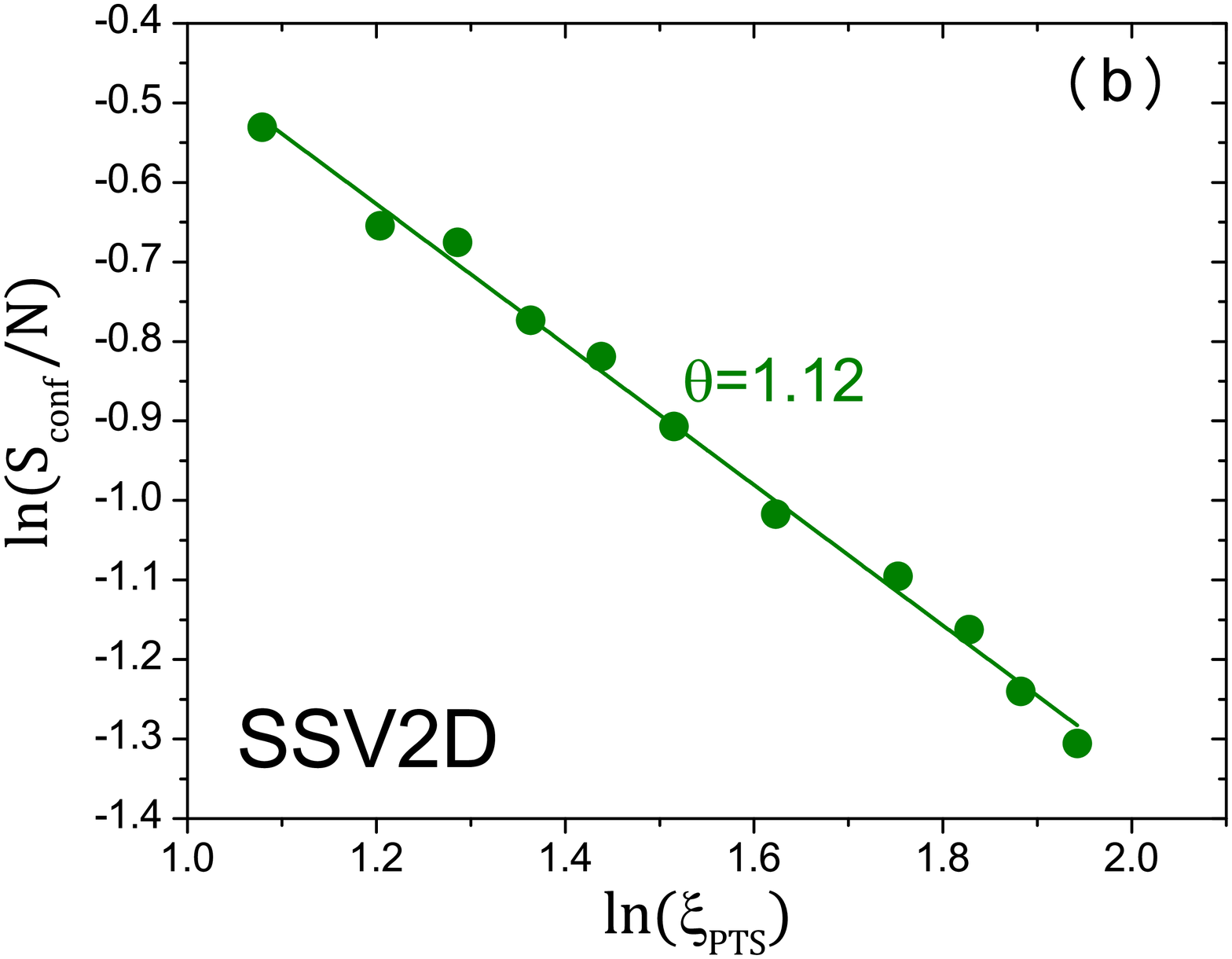}
\caption{
$S_{\rm conf}$ vs. $\xi_{\rm pts}$ plot in $d=3$ hard spheres (HS3D) (a) and $d=2$ soft disks (SSV2D) (b).
The straight lines are power law fits.
For HS3D, we show two independent estimates of $S_{\rm conf}$ obtained from: the generalized Frenkel-Ladd (GFL) method and the Franz-Parisi (FP) free energy approach. 
}
\label{fig:theta}
\end{figure}  

Figure~\ref{fig:theta} shows a log-log plot of $S_{\rm conf}$ versus 
$\xi_{\rm pts}$ for three dimensional polydisperse hard spheres (HS3D) (a) and two dimensional soft disks (SSV2D) (b).
\MO{We use the fitted functional form for $S_{\rm conf}$ (obtained in Fig.~\ref{fig:s_conf_data}), whereas the actual data points are used for $\xi_{\rm pts}$.}
We emphasize that while temperature is a running parameter in this plot, the data point in Fig.~\ref{fig:theta} correspond to the regime of interest $T<T_{\rm mct}$. Such results have never been achieved, as earlier numerical work were all performed for $T > T_{\rm mct}$, or only slightly below $T_{\rm mct}$~\cite{cammarota2009numerical}. Despite the larger temperature range explored in this work, we are fully aware the relative variation of $\xi_{\rm pts}$ and $S_{\rm conf}$ remain fairly modest, which makes the determination of a critical exponent quite difficult.    

For HS3D, we report two estimates for $S_{\rm conf}$, obtained from different schemes. One is a generalized Frenkel-Ladd (GFL) method~\cite{Frenkelbook,ozawa2018configurational}, and the other is the Franz-Parisi (FP) free energy method proposed earlier~\cite{FP97,BC14,ceiling17}. The exponent 
$\theta$ is extracted by fits to straight lines, whose slope gives $\theta-d$, see Eq.~(\ref{eq:RFOT2}). We obtain $\theta = 1.35 \pm 0.06$ for GFL and $\theta \simeq 1.84 \pm 0.09$ for FP.
These values are compatible with either the theoretical prediction $\theta=d/2$ by Kirkpatrick {\it et al.}~\cite{KTW89}, or with that of Franz $\theta = d-1$~\cite{effort2}.
 
We obtain $\theta=1.12  \pm 0.02$ for SSV2D. 
This value is close to both theoretical predictions, $\theta=d/2$ and $\theta=d-1$, which coincide in $d=2$, giving $\theta=1$ . Obviously, one cannot discriminate between the two predictions. 

Overall, we find that for $d=3$ the value measured for $\theta$ conforms with the two available predictions, which is an encouraging result from the viewpoint of RFOT theory. Unfortunately, the obtained values fall in-between the two predictions, which are too close to be discriminated. We suggest that performing point-to-set and configurational entropy measurements in $d=4$, combining recently developed tools~\cite{berthier2016efficient,kundu,ozawa2018configurational}, would be very useful to conclude on this point. Indeed, when  $d=4$, the two predictions yield $\theta=d/2=2$ and $\theta=d-1=3$, which are further appart than in $d=3$.

\subsection{Breakdown of the Adam-Gibbs relation and numerical estimation of $\alpha$}

\begin{figure}
\includegraphics[width=0.48\columnwidth]{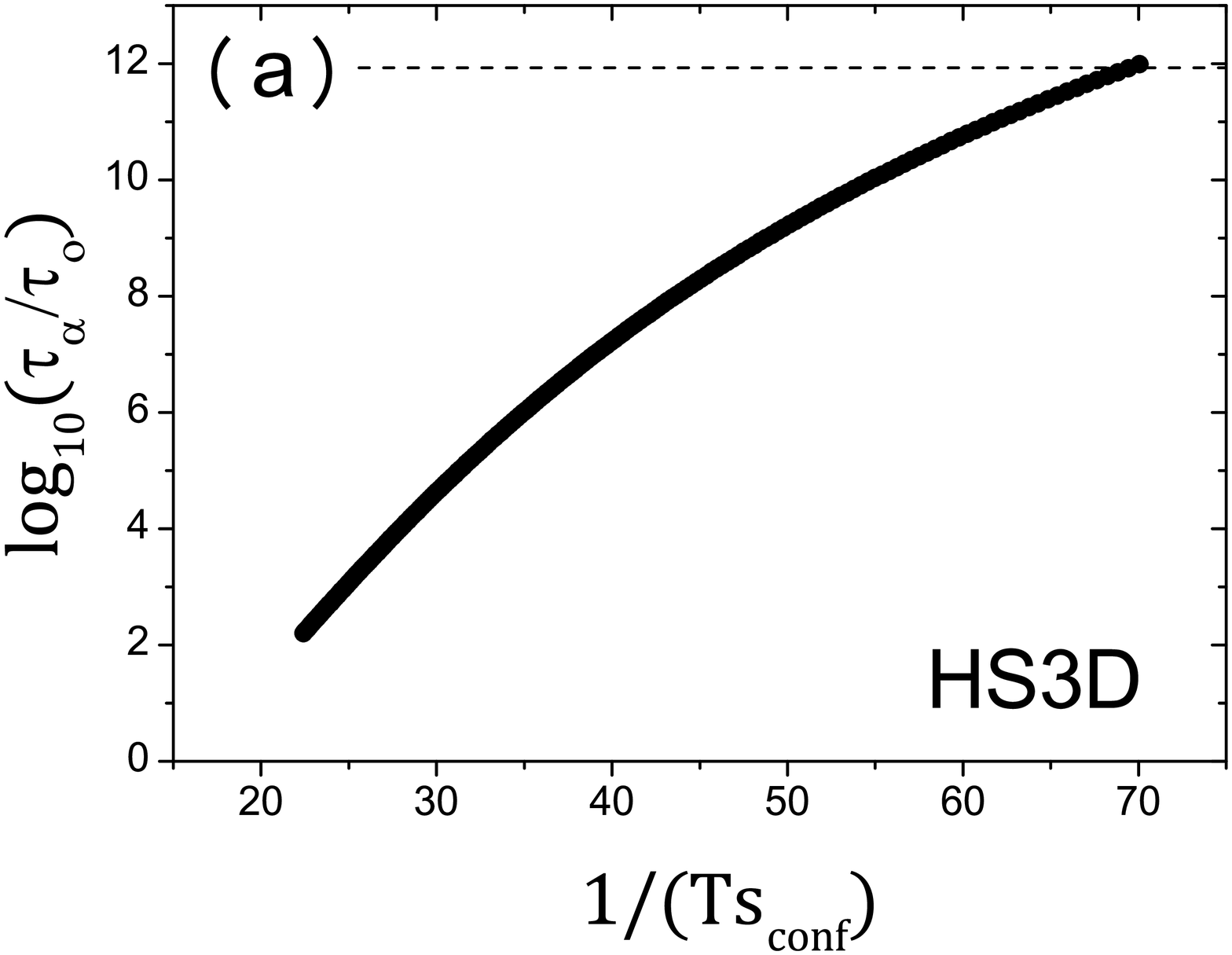}
\includegraphics[width=0.48\columnwidth]{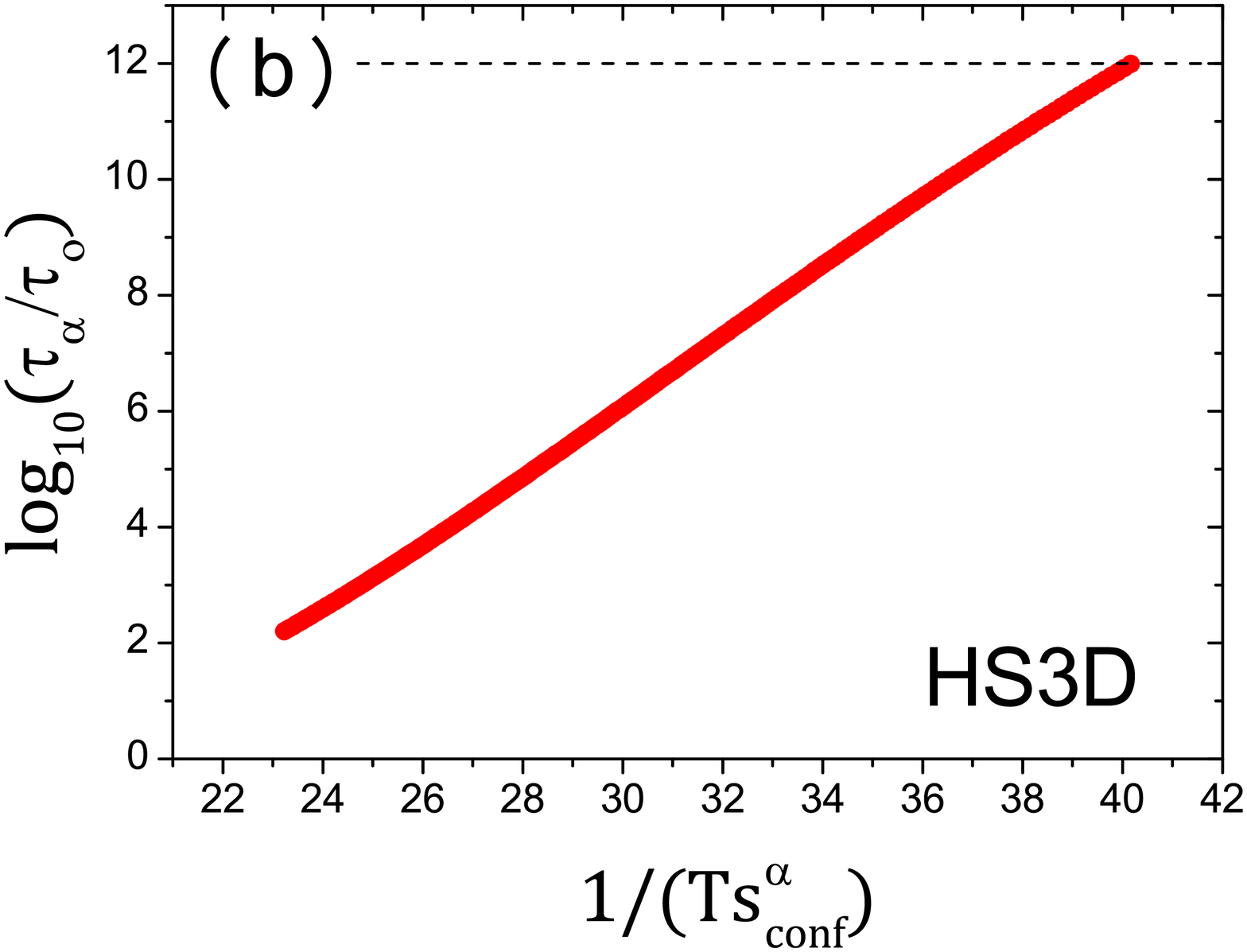}
\includegraphics[width=0.48\columnwidth]{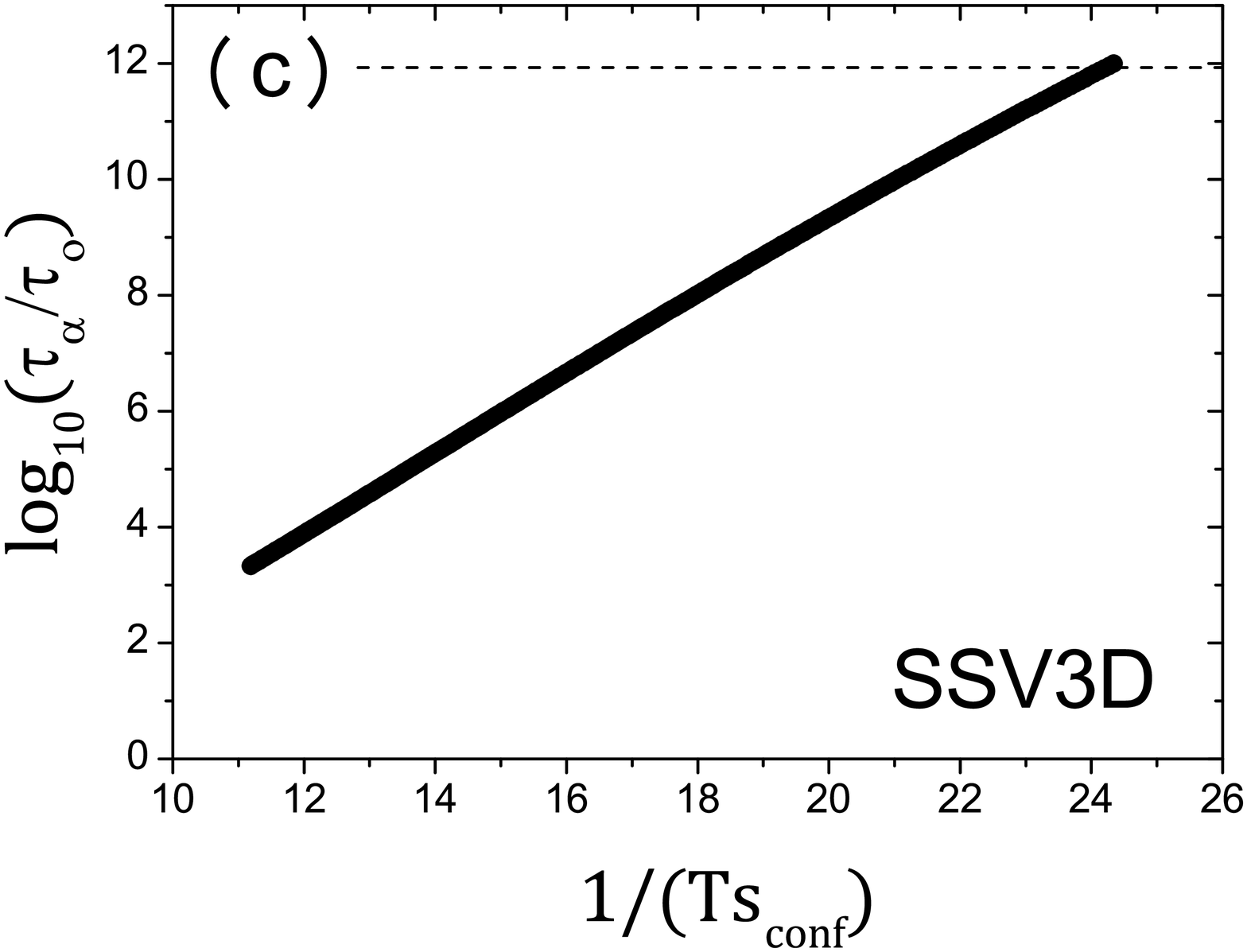}
\includegraphics[width=0.48\columnwidth]{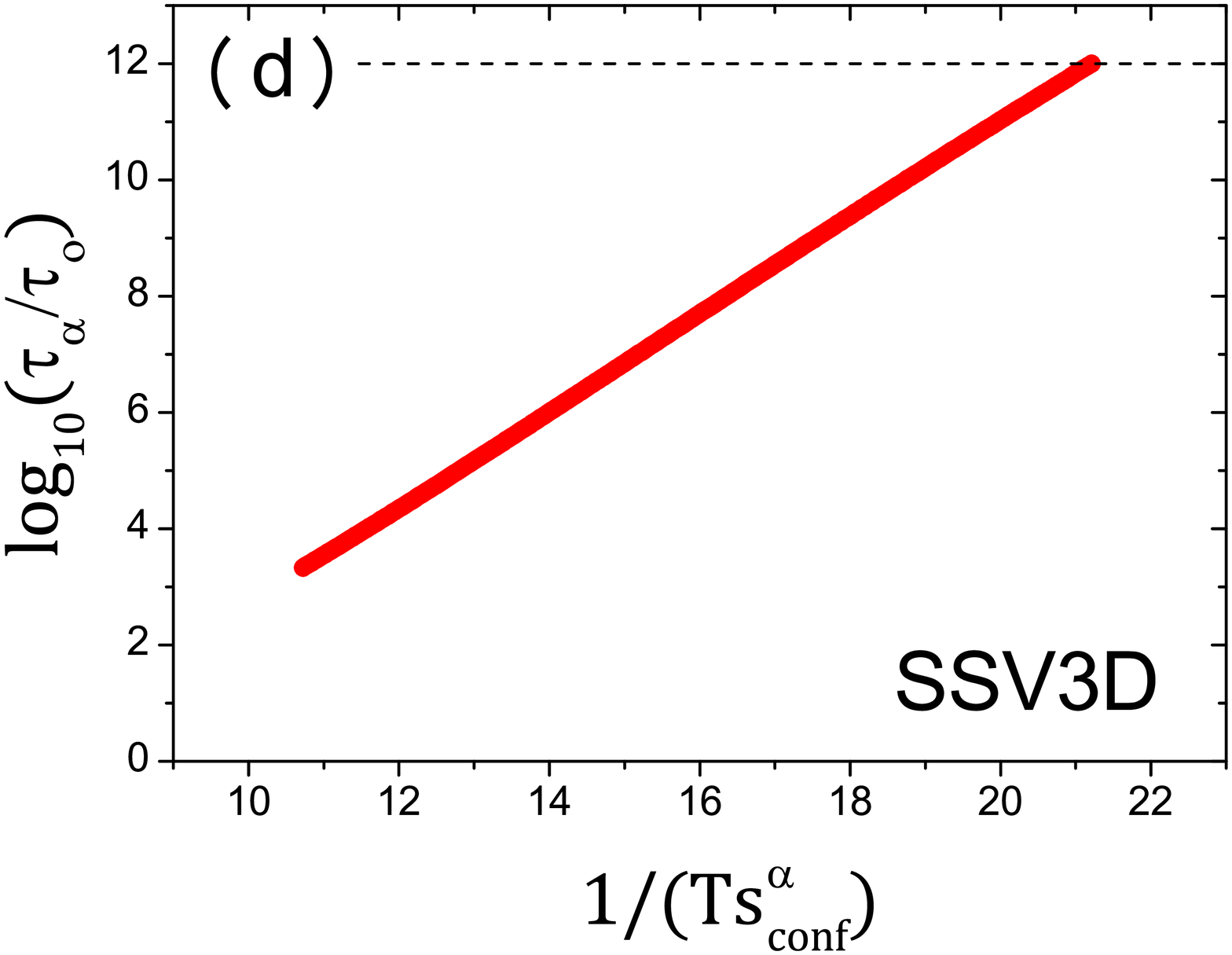}
\includegraphics[width=0.48\columnwidth]{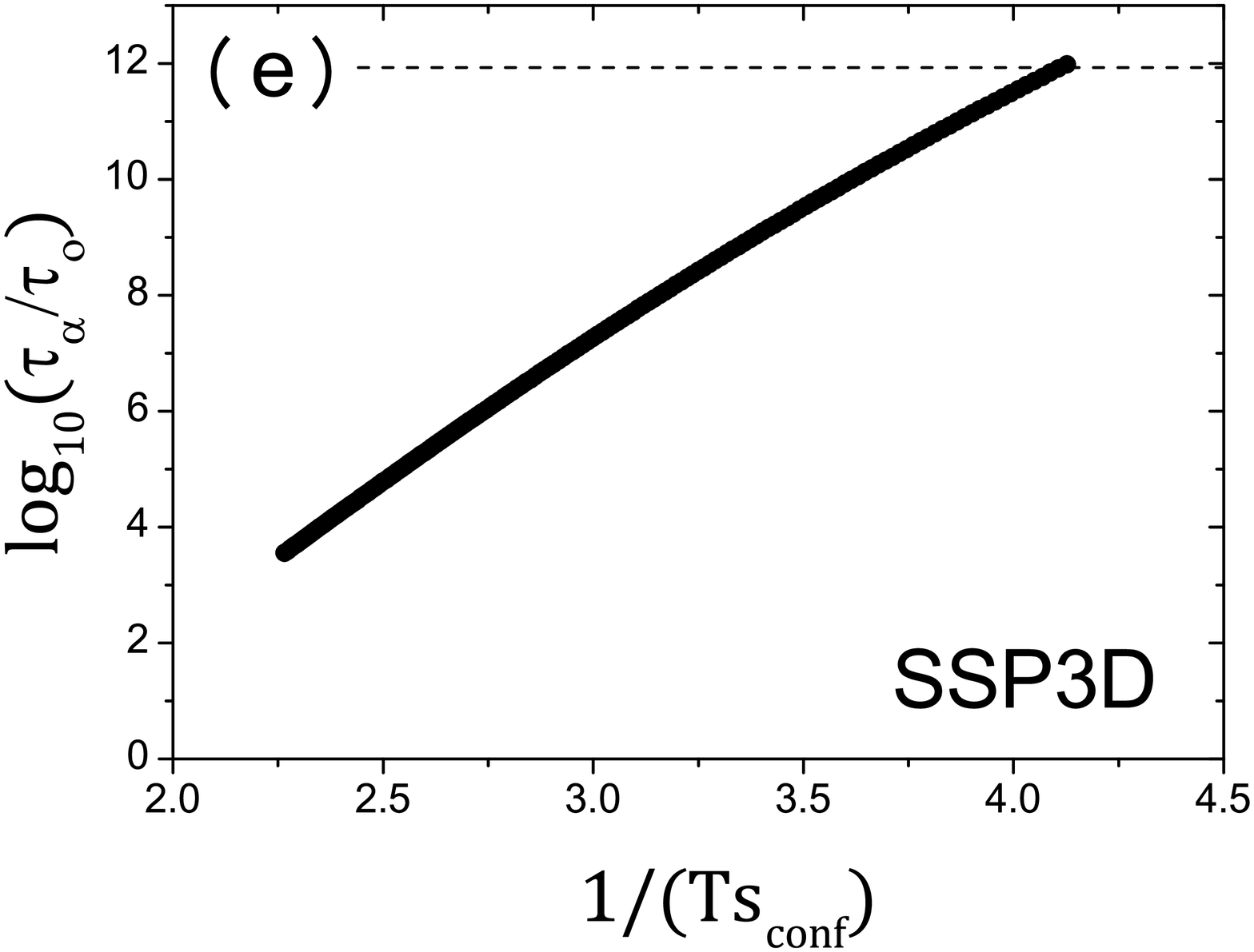}
\includegraphics[width=0.48\columnwidth]{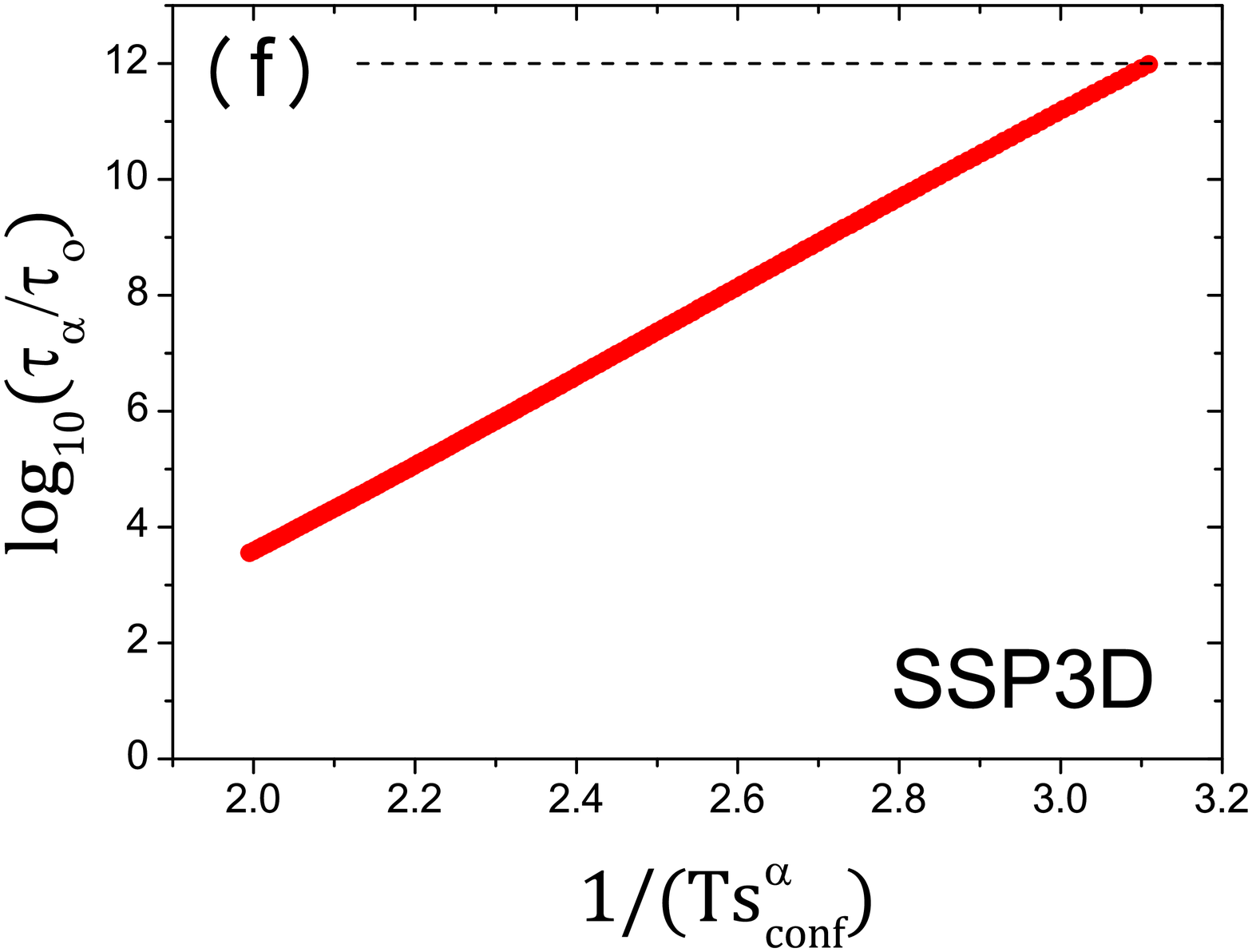}
\includegraphics[width=0.48\columnwidth]{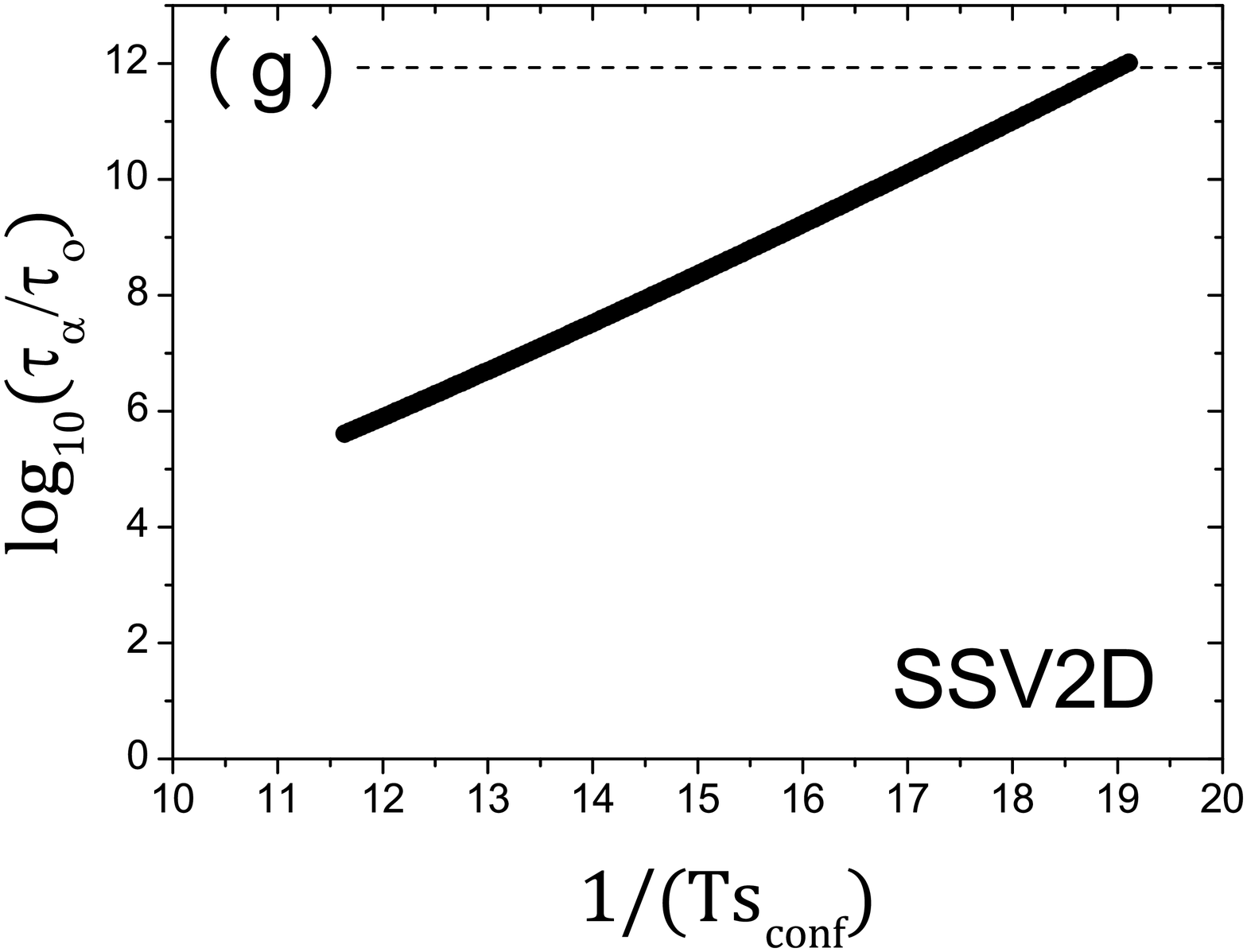}
\includegraphics[width=0.48\columnwidth]{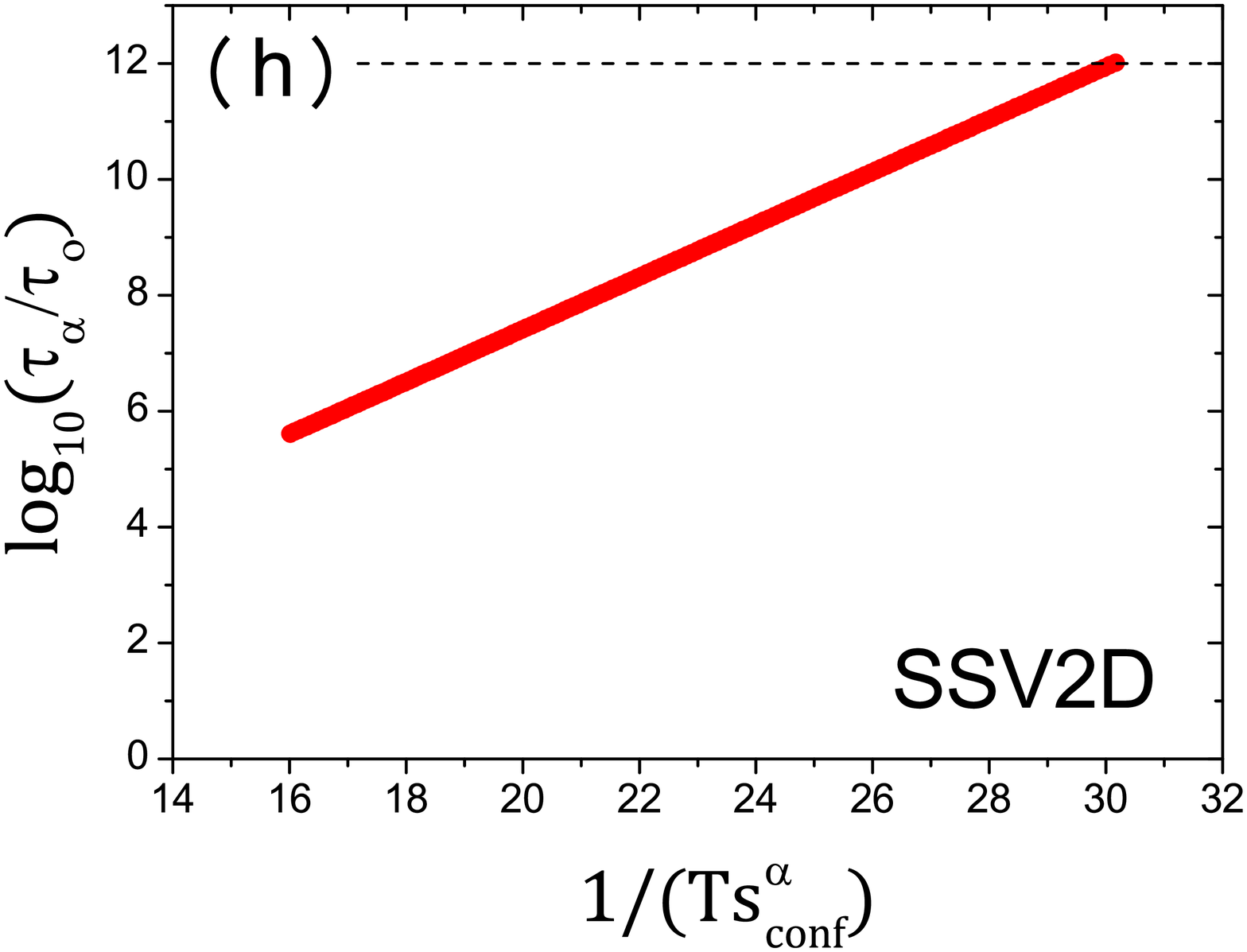}
\caption{Left panels: Standard Adam-Gibbs plot for the $d=3$ hard spheres (HS3D) (a), $d=3$ soft spheres along the isochoric path (SSV3D) (b), the isobaric path (SSP3D) (c), and $d=2$ soft disks (SSV2D) (d).
Right panels: Generalized Adam Gibbs plots with the fitted $\alpha$ value for  each model for HS3D (b), SSV3D (d), SSP3D (f), and SSV2D (h). The horizontal dashed lines correspond to the timescale for the experimental glass transition $T_{\rm g}$.}
\label{fig:GAG_sim}
\end{figure}  

We next examine the validity of Eq.~(\ref{eq:GAG}) by connecting $\tau_{\alpha}$ and $S_{\rm conf}$, and estimating the exponent $\alpha$. When $\alpha=1$, the Adam-Gibbs relation in Eq.~(\ref{eq:AG}) is recovered. 

In Fig.~\ref{fig:GAG_sim}(a,c,e,g) we show conventional Adam-Gibbs plots where the evolution of $\log_{10}(\tau_{\alpha}/\tau_o)$ is represented as a function of $1/(Ts_{\rm conf})$, where $s_{\rm conf}=S_{\rm conf}/N$, for hard spheres (HS3D) (a), soft spheres along the isochoric path (SSV3D) (c), along the isobaric path (SSP3D) (e), and the soft disks (SSV2D) (g). We combine the dynamic and thermodynamic data described in Sec.~\ref{sec:Methods}, restricted to the experimental time window ($\tau_{\alpha}/\tau_o \in [10^3 - 10^{12}]$). 
\MO{We use the fitted functional forms for both $\tau_{\alpha}$ (estimated in Fig.~\ref{fig:tau_data}) and $S_{\rm conf}$ (obtained in Fig.~\ref{fig:s_conf_data}), which produces ``continuous curves'' instead of a discrete data points.}
To our knowledge, this is the first time that the Adam-Gibbs relation is tested for computer models over the time window where it is actually supposed to apply. 

For all three-dimensional models, we find that $\log_{10} (\tau_{\alpha}/\tau_o)$ is a concave function of $1/T s_{\rm conf}$, whereas it is convex for the two-dimensional model. If tested over a narrow time window close to $T_{\rm mct}$, an acceptable linear behaviour could possibly be observed, that would suggest the validity of the Adam-Gibbs relation, in agreement with many earlier findings~\cite{sastry2001relationship,mossa2002dynamics,sciortinoPEL,saika2001fragile,foffi2005,sengupta2012adam,starr2013relationship,parmar2017length,handle2018adam}. The trend that we report here appears to contrast with recent results obtained in the Kob-Andersen model, where slight convexity and concavity are respectively observed in $d=3$~\cite{parmar2017length} and $d=2$~\cite{sengupta2012adam}. These results were however obtained in the numerical time window, above $T_{\rm mct}$. 
Our results demonstrate that when observed over a much broader range, and closer to $T_g$, the Adam-Gibbs relation is actually not obeyed for any of the numerical models studied here. 

The clear violations of the standard Adam-Gibbs relation that we find over the experimental time window imply that the exponent $\alpha$ must deviate from the value $\alpha=1$. We varied its value around unity and used it as a free parameter to obtain generalised Adam-Gibbs plots, which are shown in Fig.~\ref{fig:GAG_sim}(b,d,f,h) for the same numerical models. All plots now show a perfect straight line, suggesting that the introduction of the parameter $\alpha$ is sufficient to describe the data.
We obtain $\alpha=0.24$, $0.49$, $0.72$, and $1.89$, for HS3D, SSP3D, SSV3D, and SSV2D, respectively, so that $\alpha < 1$ for the three dimensional models, whereas $\alpha>1$ for the two dimensional model.

Since the four models we have simulated all display violations of the Adam-Gibbs relation, we conclude that Eq.~(\ref{eq:AG}) does not describe well the physics of simulated supercooled liquids when analysed over the experimental time window. Additional models should be studied and analysed before concluding about the possible universality of the exponent $\alpha$, but our initial results do not point towards a constant value. Once more, it would be very valuable to obtain data in $d=4$ to see if a different value for $\alpha$ is found in larger spatial dimensions. 

\subsection{Breakdown of the Adam-Gibbs relation and experimental estimation of $\alpha$}

Before starting this study, we felt that there was a general consensus in the community that the Adam-Gibbs relation is well-obeyed in real materials analysed near the experimental glass transition $T_g$. Thus, the outcome of the computer simulations showing deviations from Eq.~(\ref{eq:AG}) appeared as a worrying disagreement between simulations and experiments. 

Therefore, we decided to collect data sets for several molecular liquids, where high-precision dynamic and thermodynamic data would be available over both simulation and experimental time windows, in order to perform a direct comparison with computer models. 

\begin{figure}
\includegraphics[width=0.95\columnwidth]{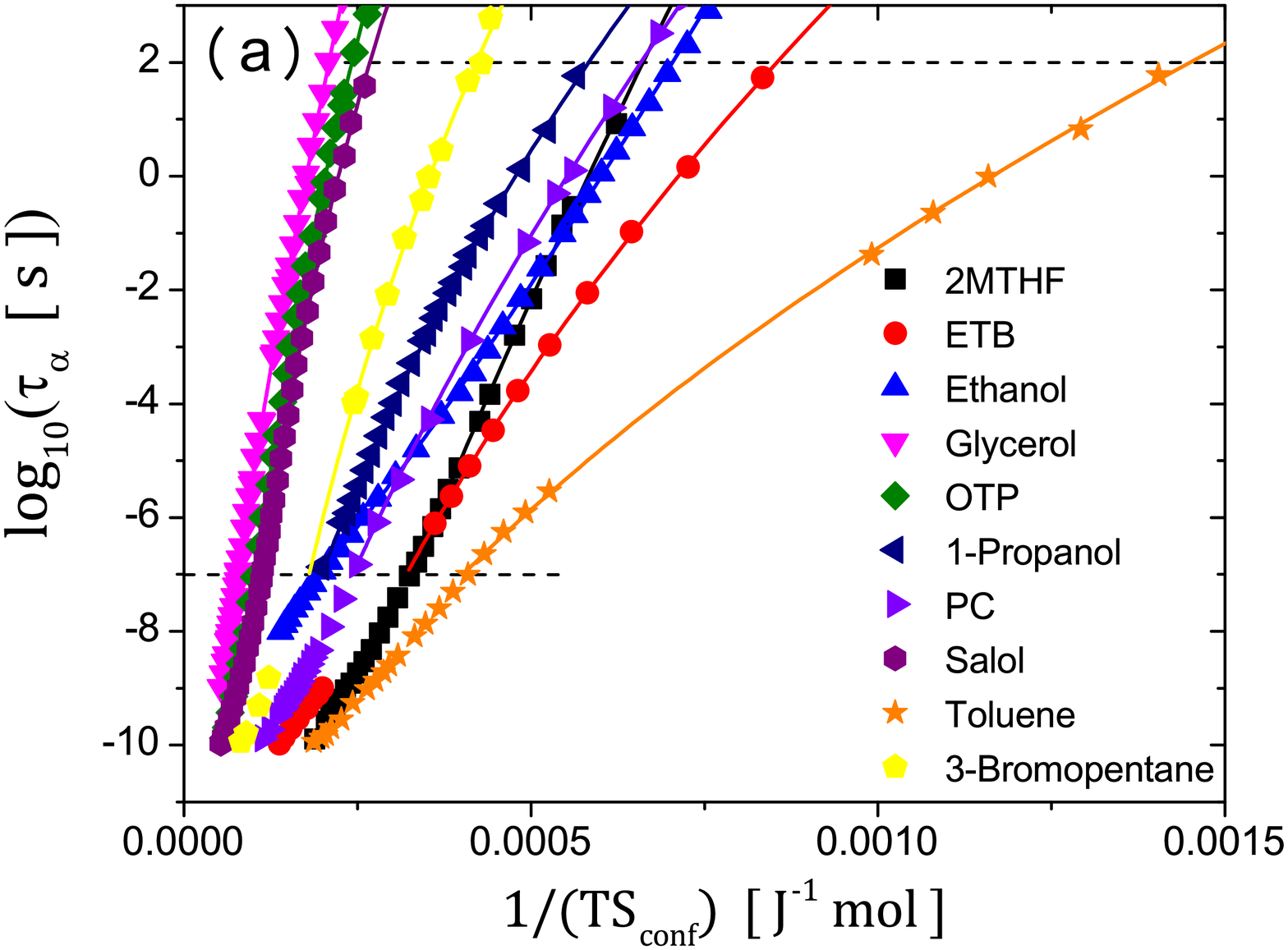}
\includegraphics[width=0.95\columnwidth]{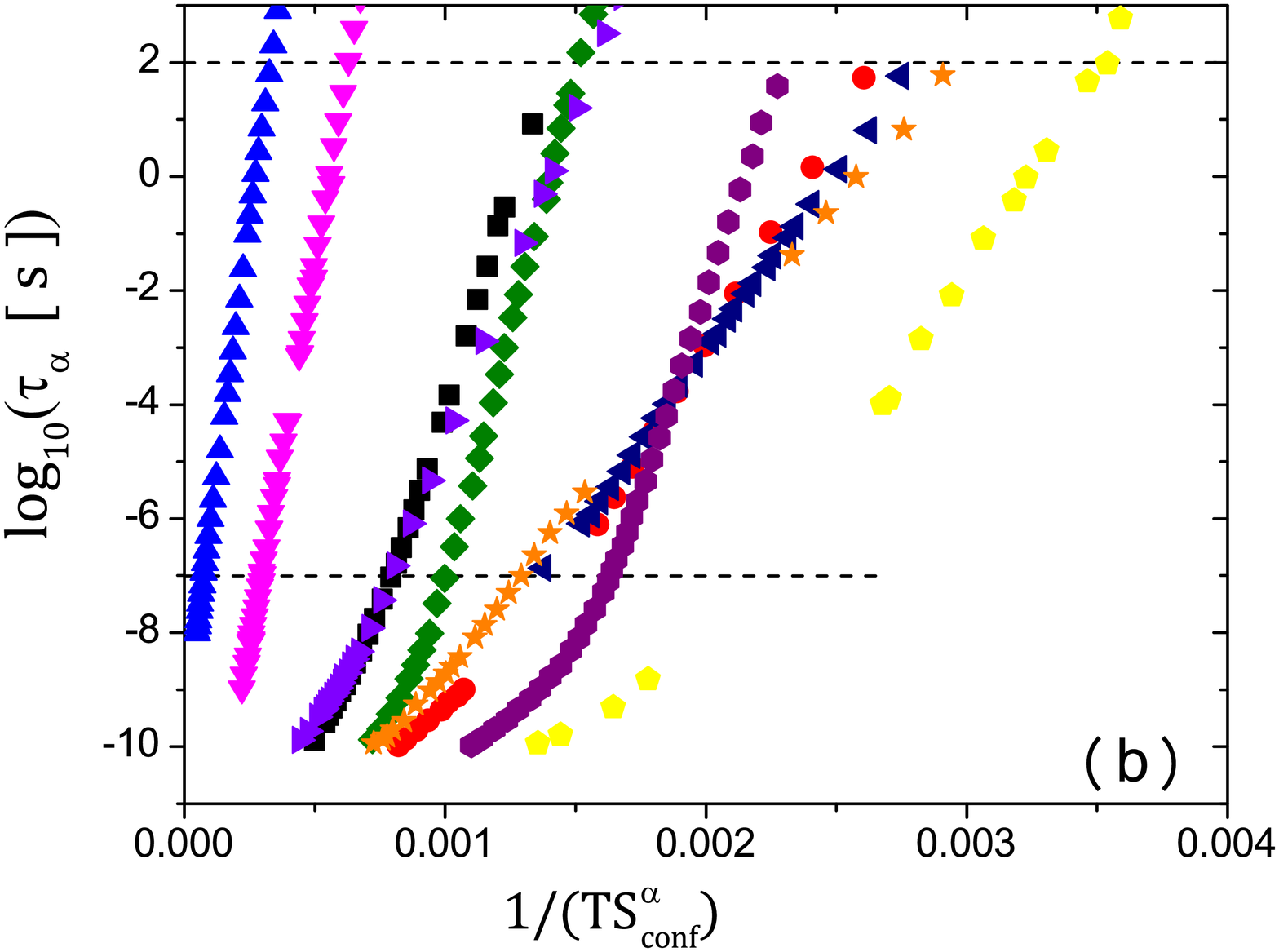}
\caption{(a) Standard Adam-Gibbs plot constructed from experimental data all except ethanol display a concave behaviour. The solid curves correspond to fits using Eq.~(\ref{eq:GAG}) using $\alpha$ as a fit parameter over the experimental time window. (b) Generalized Adam Gibbs plots with the fitted $\alpha$ value for each material. The horizontal dashed lines indicate the timescale of the experimental glass transition, $\tau_{\alpha}=100$ s, and the lower bound of the experimental time window, $\tau_{\alpha}=10^{-7}$ s.}
\label{fig:AG_Exp}
\end{figure}  

We present the results of our data collection in Fig.~\ref{fig:AG_Exp}(a) using again the representation where the standard Adam-Gibbs relation would yield a straight line. 
\MO{We use the fitted functional form for $S_{\rm conf}$ (obtained in Fig.~\ref{fig:experiments}(a)), whereas the actual data points are used for $\tau_{\alpha}$.}
When analysed over the entire experimental time window, defined above, we again observe a clear concavity for most materials. The Adam-Gibbs relation in Eq.~(\ref{eq:AG}) is violated over this regime, although of course it holds if observed over a restricted time window close to $T_g$~\cite{richert1998dynamics} (almost by definition--the data is continuous!).  

As for the simulations, we fit the experimental data using the exponent $\alpha$ as an additional free parameter. From the experimental data, we determine two distinct values for $\alpha$, obtained by fitting either over the simulation or the experimental time window. The typical trend that we observe is that $\alpha > 1$ over the simulation time window, but $\alpha < 1$ over the experimental time window. The latter fits are included in Fig.~\ref{fig:AG_Exp}(a), and they describe well the data over the entire experimental time window. 
As with the case for the simulation models, in Fig.~\ref{fig:AG_Exp}(b), we also present the generalised Adam-Gibbs plot with the fitted $\alpha$ value for each material in the experimental time window.
We confirm that the linear behavior is recovered in this plot.

We notice that the concavity in the Adam-Gibbs plot in the experimental time window was already reported~\cite{ngai1999modification,roland2004adam}.
However, the concavity would be overlooked as it is less pronounced than the convexity found at much higher temperature, close to $T_{\rm mct}$ and above~\cite{ngai1999modification}. Moreover, Ref.~\cite{roland2004adam} concluded that the observed concavity was attributed to an imprecise estimate of the configurational entropy. Our results obtained from simulation data with accurate configurational entropy measurements and recent high-quality experimental data suggest instead that the observed concavity is a generic physical phenomenon reflecting the nature of glassy dynamics over the experimental time window.

\section{Discussion}
\label{sec:discussion}

Our central conclusion from both simulations and experiments considered over a broad time regime $\tau_\alpha / \tau_o \in [10^3, 10^{12}]$ (defined to be both experimentally accessible and theoretically relevant) is that the conventional Adam-Gibbs relation in Eq.~(\ref{eq:AG}) is not obeyed. Instead, the general form predicted by RFOT theory in Eq.~(\ref{eq:GAG}) describes numerical and experimental data well. This is maybe not so surprising, from an empirical viewpoint, given that the generalised relation has one more free fitting parameter. 

\begin{figure}
\includegraphics[width=0.95\columnwidth]{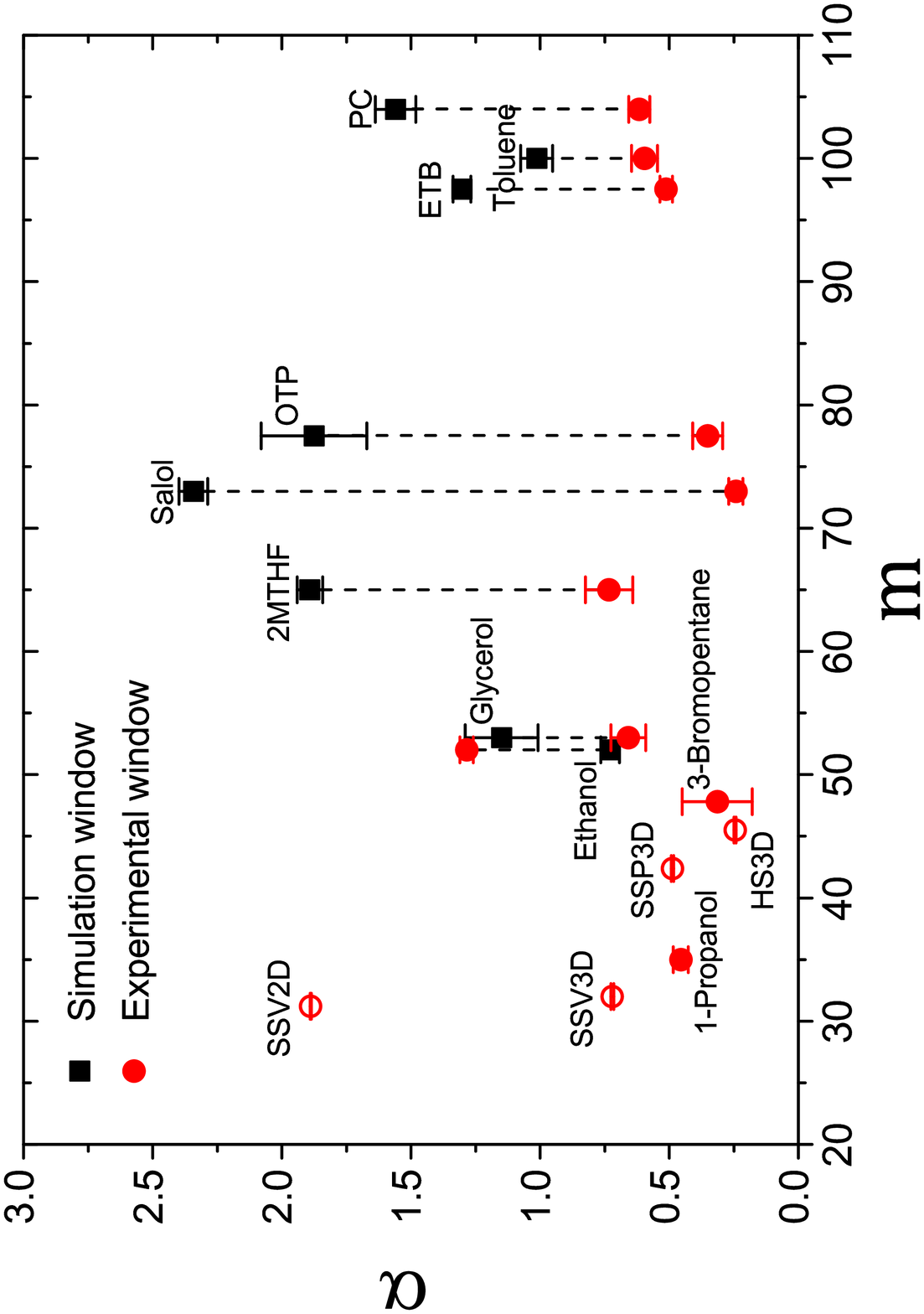}
\caption{The measured values of $\alpha$ presented, for convenience, as a function of the kinetic fragility index $m$ for the simulation (black squares) and experimental (red circles) time windows for various materials. 
The errorbars correspond to the standard error for the fit.
The vertical dashed lines connects the two values for each material.
Empty circles correspond to simulation models where $\alpha$ is measured over the experimental time window only.}
\label{fig:alpha_plot}
\end{figure}

We compile all our results for the values of $\alpha$ from simulations (empty points) and experiments (filled points) in Fig.~\ref{fig:alpha_plot}. To organise the data, we use the kinetic fragility index $m$ as the horizontal axis. This is simply a matter of convenience (as a matter of fact, no strong trend is observed). Note that, somewhat paradoxically, we do not have values for $\alpha$ in the computer models over the simulation time window because our computational schemes to measure $S_{\rm conf}$ only become applicable for low enough temperatures, typically $T \lesssim T_{\rm mct}$~\cite{BC14,ozawa2018configurational}. 

The experimental data in Fig.~\ref{fig:alpha_plot} obtained by considering the simulation time window are dispersed, $\alpha = 0.73 - 2.34$, and tend to be characterised by rather large values $\alpha>1$. By contrast, considering a broader and physically better justified experimental time window, data for both simulations and experiments are much less scattered, $\alpha \simeq 0.24 - 1.28$, with a preferred average value $\alpha \simeq 0.5 - 0.6$, except for ethanol. 

Before concluding, we make a further caveat regarding the above analysis of the RFOT theory predictions. In principle, we could have introduced additional subdominant physical prefactors into the scaling relations in Eqs.~(\ref{eq:RFOT2}, \ref{eq:RFOT1}) that could also be temperature dependent quantities. 
In particular, a surface tension could enter the relation between $\SC$ and $\xi_{\rm pts}$~\cite{BB09,lubchenko2014mechanism}, and an energy scale could enter the activated scaling relation in Eq.~(\ref{eq:RFOT1}). These prefactors would become irrelevant if some asymptotic regime could be reached with extremely long relaxation times and very small configurational entropy values, but it is understood that experimental glasses are not in this regime~\cite{Ta11}.     
In the absence of strong theoretical insights into these quantities, we decided to ignore them. They could of course very well affect the measured values of the reported exponents.  
Thus, a better determination of these quantities is an important research goal~\cite{cammarota2009numerical,cammarota2009evidence,ganapathi2018measurements}, in particular in the experimental time window.

\MO{
We also discuss potential sources of uncertainty in terms of experimental measurements of $\tau_{\alpha}$ and $S_{\rm conf}$, whose accuracy would affect the determination of the scaling exponent $\alpha$.
Regarding $\tau_{\alpha}$, we note that the dielectric relaxation measurement for ethanol involves a Debye relaxation process which is distinct from the structural $\alpha$ relaxation process, as recently clarified in an experiment~\cite{chua2017dynamics}. 
Indeed, the relaxation time extracted from the main peak that we used in this paper~\cite{brand2000excess} corresponds to the former process in ethanol whereas instead we should use the $\alpha$ relaxation process, but it is also found that the overall temperature dependence of two relaxation processes are very similar~\cite{chua2017dynamics}. 
It could be that the unusual behavior in ethanol, showing $\alpha>1$, is  related to this issue.
}

\MO{
The experimental measurement of $S_{\rm conf}$ also involves approximations.
First, using the excess entropy, $S_{\rm exe}=S_{\rm liq}-S_{\rm cry}$, instead of $S_{\rm conf}=S_{\rm liq}-S_{\rm glass}$ is an approximation, in general. 
The validity of $S_{\rm glass} \approx S_{\rm cry}$ has been widely studied~\cite{gold76,yamamuro98,J00,martinez2001thermodynamic,angell2002specific,smith2017separating,alvarez2019vibrational,han2019structural} and its validity seems to be non-universal~\cite{smith2017separating}. Typically, $S_{\rm glass}$ is determined from the heat capacity of the non-equilibrium glass state, and so it still involves some approximations compared to its theoretical definition~\cite{yoshimori2011configurational}.}
\MO{
Second, the measurements of $S_{\rm conf}$ in Ref.~\onlinecite{tatsumi12} were performed by doping the different materials to avoid crystallization. In particular,  measurements for toluene and ETB involve $10$ wt \% doping with benzene. Therefore, mixing effects possibly contribute to the absolute value of $S_{\rm conf}$, and to its temperature dependence~\cite{OB17}.}

To summarize our results in terms of numerical values for the critical exponents introduced within RFOT theory, we observe in 
$d=3$ that the combination $\theta \simeq 3/2$ and $\alpha \simeq 0.5-0.6$ works well, which would then result in $\psi$ falling in the range $\psi \simeq 0.75-0.90$. If we use instead the value $\theta=2$, we would obtain a somewhat larger value for the dynamic exponent $\psi \simeq 1.0 - 1.2$, which agrees well with earlier indirect analysis~\cite{capaccioli2008dynamically,brun2012evidence}.
Both values violate the general bound $\psi \geq \theta$ discussed
in the context of spin glasses~\cite{fisher1988nonequilibrium}, the equality $\psi = \theta$ found for the random field Ising model~\cite{psirfim}, and the prediction $\psi = \theta = d/2$ in Ref.~\cite{KTW89}. In the absence of stronger theoretical constraints, we tentatively conclude that the measured $\psi$ value that we observe appears somewhat small, {\it i.e.}, smaller than all known theoretical predictions. In $d=2$, we get $\theta \simeq 1.1$ and $\alpha \simeq 1.9$, which in turns implies that $\psi \simeq 1.7$, which appears somewhat large, by contrast with $d=3$.   

Our conclusion that $\alpha<1$ is favored by the data over the experimental time window sheds some new light on an old debate in the glass literature~\cite{tanaka2003relation,LW07,hecksher2008little,elmatad2010corresponding}. Assuming the existence of an ideal glass transition at equilibrium where $\SC \to 0$ and $\tau_\alpha \to \infty$, one is naturally led to the determination of two critical temperatures: the Kauzmann temperature $T_K$ where $\SC$ vanishes, and the critical temperature $T_0$ where the relaxation time diverges (not to be confused with onset temperature $T_o$ used above). Typically, the latter is obtained from a Vogel-Fulcher-Tammann fit ($T_0 = T_{\rm VFT}$ in Eq.~\eqref{eq:VFT}) to the relaxation time. The possible equality $T_0 = T_K$ would provide a strong empirical sign for the existence of an ideal glass transition underlying glass formation~\cite{LW07}. A large data set collected by Tanaka suggests the existence of systematic differences between the two temperatures~\cite{tanaka2003relation}, with the tendency that $T_K> T_0$, and an apparent correlation with kinetic fragility. 
In our analysis using Eq.~(\ref{eq:GAG}) to describe the data, the connection between thermodynamics and dynamics becomes automatically satisfied, and thus by construction thermodynamic and dynamic singularities necessarily coincide. Assuming that the determination of $T_K$ is the most robust one, we conclude that it is the experimental determination of $T_0$ which should be questioned. In particular, using $\alpha<1$ in Eqs.~(\ref{eq:GAG}) and assuming an asymptotically linear vanishing of $\SC$, one would predict that $\log (\tau_\alpha /\tau_0) \propto (T-T_0)^{-\alpha}$, which is distinct from the standard Vogel-Fulcher-Tamman fit and would automatically produce the equality $T_K=T_0$. 
 
From a broader perspective, we conclude that the Adam-Gibbs relation, which is an important milestone in the field of glass transition studies, is generally violated in both computer models and real materials when tested over a broad, experimentally-relevant temperature range. We nevertheless argued that the failure of Eq.~(\ref{eq:AG}) cannot be taken as evidence that thermodynamic theories of the glass transition are incorrect. The RFOT theory prediction of a connexion between statics and dynamics in Eq.~(\ref{eq:GAG}) is obeyed by all materials, with exponent values that are reasonable, but remain to be predicted from first principles. A larger concern, perhaps, is the apparent lack of universality in the data shown in Fig.~\ref{fig:alpha_plot} which clearly display variations from one system to another. This may still be rationalised by invoking the fact that $\alpha$ is obtained from the analysis of a finite time window where additional preasymptotic effects and temperature dependent prefactors may influence the reported results.  

Taking an orthogonal perspective, we finally ask: Do our results validate or invalidate some theories of the glass transition? After all, we just established that a slightly generalised version of the Adam-Gibbs relation with $\alpha \simeq 0.6$ describes simulations and experiments over 9 orders of magnitude in the experimentally relevant regime. This is not a small accomplishment. One can take the alternative view that the deviations from the canonical exponent values should be taken as an indirect sign that thermodynamics only contributes some part of the slowing down, in addition to other physical factors~\cite{tarjus2005frustration,doi:10.1021/jp409502k,PhysRevLett.119.195501,dyre2006colloquium,tanaka2012bond,ikeda2017mean}. This view is sometimes also invoked to rationalise the ``modest'' growth of static correlation lengthscale observed numerically and experimentally~\cite{Ta11,PhysRevE.94.032605}. Our finding that $\alpha < 1$ suggests instead that it is the growth of the relaxation time that is actually too modest! It is therefore difficult to rationalise how another physical factor working in addition to the entropy could be invoked to explain our findings. The most radical view is in fact that thermodynamics is just a spectator to the glassy dynamics~\cite{chandler2010dynamics}, in which case our findings should be interpreted as purely coincidental since entropy plays in fact no role. We have no strong argument to oppose to this view, which remains perfectly admissible.

\begin{acknowledgments}

We thank J.-P. Bouchaud, D. Coslovich and F. Zamponi for insightful discussions. We also thank S. Tatsumi and O. Yamamuro for sharing their high-quality experimental data for the configurational entropy.
The research leading to these results has received funding from the Simons Foundation (\#454933, Ludovic Berthier). 

\end{acknowledgments}

\appendix

\section{Configurational entropy along an isobaric path}
\label{sec:s_conf_NPT}

We wish to measure the configurational entropy $S_{\rm conf}(T, P)$ along an isobaric (constant pressure) path. It
is computed as $S_{\rm conf}(T, P)=S_{\rm tot}(T,P)-S_{\rm glass}(T,P)$, where $S_{\rm tot}(T,P)$ and $S_{\rm glass}(T,P)$ are the total and glass entropies at the temperature $T$ and pressure $P$.
We explain how to get $S_{\rm conf}(T, P)$ from $NPT$ simulation trajectories.

\subsection{Notations}

We consider the Helmholtz free energy $-\beta F(T,V)=\ln Z(T,V)$, where $\beta=1/T$ and $Z(T,V)$ is the partition function of the $NVT$ ensemble.
We also consider the Gibbs free energy $-\beta G(T, P)=\ln Y(T, P)$, where $Y(T, P)$ is the partition function of the $NPT$ ensemble, given by 
\begin{equation}
Y(T, P) = \int_0^{\infty} \mathrm{d} V e^{-\beta \left(PV+F(T,V) \right)}.
\end{equation}
We introduce the probability distribution of the volume $V$ for a given $T$ and $P$,
\begin{equation}
\rho(V|T,P)=\frac{e^{-\beta \left(PV+F(T,V)\right)}}{Y(T,P)}.
\label{eq:rho}
\end{equation}
In equilibrium, $\rho(V|T,P)$ is given by Gaussian distribution,
\begin{equation}
\rho(V|T,P) = \frac{1}{\sqrt{2 \pi \sigma_V^2}} \exp \left[ -\frac{(V-V_*)^2}{2 \sigma_V^2} \right],
\label{eq:rho_gauss}
\end{equation}
where $V_*$ and $\sigma_V^2$ are the mean and variance of the volume, respectively.
We define $\langle(\cdots)\rangle_{T, P}=\int_0^{\infty} \mathrm{d}V \rho(V|T,P) (\cdots)$. Using this average, we can write $V_*=\langle V \rangle_{T, P}$ and $\sigma_V^2=\langle (V-V_*)^2 \rangle_{T, P}$.

\subsection{Total entropy}

The total entropy $S_{\rm tot}(T, P)$ is obtained by a thermodynamic integration of the isobaric heat capacity from a reference temperature $T_{\rm ref}=1/\beta_{\rm ref}$, to the target temperature $T=1/\beta$,
\begin{eqnarray}
S_{\rm tot}(T,P) &=& S_{\rm tot}(T_{\rm ref},P) - \frac{Nd}{2}\left( \ln \beta - \ln \beta_{\rm ref}\right) \nonumber \\
&\mathrm{}& + \beta U_*(T, P) - \beta_{\rm ref} U_*(T_{\rm ref},P) \nonumber \\
&\mathrm{}& - \int_{\beta_{\rm ref}}^{\beta} \mathrm{d} \beta' U_*(T', P) \nonumber \\
&\mathrm{}& + P \left( \beta V_*(T, P) -\beta_{\rm ref} V_*(T_{\rm ref},P)\right) \nonumber \\
&\mathrm{}& - P \int_{\beta_{\rm ref}}^{\beta} \mathrm{d} \beta' V_*(T', P),
\label{eq:S_tot_final}
\end{eqnarray}
where $U_*(T, P)$ is the mean potential energy, and $V_*(T, P)$ is the mean volume;
$U_*(T, P)$ and $V_*(T,P)$ are measured by constant pressure simulations.
The entropy at the reference state is obtained by $S_{\rm tot}(T_{\rm ref}, P)=\langle S_{\rm tot}(T_{\rm ref}, V)\rangle_{T, P}$ using the $NVT$ ensemble scheme~\cite{ceiling17}.
This treatment for the reference state will be justified below.

\subsection{Glass entropy}

To get the glass entropy, we use the generalised Frenkel-Ladd method which relies on the $NVT$ ensemble~\cite{ozawa2018configurational}.
In general, one can smoothly connect $NVT$ and $NPT$ ensembles in terms of mean values.
For example, thermodynamics guarantees that $S(T,P)=S(T, \langle V\rangle_{T, P})$.
However, special attention should be paid if one uses the $NVT$ ensemble scheme with trajectories generated by the $NPT$ ensemble for finite system size~\cite{PhysRev.153.250}. 
A related issue is discussed in Ref.~\cite{cheng2018computing}.
Indeed, what we can compute is $\langle S(T, V)\rangle_{T, P}$.
In general,
\begin{equation}
S(T,P) = \langle S(T,V)\rangle_{T, P} - \langle \ln \rho(V|T,P) \rangle_{T, P}.
\label{eq:final}
\end{equation}

Therefore, we need to consider the second term in Eq.~(\ref{eq:final}) as a correction term.
We can evalute this term with Eq.~(\ref{eq:rho_gauss}):
\begin{equation}
- \frac{1}{N} \langle \ln \rho(V|T,P) \rangle_{T, P} = \frac{1}{N} \ln \sqrt{2 \pi e \sigma_V^2}.
\end{equation}
Since $\sigma_V^2 \sim N$, this term vanishes in the thermodynamic limit, as expected.
Indeed, for $N=1500$ systems, we get negligible values, $\frac{1}{N} \ln \sqrt{2 \pi e \sigma_V^2} \simeq 0.0026$ and $0.0013$ at $T_{\rm ref}=7.0$ and $T=0.37$, respectively.
These values are small compared to the absolute value of $S_{\rm conf}/N \simeq 0.36 - 0.80$.}
Thus we can safely use $S(T,P)=\langle S(T,V)\rangle_{T, P}$.
Especially we use the following equation, $S_{\rm glass}(T, P)=\langle S_{\rm glass}(T, V)\rangle_{T, P}$.

\section{Extrapolation of relaxation times towards $T_g$}

\label{sec:extrapolation}

\begin{figure}
\includegraphics[width=0.48\columnwidth]{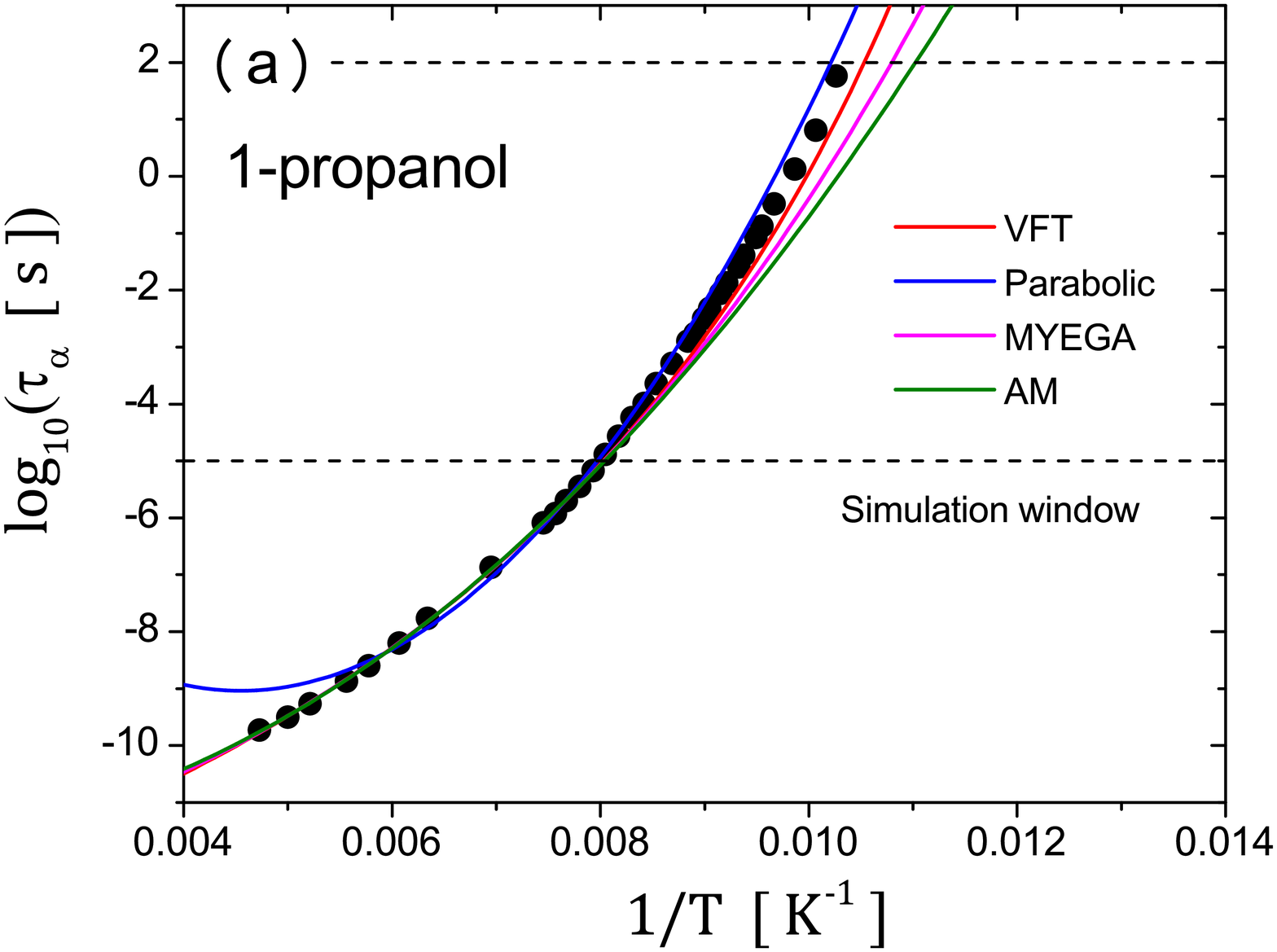}
\includegraphics[width=0.48\columnwidth]{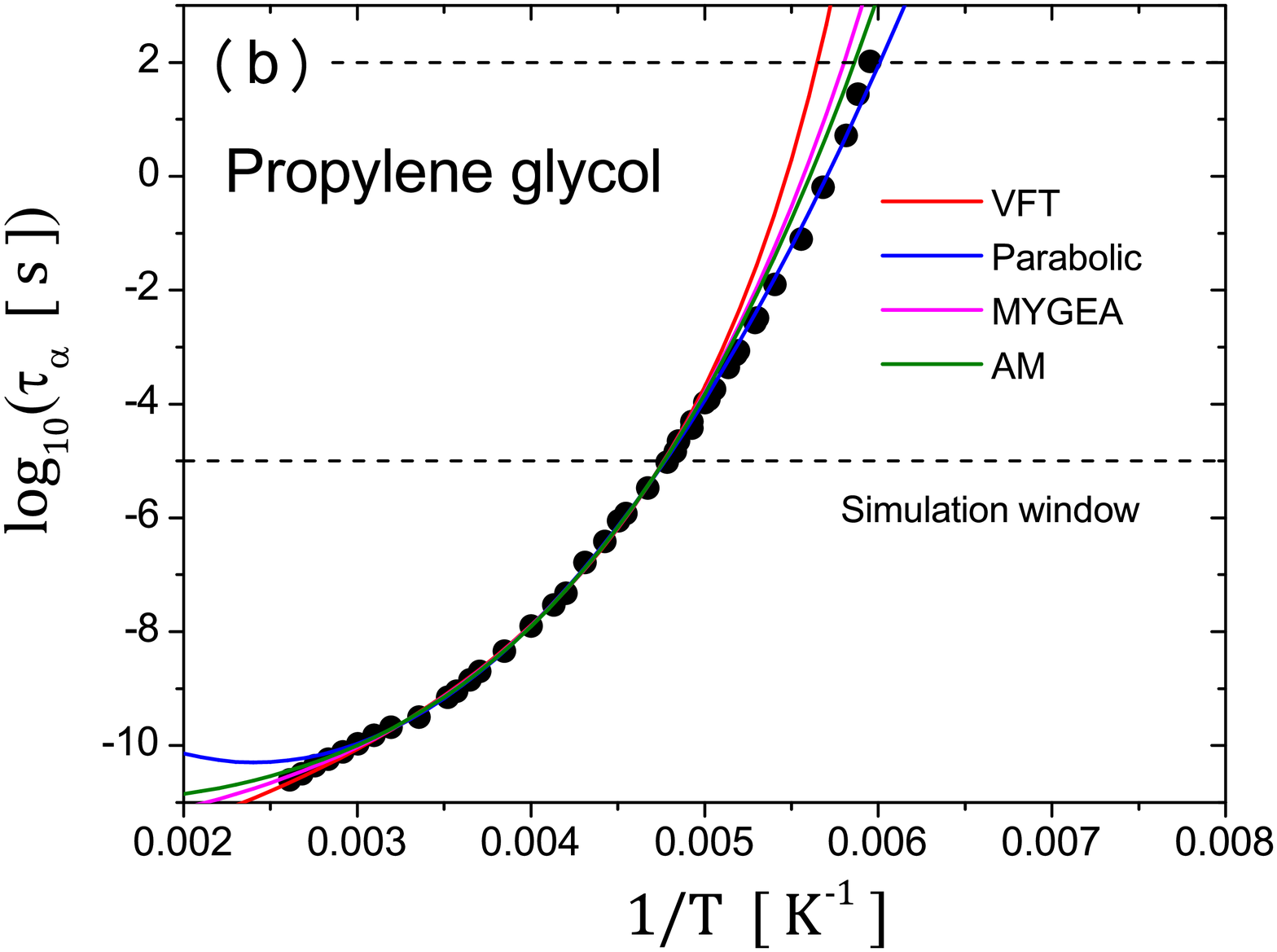}
\includegraphics[width=0.48\columnwidth]{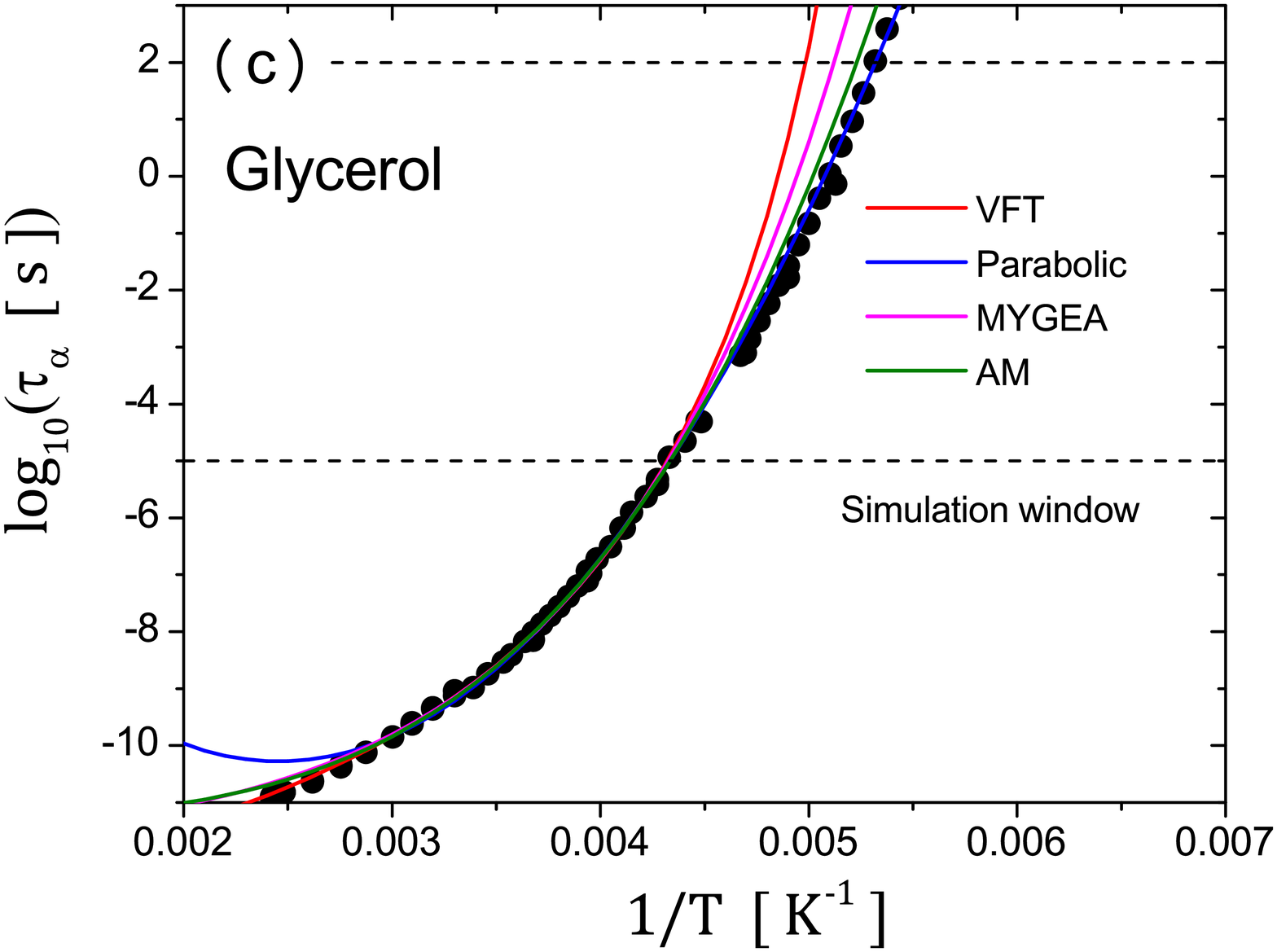}
\includegraphics[width=0.48\columnwidth]{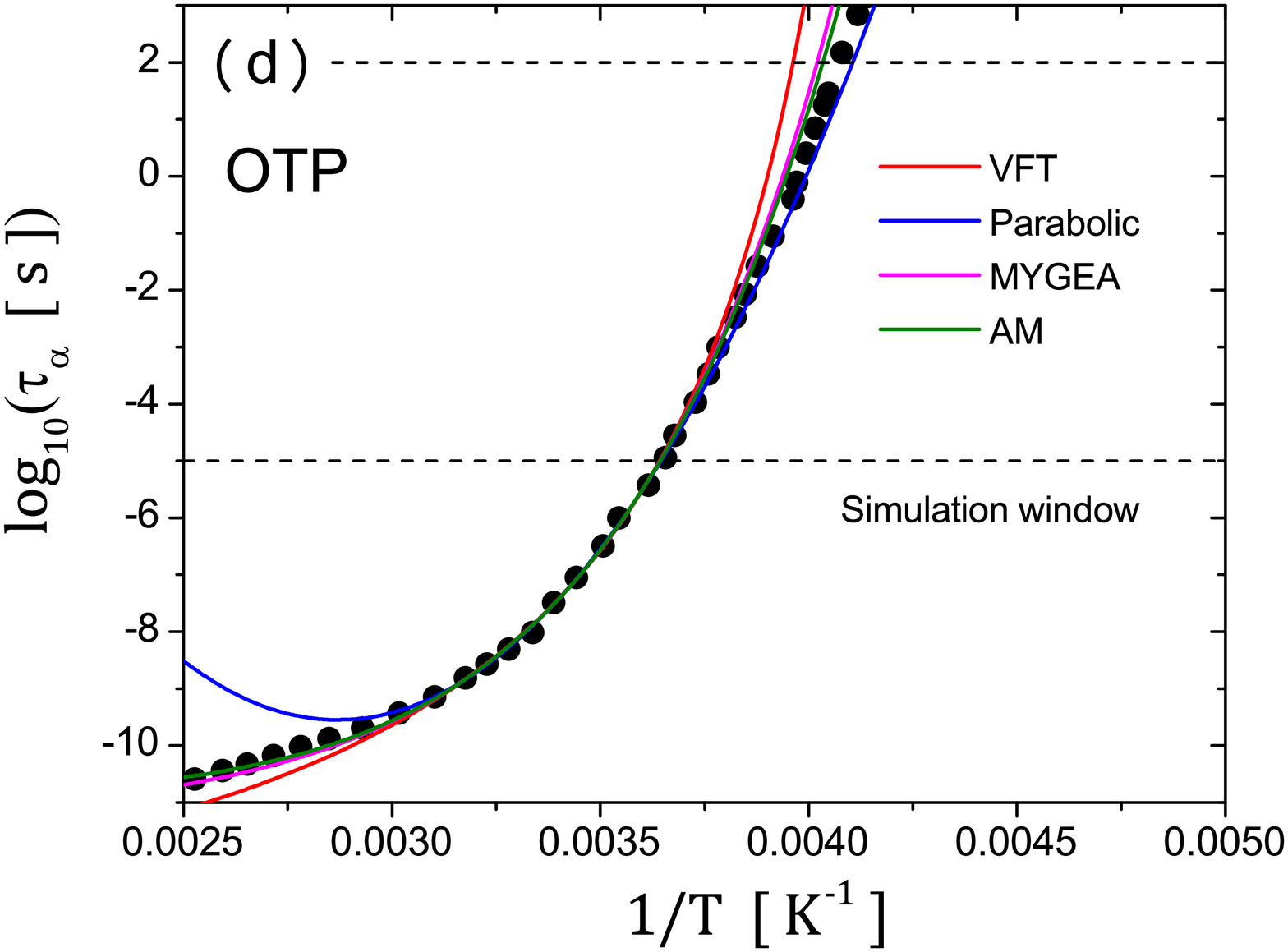}
\includegraphics[width=0.48\columnwidth]{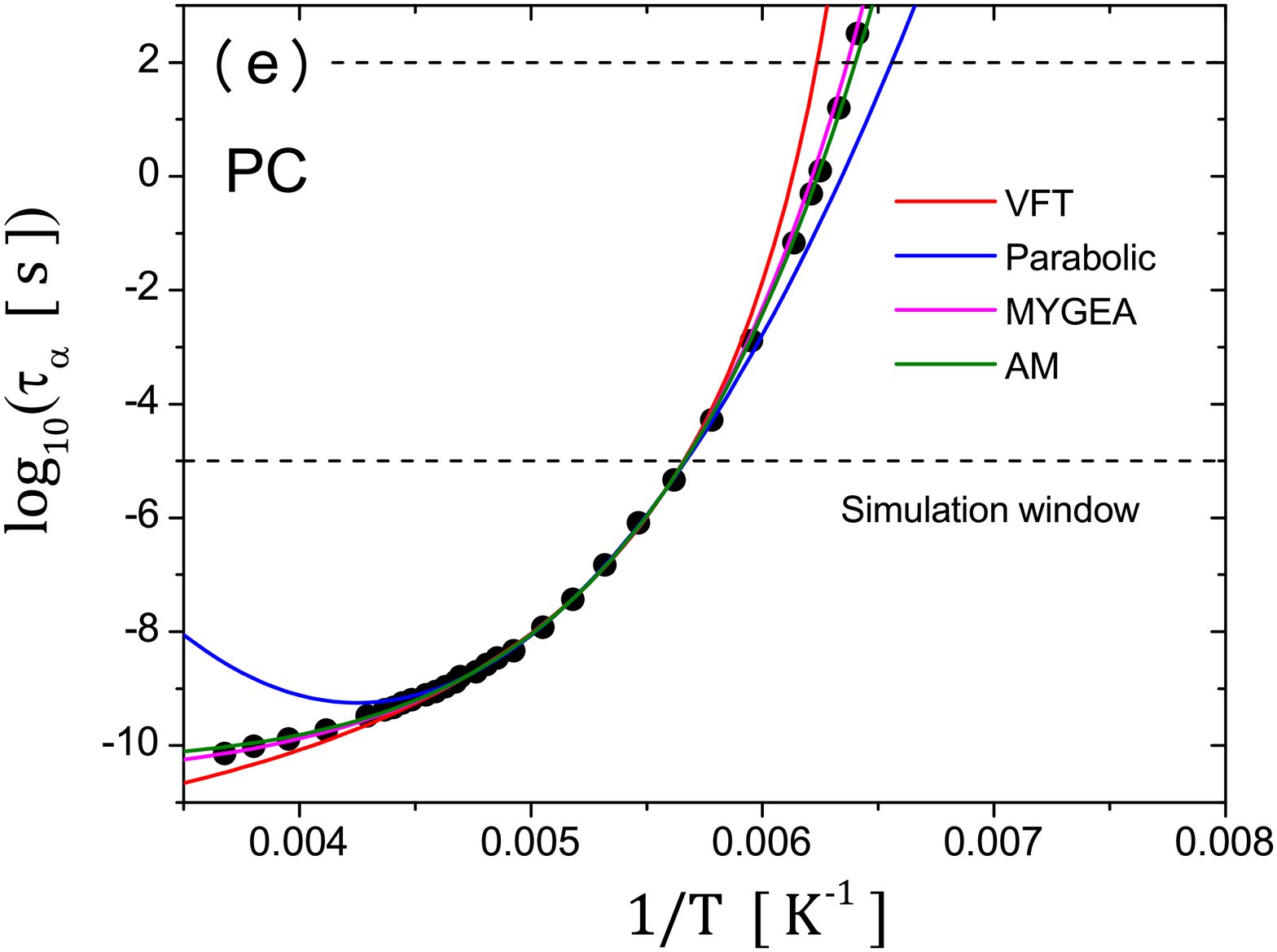}
\caption{
\MO{Extrapolation from simulation timescale ($\tau_{\alpha} \leq 10^{-5}$ sec.) to experimental timescale for 1-propanol  (a), propylene glycol (b), glycerol (c), OTP (d), and PC (e), whose kinetic fragility takes $m=35$, $48$, $53$, $78$, $104$, respectively.
The values of $m$ for 1-propanol, propylene glycol, and glycerol are comparable to the simulation models employed in this paper.}
 }
\label{fig:extrapolation}
\end{figure}  

Here we test the validity of the extrapolation of relaxation time from the numerical to the experimental timescale using various fitting functions. We employ 1-propanol, propylene glycol, glycerol, OTP, and PC. Among these, 1-propanol, propylene glycol, and glycerol have kinetic fragility indexes similar to the simulation models.

Figure~\ref{fig:extrapolation} shows various fits of the data performed over the simulation time window, $\tau_{\alpha} \leq 10^{-5}$ s, and then extrapolated to lower temperatures down to $T_g$ where $\tau_\alpha = 100$ s. 
In the cases for 1-propanol, propylene glycol, glycerol, and OTP, shown in Fig.~\ref{fig:extrapolation}, the parabolic law is the best functional form that predicts the actual data well over the experimental time window. All other functional forms, when fitted over the simulation time window, tend to deviate from the actual data at low temperatures. 
\MO{For the most fragile material, PC, underestimates the actual data when the parabolic law is applied, whereas MYGEA and AM predict the data better.
Notice that the uncertainty on the determination of $T_g$ using the numerical time window and a parabolic fit is very small for the systems whose fragility is comparable to typical simulations models.} This is the strategy we have used in previous numerical studies~\cite{NBC17,ceiling17,berthier2018zero}.  

\bibliography{biblio.bib}

\begin{thebibliography}{114}%
\makeatletter
\providecommand \@ifxundefined [1]{%
 \@ifx{#1\undefined}
}%
\providecommand \@ifnum [1]{%
 \ifnum #1\expandafter \@firstoftwo
 \else \expandafter \@secondoftwo
 \fi
}%
\providecommand \@ifx [1]{%
 \ifx #1\expandafter \@firstoftwo
 \else \expandafter \@secondoftwo
 \fi
}%
\providecommand \natexlab [1]{#1}%
\providecommand \enquote  [1]{``#1''}%
\providecommand \bibnamefont  [1]{#1}%
\providecommand \bibfnamefont [1]{#1}%
\providecommand \citenamefont [1]{#1}%
\providecommand \href@noop [0]{\@secondoftwo}%
\providecommand \href [0]{\begingroup \@sanitize@url \@href}%
\providecommand \@href[1]{\@@startlink{#1}\@@href}%
\providecommand \@@href[1]{\endgroup#1\@@endlink}%
\providecommand \@sanitize@url [0]{\catcode `\\12\catcode `\$12\catcode
  `\&12\catcode `\#12\catcode `\^12\catcode `\_12\catcode `\%12\relax}%
\providecommand \@@startlink[1]{}%
\providecommand \@@endlink[0]{}%
\providecommand \url  [0]{\begingroup\@sanitize@url \@url }%
\providecommand \@url [1]{\endgroup\@href {#1}{\urlprefix }}%
\providecommand \urlprefix  [0]{URL }%
\providecommand \Eprint [0]{\href }%
\providecommand \doibase [0]{http://dx.doi.org/}%
\providecommand \selectlanguage [0]{\@gobble}%
\providecommand \bibinfo  [0]{\@secondoftwo}%
\providecommand \bibfield  [0]{\@secondoftwo}%
\providecommand \translation [1]{[#1]}%
\providecommand \BibitemOpen [0]{}%
\providecommand \bibitemStop [0]{}%
\providecommand \bibitemNoStop [0]{.\EOS\space}%
\providecommand \EOS [0]{\spacefactor3000\relax}%
\providecommand \BibitemShut  [1]{\csname bibitem#1\endcsname}%
\let\auto@bib@innerbib\@empty
\bibitem [{\citenamefont {Adam}\ and\ \citenamefont
  {Gibbs}(1965)}]{adam1965temperature}%
  \BibitemOpen
  \bibfield  {author} {\bibinfo {author} {\bibfnamefont {Gerold}\ \bibnamefont
  {Adam}}\ and\ \bibinfo {author} {\bibfnamefont {Julian~H}\ \bibnamefont
  {Gibbs}},\ }\bibfield  {title} {\enquote {\bibinfo {title} {On the
  temperature dependence of cooperative relaxation properties in glass-forming
  liquids},}\ }\href {https://doi.org/10.1063/1.1696442} {\bibfield  {journal}
  {\bibinfo  {journal} {J. Chem. Phys.}\ }\textbf {\bibinfo {volume} {43}},\
  \bibinfo {pages} {139} (\bibinfo {year} {1965})}\BibitemShut {NoStop}%
\bibitem [{\citenamefont {Berthier}\ and\ \citenamefont
  {Biroli}(2011)}]{berthier2011theoretical}%
  \BibitemOpen
  \bibfield  {author} {\bibinfo {author} {\bibfnamefont {Ludovic}\ \bibnamefont
  {Berthier}}\ and\ \bibinfo {author} {\bibfnamefont {Giulio}\ \bibnamefont
  {Biroli}},\ }\bibfield  {title} {\enquote {\bibinfo {title} {Theoretical
  perspective on the glass transition and amorphous materials},}\ }\href
  {https://doi.org/10.1103/RevModPhys.83.587} {\bibfield  {journal} {\bibinfo
  {journal} {Rev. Mod. Phys.}\ }\textbf {\bibinfo {volume} {83}},\ \bibinfo
  {pages} {587} (\bibinfo {year} {2011})}\BibitemShut {NoStop}%
\bibitem [{\citenamefont {Kirkpatrick}\ \emph {et~al.}(1989)\citenamefont
  {Kirkpatrick}, \citenamefont {Thirumalai},\ and\ \citenamefont
  {Wolynes}}]{KTW89}%
  \BibitemOpen
  \bibfield  {author} {\bibinfo {author} {\bibfnamefont {T.~R.}\ \bibnamefont
  {Kirkpatrick}}, \bibinfo {author} {\bibfnamefont {D.}~\bibnamefont
  {Thirumalai}}, \ and\ \bibinfo {author} {\bibfnamefont {P.~G.}\ \bibnamefont
  {Wolynes}},\ }\bibfield  {title} {\enquote {\bibinfo {title} {Scaling
  concepts for the dynamics of viscous liquids near an ideal glassy state},}\
  }\href {\doibase 10.1103/PhysRevA.40.1045} {\bibfield  {journal} {\bibinfo
  {journal} {Phys. Rev. A}\ }\textbf {\bibinfo {volume} {40}},\ \bibinfo
  {pages} {1045--1054} (\bibinfo {year} {1989})}\BibitemShut {NoStop}%
\bibitem [{\citenamefont {Bouchaud}\ and\ \citenamefont {Biroli}(2004)}]{BB04}%
  \BibitemOpen
  \bibfield  {author} {\bibinfo {author} {\bibfnamefont {Jean-Philippe}\
  \bibnamefont {Bouchaud}}\ and\ \bibinfo {author} {\bibfnamefont {Giulio}\
  \bibnamefont {Biroli}},\ }\bibfield  {title} {\enquote {\bibinfo {title} {On
  the adam-gibbs-kirkpatrick-thirumalai-wolynes scenario for the viscosity
  increase in glasses},}\ }\href {\doibase 10.1063/1.1796231} {\bibfield
  {journal} {\bibinfo  {journal} {J. Chem. Phys.}\ }\textbf {\bibinfo {volume}
  {121}},\ \bibinfo {pages} {7347--7354} (\bibinfo {year} {2004})}\BibitemShut
  {NoStop}%
\bibitem [{\citenamefont {Lubchenko}\ and\ \citenamefont
  {Wolynes}(2007)}]{LW07}%
  \BibitemOpen
  \bibfield  {author} {\bibinfo {author} {\bibfnamefont {Vassiliy}\
  \bibnamefont {Lubchenko}}\ and\ \bibinfo {author} {\bibfnamefont {Peter~G.}\
  \bibnamefont {Wolynes}},\ }\bibfield  {title} {\enquote {\bibinfo {title}
  {Theory of structural glasses and supercooled liquids},}\ }\href
  {https://doi.org/10.1146/annurev.physchem.58.032806.104653} {\bibfield
  {journal} {\bibinfo  {journal} {Annual Review of Physical Chemistry}\
  }\textbf {\bibinfo {volume} {58}},\ \bibinfo {pages} {235--266} (\bibinfo
  {year} {2007})}\BibitemShut {NoStop}%
\bibitem [{\citenamefont {Biroli}\ and\ \citenamefont {Bouchaud}(2012)}]{BB09}%
  \BibitemOpen
  \bibfield  {author} {\bibinfo {author} {\bibfnamefont {G.}~\bibnamefont
  {Biroli}}\ and\ \bibinfo {author} {\bibfnamefont {J.P.}\ \bibnamefont
  {Bouchaud}},\ }\bibfield  {title} {\enquote {\bibinfo {title} {{The Random
  First-Order Transition Theory of Glasses: a critical assessment}},}\ }in\
  \href@noop {} {\emph {\bibinfo {booktitle} {Structural Glasses and
  Supercooled Liquids: Theory, Experiment and Applications}}},\ \bibinfo
  {editor} {edited by\ \bibinfo {editor} {\bibnamefont {P.G.Wolynes}}\ and\
  \bibinfo {editor} {\bibnamefont {V.Lubchenko}}}\ (\bibinfo  {publisher}
  {Wiley \& Sons},\ \bibinfo {year} {2012})\ \Eprint {http://arxiv.org/abs/{\tt
  arXiv:0912.2542}} {{\tt arXiv:0912.2542}} \BibitemShut {NoStop}%
\bibitem [{\citenamefont {Lubchenko}(2015)}]{lubchenko2015theory}%
  \BibitemOpen
  \bibfield  {author} {\bibinfo {author} {\bibfnamefont {Vassiliy}\
  \bibnamefont {Lubchenko}},\ }\bibfield  {title} {\enquote {\bibinfo {title}
  {Theory of the structural glass transition: a pedagogical review},}\ }\href
  {\doibase 10.1080/00018732.2015.1057979} {\bibfield  {journal} {\bibinfo
  {journal} {Advances in Physics}\ }\textbf {\bibinfo {volume} {64}},\ \bibinfo
  {pages} {283--443} (\bibinfo {year} {2015})}\BibitemShut {NoStop}%
\bibitem [{\citenamefont {Charbonneau}\ \emph {et~al.}(2017)\citenamefont
  {Charbonneau}, \citenamefont {Kurchan}, \citenamefont {Parisi}, \citenamefont
  {Urbani},\ and\ \citenamefont {Zamponi}}]{CKPUZ16}%
  \BibitemOpen
  \bibfield  {author} {\bibinfo {author} {\bibfnamefont {Patrick}\ \bibnamefont
  {Charbonneau}}, \bibinfo {author} {\bibfnamefont {Jorge}\ \bibnamefont
  {Kurchan}}, \bibinfo {author} {\bibfnamefont {Giorgio}\ \bibnamefont
  {Parisi}}, \bibinfo {author} {\bibfnamefont {Pierfrancesco}\ \bibnamefont
  {Urbani}}, \ and\ \bibinfo {author} {\bibfnamefont {Francesco}\ \bibnamefont
  {Zamponi}},\ }\bibfield  {title} {\enquote {\bibinfo {title} {Glass and
  jamming transitions: From exact results to finite-dimensional
  descriptions},}\ }\href {\doibase 10.1146/annurev-conmatphys-031016-025334}
  {\bibfield  {journal} {\bibinfo  {journal} {Annual Review of Condensed Matter
  Physics}\ }\textbf {\bibinfo {volume} {8}},\ \bibinfo {pages} {265--288}
  (\bibinfo {year} {2017})}\BibitemShut {NoStop}%
\bibitem [{\citenamefont {Dudowicz}\ \emph {et~al.}(2008)\citenamefont
  {Dudowicz}, \citenamefont {Freed},\ and\ \citenamefont
  {Douglas}}]{dudowicz2008generalized}%
  \BibitemOpen
  \bibfield  {author} {\bibinfo {author} {\bibfnamefont {Jacek}\ \bibnamefont
  {Dudowicz}}, \bibinfo {author} {\bibfnamefont {Karl~F.}\ \bibnamefont
  {Freed}}, \ and\ \bibinfo {author} {\bibfnamefont {Jack~F.}\ \bibnamefont
  {Douglas}},\ }\bibfield  {title} {\enquote {\bibinfo {title} {Generalized
  entropy theory of polymer glass formation},}\ }\href {\doibase
  10.1002/9780470238080.ch3} {\bibfield  {journal} {\bibinfo  {journal}
  {Advances in Chemical Physics}\ ,\ \bibinfo {pages} {125--222}} (\bibinfo
  {year} {2008})}\BibitemShut {NoStop}%
\bibitem [{\citenamefont {Angell}(1997)}]{angell1997entropy}%
  \BibitemOpen
  \bibfield  {author} {\bibinfo {author} {\bibfnamefont {CA}~\bibnamefont
  {Angell}},\ }\bibfield  {title} {\enquote {\bibinfo {title} {Entropy and
  fragility in supercooling liquids},}\ }\href
  {https://www.ncbi.nlm.nih.gov/pmc/articles/PMC4900877/} {\bibfield  {journal}
  {\bibinfo  {journal} {Journal of research of the National Institute of
  Standards and Technology}\ }\textbf {\bibinfo {volume} {102}},\ \bibinfo
  {pages} {171} (\bibinfo {year} {1997})}\BibitemShut {NoStop}%
\bibitem [{\citenamefont {Richert}\ and\ \citenamefont
  {Angell}(1998)}]{richert1998dynamics}%
  \BibitemOpen
  \bibfield  {author} {\bibinfo {author} {\bibfnamefont {R.}~\bibnamefont
  {Richert}}\ and\ \bibinfo {author} {\bibfnamefont {C.~A.}\ \bibnamefont
  {Angell}},\ }\bibfield  {title} {\enquote {\bibinfo {title} {Dynamics of
  glass-forming liquids. v. on the link between molecular dynamics and
  configurational entropy},}\ }\href {\doibase 10.1063/1.476348} {\bibfield
  {journal} {\bibinfo  {journal} {The Journal of Chemical Physics}\ }\textbf
  {\bibinfo {volume} {108}},\ \bibinfo {pages} {9016--9026} (\bibinfo {year}
  {1998})}\BibitemShut {NoStop}%
\bibitem [{\citenamefont {Sastry}(2001)}]{sastry2001relationship}%
  \BibitemOpen
  \bibfield  {author} {\bibinfo {author} {\bibfnamefont {Srikanth}\
  \bibnamefont {Sastry}},\ }\bibfield  {title} {\enquote {\bibinfo {title} {The
  relationship between fragility, configurational entropy and the potential
  energy landscape of glass-forming liquids},}\ }\href
  {https://doi.org/10.1038/35051524} {\bibfield  {journal} {\bibinfo  {journal}
  {Nature}\ }\textbf {\bibinfo {volume} {409}},\ \bibinfo {pages} {164}
  (\bibinfo {year} {2001})}\BibitemShut {NoStop}%
\bibitem [{\citenamefont {Dyre}\ \emph {et~al.}(2009)\citenamefont {Dyre},
  \citenamefont {Hechsher},\ and\ \citenamefont {Niss}}]{dyre2009brief}%
  \BibitemOpen
  \bibfield  {author} {\bibinfo {author} {\bibfnamefont {Jeppe~C}\ \bibnamefont
  {Dyre}}, \bibinfo {author} {\bibfnamefont {Tina}\ \bibnamefont {Hechsher}}, \
  and\ \bibinfo {author} {\bibfnamefont {Kristine}\ \bibnamefont {Niss}},\
  }\bibfield  {title} {\enquote {\bibinfo {title} {A brief critique of the
  adam--gibbs entropy model},}\ }\href
  {https://doi.org/10.1016/j.jnoncrysol.2009.01.039} {\bibfield  {journal}
  {\bibinfo  {journal} {Journal of Non-Crystalline Solids}\ }\textbf {\bibinfo
  {volume} {355}},\ \bibinfo {pages} {624--627} (\bibinfo {year}
  {2009})}\BibitemShut {NoStop}%
\bibitem [{\citenamefont {Sciortino}\ \emph {et~al.}(1999)\citenamefont
  {Sciortino}, \citenamefont {Kob},\ and\ \citenamefont {Tartaglia}}]{SKT99}%
  \BibitemOpen
  \bibfield  {author} {\bibinfo {author} {\bibfnamefont {F.}~\bibnamefont
  {Sciortino}}, \bibinfo {author} {\bibfnamefont {W.}~\bibnamefont {Kob}}, \
  and\ \bibinfo {author} {\bibfnamefont {P.}~\bibnamefont {Tartaglia}},\
  }\bibfield  {title} {\enquote {\bibinfo {title} {Inherent structure entropy
  of supercooled liquids},}\ }\href {\doibase 10.1103/PhysRevLett.83.3214}
  {\bibfield  {journal} {\bibinfo  {journal} {Phys. Rev. Lett.}\ }\textbf
  {\bibinfo {volume} {83}},\ \bibinfo {pages} {3214--3217} (\bibinfo {year}
  {1999})}\BibitemShut {NoStop}%
\bibitem [{\citenamefont {Sastry}(2000)}]{sastry2000evaluation}%
  \BibitemOpen
  \bibfield  {author} {\bibinfo {author} {\bibfnamefont {Srikanth}\
  \bibnamefont {Sastry}},\ }\bibfield  {title} {\enquote {\bibinfo {title}
  {Evaluation of the configurational entropy of a model liquid from computer
  simulations},}\ }\href {\doibase 10.1088/0953-8984/12/29/323} {\bibfield
  {journal} {\bibinfo  {journal} {Journal of Physics: Condensed Matter}\
  }\textbf {\bibinfo {volume} {12}},\ \bibinfo {pages} {6515--6523} (\bibinfo
  {year} {2000})}\BibitemShut {NoStop}%
\bibitem [{\citenamefont {Berthier}\ \emph
  {et~al.}(2019{\natexlab{a}})\citenamefont {Berthier}, \citenamefont {Ozawa},\
  and\ \citenamefont {Scalliet}}]{reviewsconf}%
  \BibitemOpen
  \bibfield  {author} {\bibinfo {author} {\bibfnamefont {Ludovic}\ \bibnamefont
  {Berthier}}, \bibinfo {author} {\bibfnamefont {Misaki}\ \bibnamefont
  {Ozawa}}, \ and\ \bibinfo {author} {\bibfnamefont {Camille}\ \bibnamefont
  {Scalliet}},\ }\bibfield  {title} {\enquote {\bibinfo {title}
  {Configurational entropy of glass-forming liquids},}\ }\href {\doibase
  10.1063/1.5091961} {\bibfield  {journal} {\bibinfo  {journal} {The Journal of
  Chemical Physics}\ }\textbf {\bibinfo {volume} {150}},\ \bibinfo {pages}
  {160902} (\bibinfo {year} {2019}{\natexlab{a}})}\BibitemShut {NoStop}%
\bibitem [{\citenamefont {Mossa}\ \emph {et~al.}(2002)\citenamefont {Mossa},
  \citenamefont {La~Nave}, \citenamefont {Stanley}, \citenamefont {Donati},
  \citenamefont {Sciortino},\ and\ \citenamefont
  {Tartaglia}}]{mossa2002dynamics}%
  \BibitemOpen
  \bibfield  {author} {\bibinfo {author} {\bibfnamefont {S.}~\bibnamefont
  {Mossa}}, \bibinfo {author} {\bibfnamefont {E.}~\bibnamefont {La~Nave}},
  \bibinfo {author} {\bibfnamefont {H.~E.}\ \bibnamefont {Stanley}}, \bibinfo
  {author} {\bibfnamefont {C.}~\bibnamefont {Donati}}, \bibinfo {author}
  {\bibfnamefont {F.}~\bibnamefont {Sciortino}}, \ and\ \bibinfo {author}
  {\bibfnamefont {P.}~\bibnamefont {Tartaglia}},\ }\bibfield  {title} {\enquote
  {\bibinfo {title} {Dynamics and configurational entropy in the
  lewis-wahnstr\"om model for supercooled orthoterphenyl},}\ }\href {\doibase
  10.1103/PhysRevE.65.041205} {\bibfield  {journal} {\bibinfo  {journal} {Phys.
  Rev. E}\ }\textbf {\bibinfo {volume} {65}},\ \bibinfo {pages} {041205}
  (\bibinfo {year} {2002})}\BibitemShut {NoStop}%
\bibitem [{\citenamefont {Sciortino}(2005)}]{sciortinoPEL}%
  \BibitemOpen
  \bibfield  {author} {\bibinfo {author} {\bibfnamefont {Francesco}\
  \bibnamefont {Sciortino}},\ }\bibfield  {title} {\enquote {\bibinfo {title}
  {Potential energy landscape description of supercooled liquids and
  glasses},}\ }\href {http://stacks.iop.org/1742-5468/2005/i=05/a=P05015}
  {\bibfield  {journal} {\bibinfo  {journal} {Journal of Statistical Mechanics:
  Theory and Experiment}\ }\textbf {\bibinfo {volume} {2005}},\ \bibinfo
  {pages} {P05015} (\bibinfo {year} {2005})}\BibitemShut {NoStop}%
\bibitem [{\citenamefont {Saika-Voivod}\ \emph {et~al.}(2001)\citenamefont
  {Saika-Voivod}, \citenamefont {Poole},\ and\ \citenamefont
  {Sciortino}}]{saika2001fragile}%
  \BibitemOpen
  \bibfield  {author} {\bibinfo {author} {\bibfnamefont {Ivan}\ \bibnamefont
  {Saika-Voivod}}, \bibinfo {author} {\bibfnamefont {Peter~H}\ \bibnamefont
  {Poole}}, \ and\ \bibinfo {author} {\bibfnamefont {Francesco}\ \bibnamefont
  {Sciortino}},\ }\bibfield  {title} {\enquote {\bibinfo {title}
  {Fragile-to-strong transition and polyamorphism in the energy landscape of
  liquid silica},}\ }\href {https://doi.org/10.1038/35087524} {\bibfield
  {journal} {\bibinfo  {journal} {Nature}\ }\textbf {\bibinfo {volume} {412}},\
  \bibinfo {pages} {514} (\bibinfo {year} {2001})}\BibitemShut {NoStop}%
\bibitem [{\citenamefont {Angelani}\ \emph {et~al.}(2005)\citenamefont
  {Angelani}, \citenamefont {Foffi}, \citenamefont {Sciortino},\ and\
  \citenamefont {Tartaglia}}]{foffi2005}%
  \BibitemOpen
  \bibfield  {author} {\bibinfo {author} {\bibfnamefont {Luca}\ \bibnamefont
  {Angelani}}, \bibinfo {author} {\bibfnamefont {Giuseppe}\ \bibnamefont
  {Foffi}}, \bibinfo {author} {\bibfnamefont {Francesco}\ \bibnamefont
  {Sciortino}}, \ and\ \bibinfo {author} {\bibfnamefont {Piero}\ \bibnamefont
  {Tartaglia}},\ }\bibfield  {title} {\enquote {\bibinfo {title} {Diffusivity
  and configurational entropy maxima in short range attractive colloids},}\
  }\href {\doibase 10.1088/0953-8984/17/12/l02} {\bibfield  {journal} {\bibinfo
   {journal} {Journal of Physics: Condensed Matter}\ }\textbf {\bibinfo
  {volume} {17}},\ \bibinfo {pages} {L113--L119} (\bibinfo {year}
  {2005})}\BibitemShut {NoStop}%
\bibitem [{\citenamefont {Sengupta}\ \emph {et~al.}(2012)\citenamefont
  {Sengupta}, \citenamefont {Karmakar}, \citenamefont {Dasgupta},\ and\
  \citenamefont {Sastry}}]{sengupta2012adam}%
  \BibitemOpen
  \bibfield  {author} {\bibinfo {author} {\bibfnamefont {Shiladitya}\
  \bibnamefont {Sengupta}}, \bibinfo {author} {\bibfnamefont {Smarajit}\
  \bibnamefont {Karmakar}}, \bibinfo {author} {\bibfnamefont {Chandan}\
  \bibnamefont {Dasgupta}}, \ and\ \bibinfo {author} {\bibfnamefont {Srikanth}\
  \bibnamefont {Sastry}},\ }\bibfield  {title} {\enquote {\bibinfo {title}
  {Adam-gibbs relation for glass-forming liquids in two, three, and four
  dimensions},}\ }\href {\doibase 10.1103/PhysRevLett.109.095705} {\bibfield
  {journal} {\bibinfo  {journal} {Phys. Rev. Lett.}\ }\textbf {\bibinfo
  {volume} {109}},\ \bibinfo {pages} {095705} (\bibinfo {year}
  {2012})}\BibitemShut {NoStop}%
\bibitem [{\citenamefont {Starr}\ \emph {et~al.}(2013)\citenamefont {Starr},
  \citenamefont {Douglas},\ and\ \citenamefont
  {Sastry}}]{starr2013relationship}%
  \BibitemOpen
  \bibfield  {author} {\bibinfo {author} {\bibfnamefont {Francis~W}\
  \bibnamefont {Starr}}, \bibinfo {author} {\bibfnamefont {Jack~F}\
  \bibnamefont {Douglas}}, \ and\ \bibinfo {author} {\bibfnamefont {Srikanth}\
  \bibnamefont {Sastry}},\ }\bibfield  {title} {\enquote {\bibinfo {title} {The
  relationship of dynamical heterogeneity to the adam-gibbs and random
  first-order transition theories of glass formation},}\ }\href
  {https://aip.scitation.org/doi/10.1063/1.4790138} {\bibfield  {journal}
  {\bibinfo  {journal} {The Journal of chemical physics}\ }\textbf {\bibinfo
  {volume} {138}},\ \bibinfo {pages} {12A541} (\bibinfo {year}
  {2013})}\BibitemShut {NoStop}%
\bibitem [{\citenamefont {Parmar}\ \emph {et~al.}(2017)\citenamefont {Parmar},
  \citenamefont {Sengupta},\ and\ \citenamefont {Sastry}}]{parmar2017length}%
  \BibitemOpen
  \bibfield  {author} {\bibinfo {author} {\bibfnamefont {Anshul~DS}\
  \bibnamefont {Parmar}}, \bibinfo {author} {\bibfnamefont {Shiladitya}\
  \bibnamefont {Sengupta}}, \ and\ \bibinfo {author} {\bibfnamefont {Srikanth}\
  \bibnamefont {Sastry}},\ }\bibfield  {title} {\enquote {\bibinfo {title}
  {Length-scale dependence of the stokes-einstein and adam-gibbs relations in
  model glass formers},}\ }\href
  {https://link.aps.org/doi/10.1103/PhysRevLett.119.056001} {\bibfield
  {journal} {\bibinfo  {journal} {Physical review letters}\ }\textbf {\bibinfo
  {volume} {119}},\ \bibinfo {pages} {056001} (\bibinfo {year}
  {2017})}\BibitemShut {NoStop}%
\bibitem [{\citenamefont {Handle}\ and\ \citenamefont
  {Sciortino}(2018)}]{handle2018adam}%
  \BibitemOpen
  \bibfield  {author} {\bibinfo {author} {\bibfnamefont {Philip~H}\
  \bibnamefont {Handle}}\ and\ \bibinfo {author} {\bibfnamefont {Francesco}\
  \bibnamefont {Sciortino}},\ }\bibfield  {title} {\enquote {\bibinfo {title}
  {The adam--gibbs relation and the tip4p/2005 model of water},}\ }\href
  {https://www.ncbi.nlm.nih.gov/pmc/articles/PMC6171618/} {\bibfield  {journal}
  {\bibinfo  {journal} {Molecular physics}\ }\textbf {\bibinfo {volume}
  {116}},\ \bibinfo {pages} {3366--3371} (\bibinfo {year} {2018})}\BibitemShut
  {NoStop}%
\bibitem [{\citenamefont {G{\"o}tze}(2008)}]{gotze2008complex}%
  \BibitemOpen
  \bibfield  {author} {\bibinfo {author} {\bibfnamefont {Wolfgang}\
  \bibnamefont {G{\"o}tze}},\ }\href@noop {} {\emph {\bibinfo {title} {Complex
  dynamics of glass-forming liquids: A mode-coupling theory}}},\ Vol.\ \bibinfo
  {volume} {143}\ (\bibinfo  {publisher} {Oxford University Press, Oxford},\
  \bibinfo {year} {2008})\BibitemShut {NoStop}%
\bibitem [{\citenamefont {Magill}(1967)}]{magill1967physical}%
  \BibitemOpen
  \bibfield  {author} {\bibinfo {author} {\bibfnamefont {JH}~\bibnamefont
  {Magill}},\ }\bibfield  {title} {\enquote {\bibinfo {title} {Physical
  properties of aromatic hydrocarbons. iii. a test of the adam--gibbs
  relaxation model for glass formers based on the heat-capacity data of 1, 3,
  5-tri-$\alpha$-naphthylbenzene},}\ }\href
  {https://aip.scitation.org/doi/10.1063/1.1712301} {\bibfield  {journal}
  {\bibinfo  {journal} {The Journal of chemical physics}\ }\textbf {\bibinfo
  {volume} {47}},\ \bibinfo {pages} {2802--2807} (\bibinfo {year}
  {1967})}\BibitemShut {NoStop}%
\bibitem [{\citenamefont {Takahara}\ \emph {et~al.}(1995)\citenamefont
  {Takahara}, \citenamefont {Yamamuro},\ and\ \citenamefont
  {Matsuo}}]{takahara1995calorimetric}%
  \BibitemOpen
  \bibfield  {author} {\bibinfo {author} {\bibfnamefont {Shuichi}\ \bibnamefont
  {Takahara}}, \bibinfo {author} {\bibfnamefont {Osamu}\ \bibnamefont
  {Yamamuro}}, \ and\ \bibinfo {author} {\bibfnamefont {Takasuke}\ \bibnamefont
  {Matsuo}},\ }\bibfield  {title} {\enquote {\bibinfo {title} {Calorimetric
  study of 3-bromopentane: correlation between structural relaxation time and
  configurational entropy},}\ }\href
  {https://pubs.acs.org/doi/pdf/10.1021/j100023a042} {\bibfield  {journal}
  {\bibinfo  {journal} {The Journal of Physical Chemistry}\ }\textbf {\bibinfo
  {volume} {99}},\ \bibinfo {pages} {9589--9592} (\bibinfo {year}
  {1995})}\BibitemShut {NoStop}%
\bibitem [{\citenamefont {Ngai}(1999)}]{ngai1999modification}%
  \BibitemOpen
  \bibfield  {author} {\bibinfo {author} {\bibfnamefont {KL}~\bibnamefont
  {Ngai}},\ }\bibfield  {title} {\enquote {\bibinfo {title} {Modification of
  the adam--gibbs model of glass transition for consistency with experimental
  data},}\ }\href {https://pubs.acs.org/doi/abs/10.1021/jp990594w} {\bibfield
  {journal} {\bibinfo  {journal} {The Journal of Physical Chemistry B}\
  }\textbf {\bibinfo {volume} {103}},\ \bibinfo {pages} {5895--5902} (\bibinfo
  {year} {1999})}\BibitemShut {NoStop}%
\bibitem [{\citenamefont {Alba-Simionesco}(2001)}]{alba2001salient}%
  \BibitemOpen
  \bibfield  {author} {\bibinfo {author} {\bibfnamefont {C}~\bibnamefont
  {Alba-Simionesco}},\ }\bibfield  {title} {\enquote {\bibinfo {title} {Salient
  properties of glassforming liquids close to the glass transition},}\ }\href
  {http://www.sciencedirect.com/science/article/pii/S1296214701011659}
  {\bibfield  {journal} {\bibinfo  {journal} {Comptes Rendus de l'Acad{\'e}mie
  des Sciences-Series IV-Physics-Astrophysics}\ }\textbf {\bibinfo {volume}
  {2}},\ \bibinfo {pages} {203--216} (\bibinfo {year} {2001})}\BibitemShut
  {NoStop}%
\bibitem [{\citenamefont {Roland}\ \emph {et~al.}(2004)\citenamefont {Roland},
  \citenamefont {Capaccioli}, \citenamefont {Lucchesi},\ and\ \citenamefont
  {Casalini}}]{roland2004adam}%
  \BibitemOpen
  \bibfield  {author} {\bibinfo {author} {\bibfnamefont {CM}~\bibnamefont
  {Roland}}, \bibinfo {author} {\bibfnamefont {Simone}\ \bibnamefont
  {Capaccioli}}, \bibinfo {author} {\bibfnamefont {Mauro}\ \bibnamefont
  {Lucchesi}}, \ and\ \bibinfo {author} {\bibfnamefont {R}~\bibnamefont
  {Casalini}},\ }\bibfield  {title} {\enquote {\bibinfo {title} {Adam--gibbs
  model for the supercooled dynamics in the ortho-terphenyl ortho-phenylphenol
  mixture},}\ }\href {https://aip.scitation.org/doi/10.1063/1.1739394}
  {\bibfield  {journal} {\bibinfo  {journal} {The Journal of chemical physics}\
  }\textbf {\bibinfo {volume} {120}},\ \bibinfo {pages} {10640--10646}
  (\bibinfo {year} {2004})}\BibitemShut {NoStop}%
\bibitem [{\citenamefont {Cangialosi}\ \emph {et~al.}(2005)\citenamefont
  {Cangialosi}, \citenamefont {Alegria},\ and\ \citenamefont
  {Colmenero}}]{cangialosi2005relationship}%
  \BibitemOpen
  \bibfield  {author} {\bibinfo {author} {\bibfnamefont {D}~\bibnamefont
  {Cangialosi}}, \bibinfo {author} {\bibfnamefont {A}~\bibnamefont {Alegria}},
  \ and\ \bibinfo {author} {\bibfnamefont {J}~\bibnamefont {Colmenero}},\
  }\bibfield  {title} {\enquote {\bibinfo {title} {Relationship between
  dynamics and thermodynamics in glass-forming polymers},}\ }\href
  {https://doi.org/10.1209/epl/i2005-10029-y} {\bibfield  {journal} {\bibinfo
  {journal} {EPL (Europhysics Letters)}\ }\textbf {\bibinfo {volume} {70}},\
  \bibinfo {pages} {614} (\bibinfo {year} {2005})}\BibitemShut {NoStop}%
\bibitem [{\citenamefont {Masiewicz}\ \emph {et~al.}(2015)\citenamefont
  {Masiewicz}, \citenamefont {Grzybowski}, \citenamefont {Grzybowska},
  \citenamefont {Pawlus}, \citenamefont {Pionteck},\ and\ \citenamefont
  {Paluch}}]{masiewicz2015adam}%
  \BibitemOpen
  \bibfield  {author} {\bibinfo {author} {\bibfnamefont {El{\.z}bieta}\
  \bibnamefont {Masiewicz}}, \bibinfo {author} {\bibfnamefont {Andrzej}\
  \bibnamefont {Grzybowski}}, \bibinfo {author} {\bibfnamefont {Katarzyna}\
  \bibnamefont {Grzybowska}}, \bibinfo {author} {\bibfnamefont {Sebastian}\
  \bibnamefont {Pawlus}}, \bibinfo {author} {\bibfnamefont {J{\"u}rgen}\
  \bibnamefont {Pionteck}}, \ and\ \bibinfo {author} {\bibfnamefont {Marian}\
  \bibnamefont {Paluch}},\ }\bibfield  {title} {\enquote {\bibinfo {title}
  {Adam-gibbs model in the density scaling regime and its implications for the
  configurational entropy scaling},}\ }\href
  {https://www.nature.com/articles/srep13998} {\bibfield  {journal} {\bibinfo
  {journal} {Scientific reports}\ }\textbf {\bibinfo {volume} {5}},\ \bibinfo
  {pages} {13998} (\bibinfo {year} {2015})}\BibitemShut {NoStop}%
\bibitem [{\citenamefont {Cammarota}\ \emph
  {et~al.}(2009{\natexlab{a}})\citenamefont {Cammarota}, \citenamefont
  {Cavagna}, \citenamefont {Gradenigo}, \citenamefont {Grigera},\ and\
  \citenamefont {Verrocchio}}]{cammarota2009numerical}%
  \BibitemOpen
  \bibfield  {author} {\bibinfo {author} {\bibfnamefont {Chiara}\ \bibnamefont
  {Cammarota}}, \bibinfo {author} {\bibfnamefont {Andrea}\ \bibnamefont
  {Cavagna}}, \bibinfo {author} {\bibfnamefont {Giacomo}\ \bibnamefont
  {Gradenigo}}, \bibinfo {author} {\bibfnamefont {Tomas}\ \bibnamefont
  {Grigera}}, \ and\ \bibinfo {author} {\bibfnamefont {Paolo}\ \bibnamefont
  {Verrocchio}},\ }\bibfield  {title} {\enquote {\bibinfo {title} {Numerical
  determination of the exponents controlling the relationship between time,
  length, and temperature in glass-forming liquids},}\ }\href {\doibase
  10.1063/1.3257739} {\bibfield  {journal} {\bibinfo  {journal} {The Journal of
  Chemical Physics}\ }\textbf {\bibinfo {volume} {131}},\ \bibinfo {pages}
  {194901} (\bibinfo {year} {2009}{\natexlab{a}})}\BibitemShut {NoStop}%
\bibitem [{\citenamefont {Karmakar}\ \emph {et~al.}(2009)\citenamefont
  {Karmakar}, \citenamefont {Dasgupta},\ and\ \citenamefont
  {Sastry}}]{karmakar2009growing}%
  \BibitemOpen
  \bibfield  {author} {\bibinfo {author} {\bibfnamefont {Smarajit}\
  \bibnamefont {Karmakar}}, \bibinfo {author} {\bibfnamefont {Chandan}\
  \bibnamefont {Dasgupta}}, \ and\ \bibinfo {author} {\bibfnamefont {Srikanth}\
  \bibnamefont {Sastry}},\ }\bibfield  {title} {\enquote {\bibinfo {title}
  {Growing length and time scales in glass-forming liquids},}\ }\href
  {https://doi.org/10.1073/pnas.0811082106} {\bibfield  {journal} {\bibinfo
  {journal} {Proceedings of the National Academy of Sciences}\ }\textbf
  {\bibinfo {volume} {106}},\ \bibinfo {pages} {3675--3679} (\bibinfo {year}
  {2009})}\BibitemShut {NoStop}%
\bibitem [{\citenamefont {Hocky}\ \emph {et~al.}(2012)\citenamefont {Hocky},
  \citenamefont {Markland},\ and\ \citenamefont {Reichman}}]{hocky2012growing}%
  \BibitemOpen
  \bibfield  {author} {\bibinfo {author} {\bibfnamefont {Glen~M.}\ \bibnamefont
  {Hocky}}, \bibinfo {author} {\bibfnamefont {Thomas~E.}\ \bibnamefont
  {Markland}}, \ and\ \bibinfo {author} {\bibfnamefont {David~R.}\ \bibnamefont
  {Reichman}},\ }\bibfield  {title} {\enquote {\bibinfo {title} {Growing
  point-to-set length scale correlates with growing relaxation times in model
  supercooled liquids},}\ }\href {\doibase 10.1103/PhysRevLett.108.225506}
  {\bibfield  {journal} {\bibinfo  {journal} {Phys. Rev. Lett.}\ }\textbf
  {\bibinfo {volume} {108}},\ \bibinfo {pages} {225506} (\bibinfo {year}
  {2012})}\BibitemShut {NoStop}%
\bibitem [{\citenamefont {Guti{\'{e}}rrez}\ \emph {et~al.}(2015)\citenamefont
  {Guti{\'{e}}rrez}, \citenamefont {Karmakar}, \citenamefont {Pollack},\ and\
  \citenamefont {Procaccia}}]{Gutierrez:2015}%
  \BibitemOpen
  \bibfield  {author} {\bibinfo {author} {\bibfnamefont {R.}~\bibnamefont
  {Guti{\'{e}}rrez}}, \bibinfo {author} {\bibfnamefont {S.}~\bibnamefont
  {Karmakar}}, \bibinfo {author} {\bibfnamefont {Y.~G.}\ \bibnamefont
  {Pollack}}, \ and\ \bibinfo {author} {\bibfnamefont {I.}~\bibnamefont
  {Procaccia}},\ }\bibfield  {title} {\enquote {\bibinfo {title} {The static
  lengthscale characterizing the glass transition at lower temperatures},}\
  }\href {\doibase 10.1209/0295-5075/111/56009} {\bibfield  {journal} {\bibinfo
   {journal} {{EPL} (Europhysics Letters)}\ }\textbf {\bibinfo {volume}
  {111}},\ \bibinfo {pages} {56009} (\bibinfo {year} {2015})}\BibitemShut
  {NoStop}%
\bibitem [{\citenamefont {Cavagna}\ \emph {et~al.}(2012)\citenamefont
  {Cavagna}, \citenamefont {Grigera},\ and\ \citenamefont
  {Verrocchio}}]{CGV12}%
  \BibitemOpen
  \bibfield  {author} {\bibinfo {author} {\bibfnamefont {Andrea}\ \bibnamefont
  {Cavagna}}, \bibinfo {author} {\bibfnamefont {Tomas~S.}\ \bibnamefont
  {Grigera}}, \ and\ \bibinfo {author} {\bibfnamefont {Paolo}\ \bibnamefont
  {Verrocchio}},\ }\bibfield  {title} {\enquote {\bibinfo {title} {Dynamic
  relaxation of a liquid cavity under amorphous boundary conditions},}\ }\href
  {\doibase 10.1063/1.4720477} {\bibfield  {journal} {\bibinfo  {journal} {The
  Journal of Chemical Physics}\ }\textbf {\bibinfo {volume} {136}},\ \bibinfo
  {pages} {204502} (\bibinfo {year} {2012})}\BibitemShut {NoStop}%
\bibitem [{\citenamefont {Capaccioli}\ \emph {et~al.}(2008)\citenamefont
  {Capaccioli}, \citenamefont {Ruocco},\ and\ \citenamefont
  {Zamponi}}]{capaccioli2008dynamically}%
  \BibitemOpen
  \bibfield  {author} {\bibinfo {author} {\bibfnamefont {Simone}\ \bibnamefont
  {Capaccioli}}, \bibinfo {author} {\bibfnamefont {Giancarlo}\ \bibnamefont
  {Ruocco}}, \ and\ \bibinfo {author} {\bibfnamefont {Francesco}\ \bibnamefont
  {Zamponi}},\ }\bibfield  {title} {\enquote {\bibinfo {title} {Dynamically
  correlated regions and configurational entropy in supercooled liquids},}\
  }\href {https://pubs.acs.org/doi/abs/10.1021/jp802097u} {\bibfield  {journal}
  {\bibinfo  {journal} {The Journal of Physical Chemistry B}\ }\textbf
  {\bibinfo {volume} {112}},\ \bibinfo {pages} {10652--10658} (\bibinfo {year}
  {2008})}\BibitemShut {NoStop}%
\bibitem [{\citenamefont {Brun}\ \emph {et~al.}(2012)\citenamefont {Brun},
  \citenamefont {Ladieu}, \citenamefont {L’H{\^o}te}, \citenamefont
  {Biroli},\ and\ \citenamefont {Bouchaud}}]{brun2012evidence}%
  \BibitemOpen
  \bibfield  {author} {\bibinfo {author} {\bibfnamefont {C}~\bibnamefont
  {Brun}}, \bibinfo {author} {\bibfnamefont {F}~\bibnamefont {Ladieu}},
  \bibinfo {author} {\bibfnamefont {D}~\bibnamefont {L’H{\^o}te}}, \bibinfo
  {author} {\bibfnamefont {G}~\bibnamefont {Biroli}}, \ and\ \bibinfo {author}
  {\bibfnamefont {JP}~\bibnamefont {Bouchaud}},\ }\bibfield  {title} {\enquote
  {\bibinfo {title} {Evidence of growing spatial correlations during the aging
  of glassy glycerol},}\ }\href
  {https://journals.aps.org/prl/abstract/10.1103/PhysRevLett.109.175702}
  {\bibfield  {journal} {\bibinfo  {journal} {Physical review letters}\
  }\textbf {\bibinfo {volume} {109}},\ \bibinfo {pages} {175702} (\bibinfo
  {year} {2012})}\BibitemShut {NoStop}%
\bibitem [{\citenamefont {Berthier}\ \emph {et~al.}(2005)\citenamefont
  {Berthier}, \citenamefont {Biroli}, \citenamefont {Bouchaud}, \citenamefont
  {Cipelletti}, \citenamefont {Masri}, \citenamefont
  {L{\textquoteright}H{\^o}te}, \citenamefont {Ladieu},\ and\ \citenamefont
  {Pierno}}]{BBBCEHLP05}%
  \BibitemOpen
  \bibfield  {author} {\bibinfo {author} {\bibfnamefont {L.}~\bibnamefont
  {Berthier}}, \bibinfo {author} {\bibfnamefont {G.}~\bibnamefont {Biroli}},
  \bibinfo {author} {\bibfnamefont {J.-P.}\ \bibnamefont {Bouchaud}}, \bibinfo
  {author} {\bibfnamefont {L.}~\bibnamefont {Cipelletti}}, \bibinfo {author}
  {\bibfnamefont {D.~El}\ \bibnamefont {Masri}}, \bibinfo {author}
  {\bibfnamefont {D.}~\bibnamefont {L{\textquoteright}H{\^o}te}}, \bibinfo
  {author} {\bibfnamefont {F.}~\bibnamefont {Ladieu}}, \ and\ \bibinfo {author}
  {\bibfnamefont {M.}~\bibnamefont {Pierno}},\ }\bibfield  {title} {\enquote
  {\bibinfo {title} {Direct experimental evidence of a growing length scale
  accompanying the glass transition},}\ }\href
  {http://science.sciencemag.org/content/310/5755/1797} {\bibfield  {journal}
  {\bibinfo  {journal} {Science}\ }\textbf {\bibinfo {volume} {310}},\ \bibinfo
  {pages} {1797--1800} (\bibinfo {year} {2005})}\BibitemShut {NoStop}%
\bibitem [{\citenamefont {Berthier}\ \emph
  {et~al.}(2016{\natexlab{a}})\citenamefont {Berthier}, \citenamefont
  {Coslovich}, \citenamefont {Ninarello},\ and\ \citenamefont
  {Ozawa}}]{berthier2016equilibrium}%
  \BibitemOpen
  \bibfield  {author} {\bibinfo {author} {\bibfnamefont {Ludovic}\ \bibnamefont
  {Berthier}}, \bibinfo {author} {\bibfnamefont {Daniele}\ \bibnamefont
  {Coslovich}}, \bibinfo {author} {\bibfnamefont {Andrea}\ \bibnamefont
  {Ninarello}}, \ and\ \bibinfo {author} {\bibfnamefont {Misaki}\ \bibnamefont
  {Ozawa}},\ }\bibfield  {title} {\enquote {\bibinfo {title} {Equilibrium
  sampling of hard spheres up to the jamming density and beyond},}\ }\href
  {https://link.aps.org/doi/10.1103/PhysRevLett.116.238002} {\bibfield
  {journal} {\bibinfo  {journal} {Phys. Rev. Lett.}\ }\textbf {\bibinfo
  {volume} {116}},\ \bibinfo {pages} {238002} (\bibinfo {year}
  {2016}{\natexlab{a}})}\BibitemShut {NoStop}%
\bibitem [{\citenamefont {Ninarello}\ \emph {et~al.}(2017)\citenamefont
  {Ninarello}, \citenamefont {Berthier},\ and\ \citenamefont
  {Coslovich}}]{NBC17}%
  \BibitemOpen
  \bibfield  {author} {\bibinfo {author} {\bibfnamefont {Andrea}\ \bibnamefont
  {Ninarello}}, \bibinfo {author} {\bibfnamefont {Ludovic}\ \bibnamefont
  {Berthier}}, \ and\ \bibinfo {author} {\bibfnamefont {Daniele}\ \bibnamefont
  {Coslovich}},\ }\bibfield  {title} {\enquote {\bibinfo {title} {Models and
  algorithms for the next generation of glass transition studies},}\ }\href
  {\doibase 10.1103/PhysRevX.7.021039} {\bibfield  {journal} {\bibinfo
  {journal} {Phys. Rev. X}\ }\textbf {\bibinfo {volume} {7}},\ \bibinfo {pages}
  {021039} (\bibinfo {year} {2017})}\BibitemShut {NoStop}%
\bibitem [{\citenamefont {Berthier}\ \emph
  {et~al.}(2019{\natexlab{b}})\citenamefont {Berthier}, \citenamefont
  {Charbonneau}, \citenamefont {Ninarello}, \citenamefont {Ozawa},\ and\
  \citenamefont {Yaida}}]{berthier2018zero}%
  \BibitemOpen
  \bibfield  {author} {\bibinfo {author} {\bibfnamefont {Ludovic}\ \bibnamefont
  {Berthier}}, \bibinfo {author} {\bibfnamefont {Patrick}\ \bibnamefont
  {Charbonneau}}, \bibinfo {author} {\bibfnamefont {Andrea}\ \bibnamefont
  {Ninarello}}, \bibinfo {author} {\bibfnamefont {Misaki}\ \bibnamefont
  {Ozawa}}, \ and\ \bibinfo {author} {\bibfnamefont {Sho}\ \bibnamefont
  {Yaida}},\ }\bibfield  {title} {\enquote {\bibinfo {title} {Zero-temperature
  glass transition in two dimensions},}\ }\href {\doibase
  10.1038/s41467-019-09512-3} {\bibfield  {journal} {\bibinfo  {journal}
  {Nature Communications}\ }\textbf {\bibinfo {volume} {10}},\ \bibinfo {pages}
  {1508} (\bibinfo {year} {2019}{\natexlab{b}})}\BibitemShut {NoStop}%
\bibitem [{\citenamefont {Berthier}\ \emph {et~al.}(2017)\citenamefont
  {Berthier}, \citenamefont {Charbonneau}, \citenamefont {Coslovich},
  \citenamefont {Ninarello}, \citenamefont {Ozawa},\ and\ \citenamefont
  {Yaida}}]{ceiling17}%
  \BibitemOpen
  \bibfield  {author} {\bibinfo {author} {\bibfnamefont {Ludovic}\ \bibnamefont
  {Berthier}}, \bibinfo {author} {\bibfnamefont {Patrick}\ \bibnamefont
  {Charbonneau}}, \bibinfo {author} {\bibfnamefont {Daniele}\ \bibnamefont
  {Coslovich}}, \bibinfo {author} {\bibfnamefont {Andrea}\ \bibnamefont
  {Ninarello}}, \bibinfo {author} {\bibfnamefont {Misaki}\ \bibnamefont
  {Ozawa}}, \ and\ \bibinfo {author} {\bibfnamefont {Sho}\ \bibnamefont
  {Yaida}},\ }\bibfield  {title} {\enquote {\bibinfo {title} {Configurational
  entropy measurements in extremely supercooled liquids that break the glass
  ceiling},}\ }\href {\doibase 10.1073/pnas.1706860114} {\bibfield  {journal}
  {\bibinfo  {journal} {Proceedings of the National Academy of Sciences}\
  }\textbf {\bibinfo {volume} {114}},\ \bibinfo {pages} {11356--11361}
  (\bibinfo {year} {2017})}\BibitemShut {NoStop}%
\bibitem [{\citenamefont {Ozawa}\ \emph {et~al.}(2018)\citenamefont {Ozawa},
  \citenamefont {Parisi},\ and\ \citenamefont
  {Berthier}}]{ozawa2018configurational}%
  \BibitemOpen
  \bibfield  {author} {\bibinfo {author} {\bibfnamefont {Misaki}\ \bibnamefont
  {Ozawa}}, \bibinfo {author} {\bibfnamefont {Giorgio}\ \bibnamefont {Parisi}},
  \ and\ \bibinfo {author} {\bibfnamefont {Ludovic}\ \bibnamefont {Berthier}},\
  }\bibfield  {title} {\enquote {\bibinfo {title} {Configurational entropy of
  polydisperse supercooled liquids},}\ }\href {\doibase 10.1063/1.5040975}
  {\bibfield  {journal} {\bibinfo  {journal} {The Journal of Chemical Physics}\
  }\textbf {\bibinfo {volume} {149}},\ \bibinfo {pages} {154501} (\bibinfo
  {year} {2018})}\BibitemShut {NoStop}%
\bibitem [{\citenamefont {Tatsumi}\ \emph {et~al.}(2012)\citenamefont
  {Tatsumi}, \citenamefont {Aso},\ and\ \citenamefont {Yamamuro}}]{tatsumi12}%
  \BibitemOpen
  \bibfield  {author} {\bibinfo {author} {\bibfnamefont {Soichi}\ \bibnamefont
  {Tatsumi}}, \bibinfo {author} {\bibfnamefont {Shintaro}\ \bibnamefont {Aso}},
  \ and\ \bibinfo {author} {\bibfnamefont {Osamu}\ \bibnamefont {Yamamuro}},\
  }\bibfield  {title} {\enquote {\bibinfo {title} {Thermodynamic study of
  simple molecular glasses: Universal features in their heat capacity and the
  size of the cooperatively rearranging regions},}\ }\href {\doibase
  10.1103/PhysRevLett.109.045701} {\bibfield  {journal} {\bibinfo  {journal}
  {Phys. Rev. Lett.}\ }\textbf {\bibinfo {volume} {109}},\ \bibinfo {pages}
  {045701} (\bibinfo {year} {2012})}\BibitemShut {NoStop}%
\bibitem [{\citenamefont {Berthier}\ \emph {et~al.}(2018)\citenamefont
  {Berthier}, \citenamefont {Flenner}, \citenamefont {Fullerton}, \citenamefont
  {Scalliet},\ and\ \citenamefont {Singh}}]{berthier2018efficient}%
  \BibitemOpen
  \bibfield  {author} {\bibinfo {author} {\bibfnamefont {Ludovic}\ \bibnamefont
  {Berthier}}, \bibinfo {author} {\bibfnamefont {Elijah}\ \bibnamefont
  {Flenner}}, \bibinfo {author} {\bibfnamefont {Christopher~J}\ \bibnamefont
  {Fullerton}}, \bibinfo {author} {\bibfnamefont {Camille}\ \bibnamefont
  {Scalliet}}, \ and\ \bibinfo {author} {\bibfnamefont {Murari}\ \bibnamefont
  {Singh}},\ }\bibfield  {title} {\enquote {\bibinfo {title} {Efficient swap
  algorithms for molecular dynamics simulations of equilibrium supercooled
  liquids},}\ }\href {https://arxiv.org/abs/1811.12837} {\bibfield  {journal}
  {\bibinfo  {journal} {arXiv preprint arXiv:1811.12837}\ } (\bibinfo {year}
  {2018})}\BibitemShut {NoStop}%
\bibitem [{\citenamefont {Berthier}\ and\ \citenamefont {Witten}(2009)}]{BW09}%
  \BibitemOpen
  \bibfield  {author} {\bibinfo {author} {\bibfnamefont {Ludovic}\ \bibnamefont
  {Berthier}}\ and\ \bibinfo {author} {\bibfnamefont {Thomas~A.}\ \bibnamefont
  {Witten}},\ }\bibfield  {title} {\enquote {\bibinfo {title} {Glass transition
  of dense fluids of hard and compressible spheres},}\ }\href {\doibase
  10.1103/PhysRevE.80.021502} {\bibfield  {journal} {\bibinfo  {journal} {Phys.
  Rev. E}\ }\textbf {\bibinfo {volume} {80}},\ \bibinfo {pages} {021502}
  (\bibinfo {year} {2009})}\BibitemShut {NoStop}%
\bibitem [{\citenamefont {Banerjee}\ \emph {et~al.}(2014)\citenamefont
  {Banerjee}, \citenamefont {Sengupta}, \citenamefont {Sastry},\ and\
  \citenamefont {Bhattacharyya}}]{banerjee2014role}%
  \BibitemOpen
  \bibfield  {author} {\bibinfo {author} {\bibfnamefont {Atreyee}\ \bibnamefont
  {Banerjee}}, \bibinfo {author} {\bibfnamefont {Shiladitya}\ \bibnamefont
  {Sengupta}}, \bibinfo {author} {\bibfnamefont {Srikanth}\ \bibnamefont
  {Sastry}}, \ and\ \bibinfo {author} {\bibfnamefont {Sarika~Maitra}\
  \bibnamefont {Bhattacharyya}},\ }\bibfield  {title} {\enquote {\bibinfo
  {title} {Role of structure and entropy in determining differences in dynamics
  for glass formers with different interaction potentials},}\ }\href {\doibase
  10.1103/PhysRevLett.113.225701} {\bibfield  {journal} {\bibinfo  {journal}
  {Phys. Rev. Lett.}\ }\textbf {\bibinfo {volume} {113}},\ \bibinfo {pages}
  {225701} (\bibinfo {year} {2014})}\BibitemShut {NoStop}%
\bibitem [{\citenamefont {Biroli}\ \emph {et~al.}(2008)\citenamefont {Biroli},
  \citenamefont {Bouchaud}, \citenamefont {Cavagna}, \citenamefont {Grigera},\
  and\ \citenamefont {Verrocchio}}]{BBCGV08}%
  \BibitemOpen
  \bibfield  {author} {\bibinfo {author} {\bibfnamefont {G.}~\bibnamefont
  {Biroli}}, \bibinfo {author} {\bibfnamefont {J.P.}\ \bibnamefont {Bouchaud}},
  \bibinfo {author} {\bibfnamefont {A.}~\bibnamefont {Cavagna}}, \bibinfo
  {author} {\bibfnamefont {TS}~\bibnamefont {Grigera}}, \ and\ \bibinfo
  {author} {\bibfnamefont {P.}~\bibnamefont {Verrocchio}},\ }\bibfield  {title}
  {\enquote {\bibinfo {title} {{Thermodynamic signature of growing amorphous
  order in glass-forming liquids}},}\ }\href
  {https://www.nature.com/articles/nphys1050} {\bibfield  {journal} {\bibinfo
  {journal} {Nature Physics}\ }\textbf {\bibinfo {volume} {4}},\ \bibinfo
  {pages} {771--775} (\bibinfo {year} {2008})}\BibitemShut {NoStop}%
\bibitem [{\citenamefont {Berthier}\ \emph
  {et~al.}(2016{\natexlab{b}})\citenamefont {Berthier}, \citenamefont
  {Charbonneau},\ and\ \citenamefont {Yaida}}]{berthier2016efficient}%
  \BibitemOpen
  \bibfield  {author} {\bibinfo {author} {\bibfnamefont {Ludovic}\ \bibnamefont
  {Berthier}}, \bibinfo {author} {\bibfnamefont {Patrick}\ \bibnamefont
  {Charbonneau}}, \ and\ \bibinfo {author} {\bibfnamefont {Sho}\ \bibnamefont
  {Yaida}},\ }\bibfield  {title} {\enquote {\bibinfo {title} {Efficient
  measurement of point-to-set correlations and overlap fluctuations in
  glass-forming liquids},}\ }\href {\doibase 10.1063/1.4939640} {\bibfield
  {journal} {\bibinfo  {journal} {The Journal of Chemical Physics}\ }\textbf
  {\bibinfo {volume} {144}},\ \bibinfo {pages} {024501} (\bibinfo {year}
  {2016}{\natexlab{b}})}\BibitemShut {NoStop}%
\bibitem [{\citenamefont {Berthier}\ and\ \citenamefont
  {Kob}(2007)}]{berthier2007monte}%
  \BibitemOpen
  \bibfield  {author} {\bibinfo {author} {\bibfnamefont {Ludovic}\ \bibnamefont
  {Berthier}}\ and\ \bibinfo {author} {\bibfnamefont {Walter}\ \bibnamefont
  {Kob}},\ }\bibfield  {title} {\enquote {\bibinfo {title} {The monte carlo
  dynamics of a binary lennard-jones glass-forming mixture},}\ }\href
  {http://iopscience.iop.org/article/10.1088/0953-8984/19/20/205130} {\bibfield
   {journal} {\bibinfo  {journal} {J. Phys.: Condens. Matter}\ }\textbf
  {\bibinfo {volume} {19}},\ \bibinfo {pages} {205130} (\bibinfo {year}
  {2007})}\BibitemShut {NoStop}%
\bibitem [{\citenamefont {Flenner}\ and\ \citenamefont
  {Szamel}(2015)}]{flenner2015fundamental}%
  \BibitemOpen
  \bibfield  {author} {\bibinfo {author} {\bibfnamefont {Elijah}\ \bibnamefont
  {Flenner}}\ and\ \bibinfo {author} {\bibfnamefont {Grzegorz}\ \bibnamefont
  {Szamel}},\ }\bibfield  {title} {\enquote {\bibinfo {title} {Fundamental
  differences between glassy dynamics in two and three dimensions},}\ }\href
  {https://www.nature.com/articles/ncomms8392} {\bibfield  {journal} {\bibinfo
  {journal} {Nature communications}\ }\textbf {\bibinfo {volume} {6}},\
  \bibinfo {pages} {7392} (\bibinfo {year} {2015})}\BibitemShut {NoStop}%
\bibitem [{\citenamefont {Stickel}\ \emph {et~al.}(1995)\citenamefont
  {Stickel}, \citenamefont {Fischer},\ and\ \citenamefont
  {Richert}}]{stickel_dynamics_1995}%
  \BibitemOpen
  \bibfield  {author} {\bibinfo {author} {\bibfnamefont {F.}~\bibnamefont
  {Stickel}}, \bibinfo {author} {\bibfnamefont {E.~W.}\ \bibnamefont
  {Fischer}}, \ and\ \bibinfo {author} {\bibfnamefont {R.}~\bibnamefont
  {Richert}},\ }\bibfield  {title} {\enquote {\bibinfo {title} {Dynamics of
  glass-forming liquids. {{I}}. {{Temperature}}-derivative analysis of
  dielectric relaxation data},}\ }\href {\doibase 10.1063/1.469071} {\bibfield
  {journal} {\bibinfo  {journal} {J. Chem. Phys.}\ }\textbf {\bibinfo {volume}
  {102}},\ \bibinfo {pages} {6251--6257} (\bibinfo {year} {1995})}\BibitemShut
  {NoStop}%
\bibitem [{\citenamefont {Blodgett}\ \emph {et~al.}(2015)\citenamefont
  {Blodgett}, \citenamefont {Egami}, \citenamefont {Nussinov},\ and\
  \citenamefont {Kelton}}]{blodgett2015proposal}%
  \BibitemOpen
  \bibfield  {author} {\bibinfo {author} {\bibfnamefont {ME}~\bibnamefont
  {Blodgett}}, \bibinfo {author} {\bibfnamefont {Takeshi}\ \bibnamefont
  {Egami}}, \bibinfo {author} {\bibfnamefont {Z}~\bibnamefont {Nussinov}}, \
  and\ \bibinfo {author} {\bibfnamefont {KF}~\bibnamefont {Kelton}},\
  }\bibfield  {title} {\enquote {\bibinfo {title} {Proposal for universality in
  the viscosity of metallic liquids},}\ }\href
  {https://www.nature.com/articles/srep13837} {\bibfield  {journal} {\bibinfo
  {journal} {Scientific reports}\ }\textbf {\bibinfo {volume} {5}},\ \bibinfo
  {pages} {13837} (\bibinfo {year} {2015})}\BibitemShut {NoStop}%
\bibitem [{\citenamefont {Elmatad}\ \emph {et~al.}(2010)\citenamefont
  {Elmatad}, \citenamefont {Chandler},\ and\ \citenamefont
  {Garrahan}}]{elmatad2010corresponding}%
  \BibitemOpen
  \bibfield  {author} {\bibinfo {author} {\bibfnamefont {Yael~S.}\ \bibnamefont
  {Elmatad}}, \bibinfo {author} {\bibfnamefont {David}\ \bibnamefont
  {Chandler}}, \ and\ \bibinfo {author} {\bibfnamefont {Juan~P.}\ \bibnamefont
  {Garrahan}},\ }\bibfield  {title} {\enquote {\bibinfo {title} {Corresponding
  states of structural glass formers. ii},}\ }\href {\doibase
  10.1021/jp1076438} {\bibfield  {journal} {\bibinfo  {journal} {The Journal of
  Physical Chemistry B}\ }\textbf {\bibinfo {volume} {114}},\ \bibinfo {pages}
  {17113--17119} (\bibinfo {year} {2010})}\BibitemShut {NoStop}%
\bibitem [{\citenamefont {Mauro}\ \emph {et~al.}(2009)\citenamefont {Mauro},
  \citenamefont {Yue}, \citenamefont {Ellison}, \citenamefont {Gupta},\ and\
  \citenamefont {Allan}}]{mauro2009viscosity}%
  \BibitemOpen
  \bibfield  {author} {\bibinfo {author} {\bibfnamefont {John~C.}\ \bibnamefont
  {Mauro}}, \bibinfo {author} {\bibfnamefont {Yuanzheng}\ \bibnamefont {Yue}},
  \bibinfo {author} {\bibfnamefont {Adam~J.}\ \bibnamefont {Ellison}}, \bibinfo
  {author} {\bibfnamefont {Prabhat~K.}\ \bibnamefont {Gupta}}, \ and\ \bibinfo
  {author} {\bibfnamefont {Douglas~C.}\ \bibnamefont {Allan}},\ }\bibfield
  {title} {\enquote {\bibinfo {title} {Viscosity of glass-forming liquids},}\
  }\href {\doibase 10.1073/pnas.0911705106} {\bibfield  {journal} {\bibinfo
  {journal} {Proceedings of the National Academy of Sciences}\ }\textbf
  {\bibinfo {volume} {106}},\ \bibinfo {pages} {19780--19784} (\bibinfo {year}
  {2009})}\BibitemShut {NoStop}%
\bibitem [{\citenamefont {Avramov}\ and\ \citenamefont
  {Milchev}(1988)}]{avramov1988effect}%
  \BibitemOpen
  \bibfield  {author} {\bibinfo {author} {\bibfnamefont {I}~\bibnamefont
  {Avramov}}\ and\ \bibinfo {author} {\bibfnamefont {A}~\bibnamefont
  {Milchev}},\ }\bibfield  {title} {\enquote {\bibinfo {title} {Effect of
  disorder on diffusion and viscosity in condensed systems},}\ }\href
  {https://www.sciencedirect.com/science/article/pii/0022309388903961}
  {\bibfield  {journal} {\bibinfo  {journal} {Journal of non-crystalline
  solids}\ }\textbf {\bibinfo {volume} {104}},\ \bibinfo {pages} {253--260}
  (\bibinfo {year} {1988})}\BibitemShut {NoStop}%
\bibitem [{\citenamefont {Coslovich}\ \emph {et~al.}(2018)\citenamefont
  {Coslovich}, \citenamefont {Ozawa},\ and\ \citenamefont
  {Kob}}]{coslovich2018dynamic}%
  \BibitemOpen
  \bibfield  {author} {\bibinfo {author} {\bibfnamefont {Daniele}\ \bibnamefont
  {Coslovich}}, \bibinfo {author} {\bibfnamefont {Misaki}\ \bibnamefont
  {Ozawa}}, \ and\ \bibinfo {author} {\bibfnamefont {Walter}\ \bibnamefont
  {Kob}},\ }\bibfield  {title} {\enquote {\bibinfo {title} {Dynamic and
  thermodynamic crossover scenarios in the kob-andersen mixture: Insights from
  multi-cpu and multi-gpu simulations},}\ }\href {\doibase
  10.1140/epje/i2018-11671-2} {\bibfield  {journal} {\bibinfo  {journal} {The
  European Physical Journal E}\ }\textbf {\bibinfo {volume} {41}},\ \bibinfo
  {pages} {62} (\bibinfo {year} {2018})}\BibitemShut {NoStop}%
\bibitem [{\citenamefont {Novikov}\ and\ \citenamefont
  {Sokolov}(2003)}]{novikov_universality_2003}%
  \BibitemOpen
  \bibfield  {author} {\bibinfo {author} {\bibfnamefont {V.~N.}\ \bibnamefont
  {Novikov}}\ and\ \bibinfo {author} {\bibfnamefont {A.~P.}\ \bibnamefont
  {Sokolov}},\ }\bibfield  {title} {\enquote {\bibinfo {title} {Universality of
  the dynamic crossover in glass-forming liquids: {{A}} ``magic'' relaxation
  time},}\ }\href {\doibase 10.1103/PhysRevE.67.031507} {\bibfield  {journal}
  {\bibinfo  {journal} {Phys. Rev. E}\ }\textbf {\bibinfo {volume} {67}},\
  \bibinfo {pages} {031507} (\bibinfo {year} {2003})}\BibitemShut {NoStop}%
\bibitem [{\citenamefont {Royall}\ \emph {et~al.}(2018)\citenamefont {Royall},
  \citenamefont {Turci}, \citenamefont {Tatsumi}, \citenamefont {Russo},\ and\
  \citenamefont {Robinson}}]{royall2018race}%
  \BibitemOpen
  \bibfield  {author} {\bibinfo {author} {\bibfnamefont {C~Patrick}\
  \bibnamefont {Royall}}, \bibinfo {author} {\bibfnamefont {Francesco}\
  \bibnamefont {Turci}}, \bibinfo {author} {\bibfnamefont {Soichi}\
  \bibnamefont {Tatsumi}}, \bibinfo {author} {\bibfnamefont {John}\
  \bibnamefont {Russo}}, \ and\ \bibinfo {author} {\bibfnamefont {Joshua}\
  \bibnamefont {Robinson}},\ }\bibfield  {title} {\enquote {\bibinfo {title}
  {The race to the bottom: approaching the ideal glass?}}\ }\href {\doibase
  10.1088/1361-648x/aad10a} {\bibfield  {journal} {\bibinfo  {journal} {Journal
  of Physics: Condensed Matter}\ }\textbf {\bibinfo {volume} {30}},\ \bibinfo
  {pages} {363001} (\bibinfo {year} {2018})}\BibitemShut {NoStop}%
\bibitem [{\citenamefont {Takahara}\ \emph {et~al.}(1994)\citenamefont
  {Takahara}, \citenamefont {Yamamuro},\ and\ \citenamefont
  {Suga}}]{takahara1994heat}%
  \BibitemOpen
  \bibfield  {author} {\bibinfo {author} {\bibfnamefont {Shuichi}\ \bibnamefont
  {Takahara}}, \bibinfo {author} {\bibfnamefont {Osamu}\ \bibnamefont
  {Yamamuro}}, \ and\ \bibinfo {author} {\bibfnamefont {Hiroshi}\ \bibnamefont
  {Suga}},\ }\bibfield  {title} {\enquote {\bibinfo {title} {Heat capacities
  and glass transitions of 1-propanol and 3-methylpentane under pressure. new
  evidence for the entropy theory},}\ }\href
  {https://www.sciencedirect.com/science/article/pii/0022309394901953}
  {\bibfield  {journal} {\bibinfo  {journal} {Journal of non-crystalline
  solids}\ }\textbf {\bibinfo {volume} {171}},\ \bibinfo {pages} {259--270}
  (\bibinfo {year} {1994})}\BibitemShut {NoStop}%
\bibitem [{\citenamefont {Yoshimori}\ and\ \citenamefont
  {Odagaki}(2011)}]{yoshimori2011configurational}%
  \BibitemOpen
  \bibfield  {author} {\bibinfo {author} {\bibfnamefont {Akira}\ \bibnamefont
  {Yoshimori}}\ and\ \bibinfo {author} {\bibfnamefont {Takashi}\ \bibnamefont
  {Odagaki}},\ }\bibfield  {title} {\enquote {\bibinfo {title} {Configurational
  entropy and heat capacity in supercooled liquids},}\ }\href
  {https://journals.jps.jp/doi/abs/10.1143/JPSJ.80.064601} {\bibfield
  {journal} {\bibinfo  {journal} {Journal of the Physical Society of Japan}\
  }\textbf {\bibinfo {volume} {80}},\ \bibinfo {pages} {064601} (\bibinfo
  {year} {2011})}\BibitemShut {NoStop}%
\bibitem [{\citenamefont {Smith}\ \emph {et~al.}(2017)\citenamefont {Smith},
  \citenamefont {Li}, \citenamefont {Hoff}, \citenamefont {Garrett},
  \citenamefont {Kim}, \citenamefont {Yang}, \citenamefont {Lucas},
  \citenamefont {Swan-Wood}, \citenamefont {Lin}, \citenamefont {Stone} \emph
  {et~al.}}]{smith2017separating}%
  \BibitemOpen
  \bibfield  {author} {\bibinfo {author} {\bibfnamefont {Hillary~L}\
  \bibnamefont {Smith}}, \bibinfo {author} {\bibfnamefont {Chen~W}\
  \bibnamefont {Li}}, \bibinfo {author} {\bibfnamefont {Andrew}\ \bibnamefont
  {Hoff}}, \bibinfo {author} {\bibfnamefont {Glenn~R}\ \bibnamefont {Garrett}},
  \bibinfo {author} {\bibfnamefont {Dennis~S}\ \bibnamefont {Kim}}, \bibinfo
  {author} {\bibfnamefont {Fred~C}\ \bibnamefont {Yang}}, \bibinfo {author}
  {\bibfnamefont {Matthew~S}\ \bibnamefont {Lucas}}, \bibinfo {author}
  {\bibfnamefont {Tabitha}\ \bibnamefont {Swan-Wood}}, \bibinfo {author}
  {\bibfnamefont {Jiao~YY}\ \bibnamefont {Lin}}, \bibinfo {author}
  {\bibfnamefont {Matthew~B}\ \bibnamefont {Stone}},  \emph {et~al.},\
  }\bibfield  {title} {\enquote {\bibinfo {title} {Separating the
  configurational and vibrational entropy contributions in metallic glasses},}\
  }\href {https://doi.org/10.1038/nphys4142} {\bibfield  {journal} {\bibinfo
  {journal} {Nature Physics}\ }\textbf {\bibinfo {volume} {13}},\ \bibinfo
  {pages} {900} (\bibinfo {year} {2017})}\BibitemShut {NoStop}%
\bibitem [{\citenamefont {Takeda}\ \emph {et~al.}(1999)\citenamefont {Takeda},
  \citenamefont {Yamamuro}, \citenamefont {Tsukushi}, \citenamefont {Matsuo},\
  and\ \citenamefont {Suga}}]{takeda1999calorimetric}%
  \BibitemOpen
  \bibfield  {author} {\bibinfo {author} {\bibfnamefont {Kiyoshi}\ \bibnamefont
  {Takeda}}, \bibinfo {author} {\bibfnamefont {Osamu}\ \bibnamefont
  {Yamamuro}}, \bibinfo {author} {\bibfnamefont {Itaru}\ \bibnamefont
  {Tsukushi}}, \bibinfo {author} {\bibfnamefont {Takasuke}\ \bibnamefont
  {Matsuo}}, \ and\ \bibinfo {author} {\bibfnamefont {Hiroshi}\ \bibnamefont
  {Suga}},\ }\bibfield  {title} {\enquote {\bibinfo {title} {Calorimetric study
  of ethylene glycol and 1, 3-propanediol: configurational entropy in
  supercooled polyalcohols},}\ }\href@noop {} {\bibfield  {journal} {\bibinfo
  {journal} {Journal of molecular structure}\ }\textbf {\bibinfo {volume}
  {479}},\ \bibinfo {pages} {227--235} (\bibinfo {year} {1999})}\BibitemShut
  {NoStop}%
\bibitem [{\citenamefont {Haida}\ \emph {et~al.}(1977)\citenamefont {Haida},
  \citenamefont {Suga},\ and\ \citenamefont {Seki}}]{haida1977calorimetric}%
  \BibitemOpen
  \bibfield  {author} {\bibinfo {author} {\bibfnamefont {Osamu}\ \bibnamefont
  {Haida}}, \bibinfo {author} {\bibfnamefont {Hiroshi}\ \bibnamefont {Suga}}, \
  and\ \bibinfo {author} {\bibfnamefont {Sy{\^u}z{\^o}}\ \bibnamefont {Seki}},\
  }\bibfield  {title} {\enquote {\bibinfo {title} {Calorimetric study of the
  glassy state xii. plural glass-transition phenomena of ethanol},}\ }\href
  {https://www.sciencedirect.com/science/article/pii/002196147790115X}
  {\bibfield  {journal} {\bibinfo  {journal} {The Journal of Chemical
  Thermodynamics}\ }\textbf {\bibinfo {volume} {9}},\ \bibinfo {pages}
  {1133--1148} (\bibinfo {year} {1977})}\BibitemShut {NoStop}%
\bibitem [{\citenamefont {Gibson}\ and\ \citenamefont
  {Giauque}(1923)}]{gibson1923third}%
  \BibitemOpen
  \bibfield  {author} {\bibinfo {author} {\bibfnamefont {G.~E.}\ \bibnamefont
  {Gibson}}\ and\ \bibinfo {author} {\bibfnamefont {W.~F.}\ \bibnamefont
  {Giauque}},\ }\bibfield  {title} {\enquote {\bibinfo {title} {The third law
  of thermodynamics. evidence from the specific heats of glycerol that the
  entropy of a glass exceeds that of a crystal at the absolute zero},}\ }\href
  {\doibase 10.1021/ja01654a014} {\bibfield  {journal} {\bibinfo  {journal}
  {Journal of the American Chemical Society}\ }\textbf {\bibinfo {volume}
  {45}},\ \bibinfo {pages} {93--104} (\bibinfo {year} {1923})}\BibitemShut
  {NoStop}%
\bibitem [{\citenamefont {Beasley}\ \emph {et~al.}()\citenamefont {Beasley},
  \citenamefont {Bishop}, \citenamefont {Kasting},\ and\ \citenamefont
  {Ediger}}]{tobemark}%
  \BibitemOpen
  \bibfield  {author} {\bibinfo {author} {\bibfnamefont {M.~S.}\ \bibnamefont
  {Beasley}}, \bibinfo {author} {\bibfnamefont {C.}~\bibnamefont {Bishop}},
  \bibinfo {author} {\bibfnamefont {B.~J.}\ \bibnamefont {Kasting}}, \ and\
  \bibinfo {author} {\bibfnamefont {M.~D.}\ \bibnamefont {Ediger}},\ }\bibfield
   {title} {\enquote {\bibinfo {title} {Vapor-deposited ethylbenzene glasses
  approach “ideal glass” density},}\ }\href@noop {} {\ }\BibitemShut
  {NoStop}%
\bibitem [{\citenamefont {Chen}\ and\ \citenamefont
  {Richert}(2011)}]{chen2011dynamics}%
  \BibitemOpen
  \bibfield  {author} {\bibinfo {author} {\bibfnamefont {Zhen}\ \bibnamefont
  {Chen}}\ and\ \bibinfo {author} {\bibfnamefont {Ranko}\ \bibnamefont
  {Richert}},\ }\bibfield  {title} {\enquote {\bibinfo {title} {Dynamics of
  glass-forming liquids. xv. dynamical features of molecular liquids that form
  ultra-stable glasses by vapor deposition},}\ }\href
  {https://aip.scitation.org/doi/abs/10.1063/1.3643332} {\bibfield  {journal}
  {\bibinfo  {journal} {The Journal of chemical physics}\ }\textbf {\bibinfo
  {volume} {135}},\ \bibinfo {pages} {124515} (\bibinfo {year}
  {2011})}\BibitemShut {NoStop}%
\bibitem [{\citenamefont {Barlow}\ \emph {et~al.}(1966)\citenamefont {Barlow},
  \citenamefont {Lamb},\ and\ \citenamefont {Matheson}}]{barlow1966viscous}%
  \BibitemOpen
  \bibfield  {author} {\bibinfo {author} {\bibfnamefont {AJ}~\bibnamefont
  {Barlow}}, \bibinfo {author} {\bibfnamefont {J}~\bibnamefont {Lamb}}, \ and\
  \bibinfo {author} {\bibfnamefont {AJ}~\bibnamefont {Matheson}},\ }\bibfield
  {title} {\enquote {\bibinfo {title} {Viscous behaviour of supercooled
  liquids},}\ }\href {https://doi.org/10.1098/rspa.1966.0138} {\bibfield
  {journal} {\bibinfo  {journal} {Proceedings of the Royal Society of London.
  Series A. Mathematical and Physical Sciences}\ }\textbf {\bibinfo {volume}
  {292}},\ \bibinfo {pages} {322--342} (\bibinfo {year} {1966})}\BibitemShut
  {NoStop}%
\bibitem [{\citenamefont {Rossini}(1953)}]{rossini1953selected}%
  \BibitemOpen
  \bibfield  {author} {\bibinfo {author} {\bibfnamefont {Frederick~Dominic}\
  \bibnamefont {Rossini}},\ }\href@noop {} {\emph {\bibinfo {title} {Selected
  values of physical and thermodynamic properties of hydrocarbons and related
  compounds: comprising the tables of the American Petroleum Institute Research
  Project 44 extant as of December 31, 1952}}},\ Vol.~\bibinfo {volume} {44}\
  (\bibinfo  {publisher} {American Petroleum Institute},\ \bibinfo {year}
  {1953})\BibitemShut {NoStop}%
\bibitem [{\citenamefont {Schneider}\ \emph {et~al.}(1998)\citenamefont
  {Schneider}, \citenamefont {Lunkenheimer}, \citenamefont {Brand},\ and\
  \citenamefont {Loidl}}]{schneider1998dielectric}%
  \BibitemOpen
  \bibfield  {author} {\bibinfo {author} {\bibfnamefont {Ulrich}\ \bibnamefont
  {Schneider}}, \bibinfo {author} {\bibfnamefont {P}~\bibnamefont
  {Lunkenheimer}}, \bibinfo {author} {\bibfnamefont {R}~\bibnamefont {Brand}},
  \ and\ \bibinfo {author} {\bibfnamefont {A}~\bibnamefont {Loidl}},\
  }\bibfield  {title} {\enquote {\bibinfo {title} {Dielectric and far-infrared
  spectroscopy of glycerol},}\ }\href
  {https://www.sciencedirect.com/science/article/pii/S0022309398005614}
  {\bibfield  {journal} {\bibinfo  {journal} {Journal of non-crystalline
  solids}\ }\textbf {\bibinfo {volume} {235}},\ \bibinfo {pages} {173--179}
  (\bibinfo {year} {1998})}\BibitemShut {NoStop}%
\bibitem [{\citenamefont {Lunkenheimer}\ \emph {et~al.}(2000)\citenamefont
  {Lunkenheimer}, \citenamefont {Schneider}, \citenamefont {Brand},\ and\
  \citenamefont {Loid}}]{lunkenheimer2000glassy}%
  \BibitemOpen
  \bibfield  {author} {\bibinfo {author} {\bibfnamefont {P.}~\bibnamefont
  {Lunkenheimer}}, \bibinfo {author} {\bibfnamefont {U.}~\bibnamefont
  {Schneider}}, \bibinfo {author} {\bibfnamefont {R.}~\bibnamefont {Brand}}, \
  and\ \bibinfo {author} {\bibfnamefont {A.}~\bibnamefont {Loid}},\ }\bibfield
  {title} {\enquote {\bibinfo {title} {Glassy dynamics},}\ }\href {\doibase
  10.1080/001075100181259} {\bibfield  {journal} {\bibinfo  {journal}
  {Contemporary Physics}\ }\textbf {\bibinfo {volume} {41}},\ \bibinfo {pages}
  {15--36} (\bibinfo {year} {2000})}\BibitemShut {NoStop}%
\bibitem [{\citenamefont {Lunkenheimer}\ \emph {et~al.}(2005)\citenamefont
  {Lunkenheimer}, \citenamefont {Wehn}, \citenamefont {Schneider},\ and\
  \citenamefont {Loidl}}]{lunkenheimer2005glassy}%
  \BibitemOpen
  \bibfield  {author} {\bibinfo {author} {\bibfnamefont {P.}~\bibnamefont
  {Lunkenheimer}}, \bibinfo {author} {\bibfnamefont {R.}~\bibnamefont {Wehn}},
  \bibinfo {author} {\bibfnamefont {U.}~\bibnamefont {Schneider}}, \ and\
  \bibinfo {author} {\bibfnamefont {A.}~\bibnamefont {Loidl}},\ }\bibfield
  {title} {\enquote {\bibinfo {title} {Glassy aging dynamics},}\ }\href
  {\doibase 10.1103/PhysRevLett.95.055702} {\bibfield  {journal} {\bibinfo
  {journal} {Phys. Rev. Lett.}\ }\textbf {\bibinfo {volume} {95}},\ \bibinfo
  {pages} {055702} (\bibinfo {year} {2005})}\BibitemShut {NoStop}%
\bibitem [{\citenamefont {Schmidtke}\ \emph {et~al.}(2012)\citenamefont
  {Schmidtke}, \citenamefont {Petzold}, \citenamefont {Kahlau}, \citenamefont
  {Hofmann},\ and\ \citenamefont {R{\"o}ssler}}]{schmidtke2012boiling}%
  \BibitemOpen
  \bibfield  {author} {\bibinfo {author} {\bibfnamefont {B}~\bibnamefont
  {Schmidtke}}, \bibinfo {author} {\bibfnamefont {N}~\bibnamefont {Petzold}},
  \bibinfo {author} {\bibfnamefont {R}~\bibnamefont {Kahlau}}, \bibinfo
  {author} {\bibfnamefont {M}~\bibnamefont {Hofmann}}, \ and\ \bibinfo {author}
  {\bibfnamefont {EA}~\bibnamefont {R{\"o}ssler}},\ }\bibfield  {title}
  {\enquote {\bibinfo {title} {From boiling point to glass transition
  temperature: Transport coefficients in molecular liquids follow
  three-parameter scaling},}\ }\href
  {http://dx.doi.org/10.1103/PhysRevE.86.041507} {\bibfield  {journal}
  {\bibinfo  {journal} {Physical Review E}\ }\textbf {\bibinfo {volume} {86}},\
  \bibinfo {pages} {041507} (\bibinfo {year} {2012})}\BibitemShut {NoStop}%
\bibitem [{\citenamefont {Hansen}\ \emph {et~al.}(1997)\citenamefont {Hansen},
  \citenamefont {Stickel}, \citenamefont {Berger}, \citenamefont {Richert},\
  and\ \citenamefont {Fischer}}]{hansen1997dynamics}%
  \BibitemOpen
  \bibfield  {author} {\bibinfo {author} {\bibfnamefont {C}~\bibnamefont
  {Hansen}}, \bibinfo {author} {\bibfnamefont {F}~\bibnamefont {Stickel}},
  \bibinfo {author} {\bibfnamefont {T}~\bibnamefont {Berger}}, \bibinfo
  {author} {\bibfnamefont {Ranko}\ \bibnamefont {Richert}}, \ and\ \bibinfo
  {author} {\bibfnamefont {Erhard~W}\ \bibnamefont {Fischer}},\ }\bibfield
  {title} {\enquote {\bibinfo {title} {Dynamics of glass-forming liquids. iii.
  comparing the dielectric $\alpha$-and $\beta$-relaxation of 1-propanol and
  o-terphenyl},}\ }\href@noop {} {\bibfield  {journal} {\bibinfo  {journal}
  {The Journal of chemical physics}\ }\textbf {\bibinfo {volume} {107}},\
  \bibinfo {pages} {1086--1093} (\bibinfo {year} {1997})}\BibitemShut {NoStop}%
\bibitem [{\citenamefont {Sillr{\'e}n}\ \emph {et~al.}(2014)\citenamefont
  {Sillr{\'e}n}, \citenamefont {Matic}, \citenamefont {Karlsson}, \citenamefont
  {Koza}, \citenamefont {Maccarini}, \citenamefont {Fouquet}, \citenamefont
  {G{\"o}tz}, \citenamefont {Bauer}, \citenamefont {Gulich}, \citenamefont
  {Lunkenheimer} \emph {et~al.}}]{sillren2014liquid}%
  \BibitemOpen
  \bibfield  {author} {\bibinfo {author} {\bibfnamefont {Per}\ \bibnamefont
  {Sillr{\'e}n}}, \bibinfo {author} {\bibfnamefont {Aleksandar}\ \bibnamefont
  {Matic}}, \bibinfo {author} {\bibfnamefont {Maths}\ \bibnamefont {Karlsson}},
  \bibinfo {author} {\bibfnamefont {M}~\bibnamefont {Koza}}, \bibinfo {author}
  {\bibfnamefont {M}~\bibnamefont {Maccarini}}, \bibinfo {author}
  {\bibfnamefont {P}~\bibnamefont {Fouquet}}, \bibinfo {author} {\bibfnamefont
  {M}~\bibnamefont {G{\"o}tz}}, \bibinfo {author} {\bibfnamefont
  {Th}~\bibnamefont {Bauer}}, \bibinfo {author} {\bibfnamefont {R}~\bibnamefont
  {Gulich}}, \bibinfo {author} {\bibfnamefont {P}~\bibnamefont {Lunkenheimer}},
   \emph {et~al.},\ }\bibfield  {title} {\enquote {\bibinfo {title} {Liquid
  1-propanol studied by neutron scattering, near-infrared, and dielectric
  spectroscopy},}\ }\href
  {https://aip.scitation.org/doi/full/10.1063/1.4868556} {\bibfield  {journal}
  {\bibinfo  {journal} {The Journal of chemical physics}\ }\textbf {\bibinfo
  {volume} {140}},\ \bibinfo {pages} {124501} (\bibinfo {year}
  {2014})}\BibitemShut {NoStop}%
\bibitem [{\citenamefont {Schneider}\ \emph {et~al.}(1999)\citenamefont
  {Schneider}, \citenamefont {Lunkenheimer}, \citenamefont {Brand},\ and\
  \citenamefont {Loidl}}]{schneider1999broadband}%
  \BibitemOpen
  \bibfield  {author} {\bibinfo {author} {\bibfnamefont {U.}~\bibnamefont
  {Schneider}}, \bibinfo {author} {\bibfnamefont {P.}~\bibnamefont
  {Lunkenheimer}}, \bibinfo {author} {\bibfnamefont {R.}~\bibnamefont {Brand}},
  \ and\ \bibinfo {author} {\bibfnamefont {A.}~\bibnamefont {Loidl}},\
  }\bibfield  {title} {\enquote {\bibinfo {title} {Broadband dielectric
  spectroscopy on glass-forming propylene carbonate},}\ }\href {\doibase
  10.1103/PhysRevE.59.6924} {\bibfield  {journal} {\bibinfo  {journal} {Phys.
  Rev. E}\ }\textbf {\bibinfo {volume} {59}},\ \bibinfo {pages} {6924--6936}
  (\bibinfo {year} {1999})}\BibitemShut {NoStop}%
\bibitem [{\citenamefont {Stickel}\ \emph {et~al.}(1996)\citenamefont
  {Stickel}, \citenamefont {Fischer},\ and\ \citenamefont
  {Richert}}]{stickel1996dynamics}%
  \BibitemOpen
  \bibfield  {author} {\bibinfo {author} {\bibfnamefont {F}~\bibnamefont
  {Stickel}}, \bibinfo {author} {\bibfnamefont {Erhard~W}\ \bibnamefont
  {Fischer}}, \ and\ \bibinfo {author} {\bibfnamefont {Ranko}\ \bibnamefont
  {Richert}},\ }\bibfield  {title} {\enquote {\bibinfo {title} {Dynamics of
  glass-forming liquids. ii. detailed comparison of dielectric relaxation,
  dc-conductivity, and viscosity data},}\ }\href
  {https://aip.scitation.org/doi/abs/10.1063/1.470961} {\bibfield  {journal}
  {\bibinfo  {journal} {The Journal of chemical physics}\ }\textbf {\bibinfo
  {volume} {104}},\ \bibinfo {pages} {2043--2055} (\bibinfo {year}
  {1996})}\BibitemShut {NoStop}%
\bibitem [{\citenamefont {Berberian}\ and\ \citenamefont
  {Cole}(1986)}]{berberian1986approach}%
  \BibitemOpen
  \bibfield  {author} {\bibinfo {author} {\bibfnamefont {John~G}\ \bibnamefont
  {Berberian}}\ and\ \bibinfo {author} {\bibfnamefont {Robert~H}\ \bibnamefont
  {Cole}},\ }\bibfield  {title} {\enquote {\bibinfo {title} {Approach to glassy
  behavior of dielectric relaxation in 3-bromopentane from 298 to 107 k},}\
  }\href@noop {} {\bibfield  {journal} {\bibinfo  {journal} {The Journal of
  chemical physics}\ }\textbf {\bibinfo {volume} {84}},\ \bibinfo {pages}
  {6921--6927} (\bibinfo {year} {1986})}\BibitemShut {NoStop}%
\bibitem [{\citenamefont {Tatsumi}\ and\ \citenamefont
  {Yamamuro}()}]{tatsumi_unpublished}%
  \BibitemOpen
  \bibfield  {author} {\bibinfo {author} {\bibfnamefont {Soichi}\ \bibnamefont
  {Tatsumi}}\ and\ \bibinfo {author} {\bibfnamefont {Osamu}\ \bibnamefont
  {Yamamuro}},\ }\bibfield  {title} {\enquote {\bibinfo {title} {Unpublished
  data},}\ }\href@noop {} {\ }\BibitemShut {NoStop}%
\bibitem [{\citenamefont {Brand}\ \emph {et~al.}(2000)\citenamefont {Brand},
  \citenamefont {Lunkenheimer}, \citenamefont {Schneider},\ and\ \citenamefont
  {Loidl}}]{brand2000excess}%
  \BibitemOpen
  \bibfield  {author} {\bibinfo {author} {\bibfnamefont {R.}~\bibnamefont
  {Brand}}, \bibinfo {author} {\bibfnamefont {P.}~\bibnamefont {Lunkenheimer}},
  \bibinfo {author} {\bibfnamefont {U.}~\bibnamefont {Schneider}}, \ and\
  \bibinfo {author} {\bibfnamefont {A.}~\bibnamefont {Loidl}},\ }\bibfield
  {title} {\enquote {\bibinfo {title} {Excess wing in the dielectric loss of
  glass-forming ethanol: A relaxation process},}\ }\href {\doibase
  10.1103/PhysRevB.62.8878} {\bibfield  {journal} {\bibinfo  {journal} {Phys.
  Rev. B}\ }\textbf {\bibinfo {volume} {62}},\ \bibinfo {pages} {8878--8883}
  (\bibinfo {year} {2000})}\BibitemShut {NoStop}%
\bibitem [{\citenamefont {Frenkel}\ and\ \citenamefont
  {Smit}(2002)}]{Frenkelbook}%
  \BibitemOpen
  \bibinfo {editor} {\bibfnamefont {Daan}\ \bibnamefont {Frenkel}}\ and\
  \bibinfo {editor} {\bibfnamefont {Berend}\ \bibnamefont {Smit}},\ eds.,\
  \href@noop {} {\emph {\bibinfo {title} {Understanding Molecular Simulation
  (Second Edition)}}},\ \bibinfo {edition} {second edition}\ ed.\ (\bibinfo
  {publisher} {Academic Press},\ \bibinfo {address} {San Diego},\ \bibinfo
  {year} {2002})\BibitemShut {NoStop}%
\bibitem [{\citenamefont {Franz}\ and\ \citenamefont {Parisi}(1997)}]{FP97}%
  \BibitemOpen
  \bibfield  {author} {\bibinfo {author} {\bibfnamefont {Silvio}\ \bibnamefont
  {Franz}}\ and\ \bibinfo {author} {\bibfnamefont {Giorgio}\ \bibnamefont
  {Parisi}},\ }\bibfield  {title} {\enquote {\bibinfo {title} {Phase diagram of
  coupled glassy systems: A mean-field study},}\ }\href {\doibase
  10.1103/PhysRevLett.79.2486} {\bibfield  {journal} {\bibinfo  {journal}
  {Phys. Rev. Lett.}\ }\textbf {\bibinfo {volume} {79}},\ \bibinfo {pages}
  {2486--2489} (\bibinfo {year} {1997})}\BibitemShut {NoStop}%
\bibitem [{\citenamefont {Berthier}\ and\ \citenamefont
  {Coslovich}(2014)}]{BC14}%
  \BibitemOpen
  \bibfield  {author} {\bibinfo {author} {\bibfnamefont {Ludovic}\ \bibnamefont
  {Berthier}}\ and\ \bibinfo {author} {\bibfnamefont {Daniele}\ \bibnamefont
  {Coslovich}},\ }\bibfield  {title} {\enquote {\bibinfo {title} {Novel
  approach to numerical measurements of the configurational entropy in
  supercooled liquids},}\ }\href {\doibase 10.1073/pnas.1407934111} {\bibfield
  {journal} {\bibinfo  {journal} {Proceedings of the National Academy of
  Sciences}\ }\textbf {\bibinfo {volume} {111}},\ \bibinfo {pages}
  {11668--11672} (\bibinfo {year} {2014})}\BibitemShut {NoStop}%
\bibitem [{\citenamefont {Franz}(2005)}]{effort2}%
  \BibitemOpen
  \bibfield  {author} {\bibinfo {author} {\bibfnamefont {Silvio}\ \bibnamefont
  {Franz}},\ }\bibfield  {title} {\enquote {\bibinfo {title} {First steps of a
  nucleation theory in disordered systems},}\ }\href {\doibase
  10.1088/1742-5468/2005/04/p04001} {\bibfield  {journal} {\bibinfo  {journal}
  {Journal of Statistical Mechanics: Theory and Experiment}\ }\textbf {\bibinfo
  {volume} {2005}},\ \bibinfo {pages} {P04001} (\bibinfo {year}
  {2005})}\BibitemShut {NoStop}%
\bibitem [{\citenamefont {Berthier}\ \emph
  {et~al.}(2019{\natexlab{c}})\citenamefont {Berthier}, \citenamefont
  {Charbonneau},\ and\ \citenamefont {Kundu}}]{kundu}%
  \BibitemOpen
  \bibfield  {author} {\bibinfo {author} {\bibfnamefont {Ludovic}\ \bibnamefont
  {Berthier}}, \bibinfo {author} {\bibfnamefont {Patrick}\ \bibnamefont
  {Charbonneau}}, \ and\ \bibinfo {author} {\bibfnamefont {Joyjit}\
  \bibnamefont {Kundu}},\ }\bibfield  {title} {\enquote {\bibinfo {title}
  {Bypassing sluggishness: Swap algorithm and glassiness in high dimensions},}\
  }\href {\doibase 10.1103/PhysRevE.99.031301} {\bibfield  {journal} {\bibinfo
  {journal} {Phys. Rev. E}\ }\textbf {\bibinfo {volume} {99}},\ \bibinfo
  {pages} {031301} (\bibinfo {year} {2019}{\natexlab{c}})}\BibitemShut
  {NoStop}%
\bibitem [{\citenamefont {Lubchenko}\ and\ \citenamefont
  {Rabochiy}(2014)}]{lubchenko2014mechanism}%
  \BibitemOpen
  \bibfield  {author} {\bibinfo {author} {\bibfnamefont {Vassiliy}\
  \bibnamefont {Lubchenko}}\ and\ \bibinfo {author} {\bibfnamefont {Pyotr}\
  \bibnamefont {Rabochiy}},\ }\bibfield  {title} {\enquote {\bibinfo {title}
  {On the mechanism of activated transport in glassy liquids},}\ }\href
  {https://dx.doi.org/10.1021/jp508635n} {\bibfield  {journal} {\bibinfo
  {journal} {The Journal of Physical Chemistry B}\ }\textbf {\bibinfo {volume}
  {118}},\ \bibinfo {pages} {13744--13759} (\bibinfo {year}
  {2014})}\BibitemShut {NoStop}%
\bibitem [{\citenamefont {Tarjus}(2011)}]{Ta11}%
  \BibitemOpen
  \bibfield  {author} {\bibinfo {author} {\bibfnamefont {G.}~\bibnamefont
  {Tarjus}},\ }\bibfield  {title} {\enquote {\bibinfo {title} {An overview of
  the theories of the glass transition},}\ }in\ \href@noop {} {\emph {\bibinfo
  {booktitle} {Dynamical Heterogeneities and Glasses}}},\ \bibinfo {editor}
  {edited by\ \bibinfo {editor} {\bibfnamefont {L.}~\bibnamefont {Berthier}},
  \bibinfo {editor} {\bibfnamefont {G.}~\bibnamefont {Biroli}}, \bibinfo
  {editor} {\bibfnamefont {J-P}\ \bibnamefont {Bouchaud}}, \bibinfo {editor}
  {\bibfnamefont {L.}~\bibnamefont {Cipelletti}}, \ and\ \bibinfo {editor}
  {\bibfnamefont {W.}~\bibnamefont {van Saarloos}}}\ (\bibinfo  {publisher}
  {Oxford University Press},\ \bibinfo {year} {2011})\ \Eprint
  {http://arxiv.org/abs/{\tt arXiv:1010.2938}} {{\tt arXiv:1010.2938}}
  \BibitemShut {NoStop}%
\bibitem [{\citenamefont {Cammarota}\ \emph
  {et~al.}(2009{\natexlab{b}})\citenamefont {Cammarota}, \citenamefont
  {Cavagna}, \citenamefont {Gradenigo}, \citenamefont {Grigera},\ and\
  \citenamefont {Verrocchio}}]{cammarota2009evidence}%
  \BibitemOpen
  \bibfield  {author} {\bibinfo {author} {\bibfnamefont {C}~\bibnamefont
  {Cammarota}}, \bibinfo {author} {\bibfnamefont {A}~\bibnamefont {Cavagna}},
  \bibinfo {author} {\bibfnamefont {G}~\bibnamefont {Gradenigo}}, \bibinfo
  {author} {\bibfnamefont {T~S}\ \bibnamefont {Grigera}}, \ and\ \bibinfo
  {author} {\bibfnamefont {P}~\bibnamefont {Verrocchio}},\ }\bibfield  {title}
  {\enquote {\bibinfo {title} {Evidence for a spinodal limit of amorphous
  excitations in glassy systems},}\ }\href {\doibase
  10.1088/1742-5468/2009/12/l12002} {\bibfield  {journal} {\bibinfo  {journal}
  {Journal of Statistical Mechanics: Theory and Experiment}\ }\textbf {\bibinfo
  {volume} {2009}},\ \bibinfo {pages} {L12002} (\bibinfo {year}
  {2009}{\natexlab{b}})}\BibitemShut {NoStop}%
\bibitem [{\citenamefont {Ganapathi}\ \emph {et~al.}(2018)\citenamefont
  {Ganapathi}, \citenamefont {Nagamanasa}, \citenamefont {Sood},\ and\
  \citenamefont {Ganapathy}}]{ganapathi2018measurements}%
  \BibitemOpen
  \bibfield  {author} {\bibinfo {author} {\bibfnamefont {Divya}\ \bibnamefont
  {Ganapathi}}, \bibinfo {author} {\bibfnamefont {K~Hima}\ \bibnamefont
  {Nagamanasa}}, \bibinfo {author} {\bibfnamefont {AK}~\bibnamefont {Sood}}, \
  and\ \bibinfo {author} {\bibfnamefont {Rajesh}\ \bibnamefont {Ganapathy}},\
  }\bibfield  {title} {\enquote {\bibinfo {title} {Measurements of growing
  surface tension of amorphous--amorphous interfaces on approaching the
  colloidal glass transition},}\ }\href
  {https://www.nature.com/articles/s41467-018-02836-6} {\bibfield  {journal}
  {\bibinfo  {journal} {Nature communications}\ }\textbf {\bibinfo {volume}
  {9}},\ \bibinfo {pages} {397} (\bibinfo {year} {2018})}\BibitemShut {NoStop}%
\bibitem [{\citenamefont {Chua}\ \emph {et~al.}(2017)\citenamefont {Chua},
  \citenamefont {Young-Gonzales}, \citenamefont {Richert}, \citenamefont
  {Ediger},\ and\ \citenamefont {Schick}}]{chua2017dynamics}%
  \BibitemOpen
  \bibfield  {author} {\bibinfo {author} {\bibfnamefont {YZ}~\bibnamefont
  {Chua}}, \bibinfo {author} {\bibfnamefont {AR}~\bibnamefont
  {Young-Gonzales}}, \bibinfo {author} {\bibfnamefont {Ranko}\ \bibnamefont
  {Richert}}, \bibinfo {author} {\bibfnamefont {MD}~\bibnamefont {Ediger}}, \
  and\ \bibinfo {author} {\bibfnamefont {C}~\bibnamefont {Schick}},\ }\bibfield
   {title} {\enquote {\bibinfo {title} {Dynamics of supercooled liquid and
  plastic crystalline ethanol: Dielectric relaxation and ac nanocalorimetry
  distinguish structural $\alpha$-and debye relaxation processes},}\
  }\href@noop {} {\bibfield  {journal} {\bibinfo  {journal} {The Journal of
  chemical physics}\ }\textbf {\bibinfo {volume} {147}},\ \bibinfo {pages}
  {014502} (\bibinfo {year} {2017})}\BibitemShut {NoStop}%
\bibitem [{\citenamefont {Goldstein}(1976)}]{gold76}%
  \BibitemOpen
  \bibfield  {author} {\bibinfo {author} {\bibfnamefont {Martin}\ \bibnamefont
  {Goldstein}},\ }\bibfield  {title} {\enquote {\bibinfo {title} {Viscous
  liquids and the glass transition. v. sources of the excess specific heat of
  the liquid},}\ }\href {\doibase 10.1063/1.432063} {\bibfield  {journal}
  {\bibinfo  {journal} {J. Chem. Phys.}\ }\textbf {\bibinfo {volume} {64}},\
  \bibinfo {pages} {4767--4774} (\bibinfo {year} {1976})}\BibitemShut {NoStop}%
\bibitem [{\citenamefont {Yamamuro}\ \emph {et~al.}(1998)\citenamefont
  {Yamamuro}, \citenamefont {Tsukushi}, \citenamefont {Lindqvist},
  \citenamefont {Takahara}, \citenamefont {Ishikawa},\ and\ \citenamefont
  {Matsuo}}]{yamamuro98}%
  \BibitemOpen
  \bibfield  {author} {\bibinfo {author} {\bibfnamefont {Osamu}\ \bibnamefont
  {Yamamuro}}, \bibinfo {author} {\bibfnamefont {Itaru}\ \bibnamefont
  {Tsukushi}}, \bibinfo {author} {\bibfnamefont {Anna}\ \bibnamefont
  {Lindqvist}}, \bibinfo {author} {\bibfnamefont {Shuichi}\ \bibnamefont
  {Takahara}}, \bibinfo {author} {\bibfnamefont {Mariko}\ \bibnamefont
  {Ishikawa}}, \ and\ \bibinfo {author} {\bibfnamefont {Takasuke}\ \bibnamefont
  {Matsuo}},\ }\bibfield  {title} {\enquote {\bibinfo {title} {Calorimetric
  study of glassy and liquid toluene and ethylbenzene:? thermodynamic approach
  to spatial heterogeneity in glass-forming molecular liquids},}\ }\href
  {\doibase 10.1021/jp973439v} {\bibfield  {journal} {\bibinfo  {journal} {The
  Journal of Physical Chemistry B}\ }\textbf {\bibinfo {volume} {102}},\
  \bibinfo {pages} {1605--1609} (\bibinfo {year} {1998})}\BibitemShut {NoStop}%
\bibitem [{\citenamefont {Johari}(2000)}]{J00}%
  \BibitemOpen
  \bibfield  {author} {\bibinfo {author} {\bibfnamefont {G.~P.}\ \bibnamefont
  {Johari}},\ }\bibfield  {title} {\enquote {\bibinfo {title} {Contributions to
  the entropy of a glass and liquid, and the dielectric relaxation time},}\
  }\href {\doibase 10.1063/1.481349} {\bibfield  {journal} {\bibinfo  {journal}
  {The Journal of Chemical Physics}\ }\textbf {\bibinfo {volume} {112}},\
  \bibinfo {pages} {7518--7523} (\bibinfo {year} {2000})}\BibitemShut {NoStop}%
\bibitem [{\citenamefont {Martinez}\ and\ \citenamefont
  {Angell}(2001)}]{martinez2001thermodynamic}%
  \BibitemOpen
  \bibfield  {author} {\bibinfo {author} {\bibfnamefont {L-M}\ \bibnamefont
  {Martinez}}\ and\ \bibinfo {author} {\bibfnamefont {CA}~\bibnamefont
  {Angell}},\ }\bibfield  {title} {\enquote {\bibinfo {title} {A thermodynamic
  connection to the fragility of glass-forming liquids},}\ }\href
  {https://doi.org/10.1038/35070517} {\bibfield  {journal} {\bibinfo  {journal}
  {Nature}\ }\textbf {\bibinfo {volume} {410}},\ \bibinfo {pages} {663}
  (\bibinfo {year} {2001})}\BibitemShut {NoStop}%
\bibitem [{\citenamefont {Angell}\ and\ \citenamefont
  {Borick}(2002)}]{angell2002specific}%
  \BibitemOpen
  \bibfield  {author} {\bibinfo {author} {\bibfnamefont {CA}~\bibnamefont
  {Angell}}\ and\ \bibinfo {author} {\bibfnamefont {S}~\bibnamefont {Borick}},\
  }\bibfield  {title} {\enquote {\bibinfo {title} {Specific heats $c_{\rm p}$,
  $c_{\rm v}$, $c_{\rm conf}$ conf and energy landscapes of glassforming
  liquids},}\ }\href
  {http://www.sciencedirect.com/science/article/pii/S0022309302015004}
  {\bibfield  {journal} {\bibinfo  {journal} {J. Non-Cryst. Solids}\ }\textbf
  {\bibinfo {volume} {307}},\ \bibinfo {pages} {393} (\bibinfo {year}
  {2002})}\BibitemShut {NoStop}%
\bibitem [{\citenamefont {Alvarez-Donado}\ and\ \citenamefont
  {Antonelli}(2019)}]{alvarez2019vibrational}%
  \BibitemOpen
  \bibfield  {author} {\bibinfo {author} {\bibfnamefont {Ren{\'e}}\
  \bibnamefont {Alvarez-Donado}}\ and\ \bibinfo {author} {\bibfnamefont {Alex}\
  \bibnamefont {Antonelli}},\ }\bibfield  {title} {\enquote {\bibinfo {title}
  {Vibrational and configurational entropy separation in bulk metallic glasses:
  A thermodynamic approach},}\ }\href@noop {} {\bibfield  {journal} {\bibinfo
  {journal} {arXiv preprint arXiv:1907.02611}\ } (\bibinfo {year}
  {2019})}\BibitemShut {NoStop}%
\bibitem [{\citenamefont {Han}\ \emph {et~al.}(2019)\citenamefont {Han},
  \citenamefont {Wei}, \citenamefont {Yang}, \citenamefont {Li}, \citenamefont
  {Jiang}, \citenamefont {Wang}, \citenamefont {Dai},\ and\ \citenamefont
  {Zaccone}}]{han2019structural}%
  \BibitemOpen
  \bibfield  {author} {\bibinfo {author} {\bibfnamefont {Dong}\ \bibnamefont
  {Han}}, \bibinfo {author} {\bibfnamefont {Dan}\ \bibnamefont {Wei}}, \bibinfo
  {author} {\bibfnamefont {Jie}\ \bibnamefont {Yang}}, \bibinfo {author}
  {\bibfnamefont {Hui-Ling}\ \bibnamefont {Li}}, \bibinfo {author}
  {\bibfnamefont {Min-Qiang}\ \bibnamefont {Jiang}}, \bibinfo {author}
  {\bibfnamefont {Yun-Jiang}\ \bibnamefont {Wang}}, \bibinfo {author}
  {\bibfnamefont {Lan-Hong}\ \bibnamefont {Dai}}, \ and\ \bibinfo {author}
  {\bibfnamefont {Alessio}\ \bibnamefont {Zaccone}},\ }\bibfield  {title}
  {\enquote {\bibinfo {title} {Structural atomistic mechanism for the glass
  transition entropic scenario},}\ }\href@noop {} {\bibfield  {journal}
  {\bibinfo  {journal} {arXiv preprint arXiv:1907.03695}\ } (\bibinfo {year}
  {2019})}\BibitemShut {NoStop}%
\bibitem [{\citenamefont {Ozawa}\ and\ \citenamefont {Berthier}(2017)}]{OB17}%
  \BibitemOpen
  \bibfield  {author} {\bibinfo {author} {\bibfnamefont {Misaki}\ \bibnamefont
  {Ozawa}}\ and\ \bibinfo {author} {\bibfnamefont {Ludovic}\ \bibnamefont
  {Berthier}},\ }\bibfield  {title} {\enquote {\bibinfo {title} {Does the
  configurational entropy of polydisperse particles exist?}}\ }\href {\doibase
  10.1063/1.4972525} {\bibfield  {journal} {\bibinfo  {journal} {The Journal of
  Chemical Physics}\ }\textbf {\bibinfo {volume} {146}},\ \bibinfo {pages}
  {014502} (\bibinfo {year} {2017})}\BibitemShut {NoStop}%
\bibitem [{\citenamefont {Fisher}\ and\ \citenamefont
  {Huse}(1988)}]{fisher1988nonequilibrium}%
  \BibitemOpen
  \bibfield  {author} {\bibinfo {author} {\bibfnamefont {Daniel~S}\
  \bibnamefont {Fisher}}\ and\ \bibinfo {author} {\bibfnamefont {David~A}\
  \bibnamefont {Huse}},\ }\bibfield  {title} {\enquote {\bibinfo {title}
  {Nonequilibrium dynamics of spin glasses},}\ }\href
  {https://link.aps.org/doi/10.1103/PhysRevB.38.373} {\bibfield  {journal}
  {\bibinfo  {journal} {Physical Review B}\ }\textbf {\bibinfo {volume} {38}},\
  \bibinfo {pages} {373} (\bibinfo {year} {1988})}\BibitemShut {NoStop}%
\bibitem [{\citenamefont {Balog}\ and\ \citenamefont {Tarjus}(2015)}]{psirfim}%
  \BibitemOpen
  \bibfield  {author} {\bibinfo {author} {\bibfnamefont {Ivan}\ \bibnamefont
  {Balog}}\ and\ \bibinfo {author} {\bibfnamefont {Gilles}\ \bibnamefont
  {Tarjus}},\ }\bibfield  {title} {\enquote {\bibinfo {title} {Activated
  dynamic scaling in the random-field ising model: A nonperturbative functional
  renormalization group approach},}\ }\href {\doibase
  10.1103/PhysRevB.91.214201} {\bibfield  {journal} {\bibinfo  {journal} {Phys.
  Rev. B}\ }\textbf {\bibinfo {volume} {91}},\ \bibinfo {pages} {214201}
  (\bibinfo {year} {2015})}\BibitemShut {NoStop}%
\bibitem [{\citenamefont {Tanaka}(2003)}]{tanaka2003relation}%
  \BibitemOpen
  \bibfield  {author} {\bibinfo {author} {\bibfnamefont {Hajime}\ \bibnamefont
  {Tanaka}},\ }\bibfield  {title} {\enquote {\bibinfo {title} {Relation between
  thermodynamics and kinetics of glass-forming liquids},}\ }\href
  {https://journals.aps.org/prl/abstract/10.1103/PhysRevLett.90.055701}
  {\bibfield  {journal} {\bibinfo  {journal} {Physical review letters}\
  }\textbf {\bibinfo {volume} {90}},\ \bibinfo {pages} {055701} (\bibinfo
  {year} {2003})}\BibitemShut {NoStop}%
\bibitem [{\citenamefont {Hecksher}\ \emph {et~al.}(2008)\citenamefont
  {Hecksher}, \citenamefont {Nielsen}, \citenamefont {Olsen},\ and\
  \citenamefont {Dyre}}]{hecksher2008little}%
  \BibitemOpen
  \bibfield  {author} {\bibinfo {author} {\bibfnamefont {Tina}\ \bibnamefont
  {Hecksher}}, \bibinfo {author} {\bibfnamefont {Albena~I}\ \bibnamefont
  {Nielsen}}, \bibinfo {author} {\bibfnamefont {Niels~Boye}\ \bibnamefont
  {Olsen}}, \ and\ \bibinfo {author} {\bibfnamefont {Jeppe~C}\ \bibnamefont
  {Dyre}},\ }\bibfield  {title} {\enquote {\bibinfo {title} {Little evidence
  for dynamic divergences in ultraviscous molecular liquids},}\ }\href
  {https://www.nature.com/articles/nphys1033} {\bibfield  {journal} {\bibinfo
  {journal} {Nature Physics}\ }\textbf {\bibinfo {volume} {4}},\ \bibinfo
  {pages} {737} (\bibinfo {year} {2008})}\BibitemShut {NoStop}%
\bibitem [{\citenamefont {Tarjus}\ \emph {et~al.}(2005)\citenamefont {Tarjus},
  \citenamefont {Kivelson}, \citenamefont {Nussinov},\ and\ \citenamefont
  {Viot}}]{tarjus2005frustration}%
  \BibitemOpen
  \bibfield  {author} {\bibinfo {author} {\bibfnamefont {G}~\bibnamefont
  {Tarjus}}, \bibinfo {author} {\bibfnamefont {S~A}\ \bibnamefont {Kivelson}},
  \bibinfo {author} {\bibfnamefont {Z}~\bibnamefont {Nussinov}}, \ and\
  \bibinfo {author} {\bibfnamefont {P}~\bibnamefont {Viot}},\ }\bibfield
  {title} {\enquote {\bibinfo {title} {The frustration-based approach of
  supercooled liquids and the glass transition: a review and critical
  assessment},}\ }\href {\doibase 10.1088/0953-8984/17/50/r01} {\bibfield
  {journal} {\bibinfo  {journal} {Journal of Physics: Condensed Matter}\
  }\textbf {\bibinfo {volume} {17}},\ \bibinfo {pages} {R1143--R1182} (\bibinfo
  {year} {2005})}\BibitemShut {NoStop}%
\bibitem [{\citenamefont {Rabochiy}\ \emph {et~al.}(2013)\citenamefont
  {Rabochiy}, \citenamefont {Wolynes},\ and\ \citenamefont
  {Lubchenko}}]{doi:10.1021/jp409502k}%
  \BibitemOpen
  \bibfield  {author} {\bibinfo {author} {\bibfnamefont {Pyotr}\ \bibnamefont
  {Rabochiy}}, \bibinfo {author} {\bibfnamefont {Peter~G.}\ \bibnamefont
  {Wolynes}}, \ and\ \bibinfo {author} {\bibfnamefont {Vassiliy}\ \bibnamefont
  {Lubchenko}},\ }\bibfield  {title} {\enquote {\bibinfo {title}
  {Microscopically based calculations of the free energy barrier and dynamic
  length scale in supercooled liquids: The comparative role of configurational
  entropy and elasticity},}\ }\href {https://doi.org/10.1021/jp409502k}
  {\bibfield  {journal} {\bibinfo  {journal} {The Journal of Physical Chemistry
  B}\ }\textbf {\bibinfo {volume} {117}},\ \bibinfo {pages} {15204--15219}
  (\bibinfo {year} {2013})}\BibitemShut {NoStop}%
\bibitem [{\citenamefont {Wyart}\ and\ \citenamefont
  {Cates}(2017)}]{PhysRevLett.119.195501}%
  \BibitemOpen
  \bibfield  {author} {\bibinfo {author} {\bibfnamefont {Matthieu}\
  \bibnamefont {Wyart}}\ and\ \bibinfo {author} {\bibfnamefont {Michael~E.}\
  \bibnamefont {Cates}},\ }\bibfield  {title} {\enquote {\bibinfo {title} {Does
  a growing static length scale control the glass transition?}}\ }\href
  {\doibase 10.1103/PhysRevLett.119.195501} {\bibfield  {journal} {\bibinfo
  {journal} {Phys. Rev. Lett.}\ }\textbf {\bibinfo {volume} {119}},\ \bibinfo
  {pages} {195501} (\bibinfo {year} {2017})}\BibitemShut {NoStop}%
\bibitem [{\citenamefont {Dyre}(2006)}]{dyre2006colloquium}%
  \BibitemOpen
  \bibfield  {author} {\bibinfo {author} {\bibfnamefont {Jeppe~C.}\
  \bibnamefont {Dyre}},\ }\bibfield  {title} {\enquote {\bibinfo {title}
  {Colloquium: The glass transition and elastic models of glass-forming
  liquids},}\ }\href {\doibase 10.1103/RevModPhys.78.953} {\bibfield  {journal}
  {\bibinfo  {journal} {Rev. Mod. Phys.}\ }\textbf {\bibinfo {volume} {78}},\
  \bibinfo {pages} {953--972} (\bibinfo {year} {2006})}\BibitemShut {NoStop}%
\bibitem [{\citenamefont {Tanaka}(2012)}]{tanaka2012bond}%
  \BibitemOpen
  \bibfield  {author} {\bibinfo {author} {\bibfnamefont {Hajime}\ \bibnamefont
  {Tanaka}},\ }\bibfield  {title} {\enquote {\bibinfo {title} {Bond
  orientational order in liquids: Towards a unified description of water-like
  anomalies, liquid-liquid transition, glass transition, and
  crystallization},}\ }\href {https://doi.org/10.1140/epje/i2012-12113-y}
  {\bibfield  {journal} {\bibinfo  {journal} {The European Physical Journal E}\
  }\textbf {\bibinfo {volume} {35}},\ \bibinfo {pages} {113} (\bibinfo {year}
  {2012})}\BibitemShut {NoStop}%
\bibitem [{\citenamefont {Ikeda}\ \emph {et~al.}(2017)\citenamefont {Ikeda},
  \citenamefont {Zamponi},\ and\ \citenamefont {Ikeda}}]{ikeda2017mean}%
  \BibitemOpen
  \bibfield  {author} {\bibinfo {author} {\bibfnamefont {Harukuni}\
  \bibnamefont {Ikeda}}, \bibinfo {author} {\bibfnamefont {Francesco}\
  \bibnamefont {Zamponi}}, \ and\ \bibinfo {author} {\bibfnamefont {Atsushi}\
  \bibnamefont {Ikeda}},\ }\bibfield  {title} {\enquote {\bibinfo {title} {Mean
  field theory of the swap monte carlo algorithm},}\ }\href@noop {} {\bibfield
  {journal} {\bibinfo  {journal} {J. Chem. Phys.}\ }\textbf {\bibinfo {volume}
  {147}},\ \bibinfo {pages} {234506} (\bibinfo {year} {2017})}\BibitemShut
  {NoStop}%
\bibitem [{\citenamefont {Yaida}\ \emph {et~al.}(2016)\citenamefont {Yaida},
  \citenamefont {Berthier}, \citenamefont {Charbonneau},\ and\ \citenamefont
  {Tarjus}}]{PhysRevE.94.032605}%
  \BibitemOpen
  \bibfield  {author} {\bibinfo {author} {\bibfnamefont {Sho}\ \bibnamefont
  {Yaida}}, \bibinfo {author} {\bibfnamefont {Ludovic}\ \bibnamefont
  {Berthier}}, \bibinfo {author} {\bibfnamefont {Patrick}\ \bibnamefont
  {Charbonneau}}, \ and\ \bibinfo {author} {\bibfnamefont {Gilles}\
  \bibnamefont {Tarjus}},\ }\bibfield  {title} {\enquote {\bibinfo {title}
  {Point-to-set lengths, local structure, and glassiness},}\ }\href {\doibase
  10.1103/PhysRevE.94.032605} {\bibfield  {journal} {\bibinfo  {journal} {Phys.
  Rev. E}\ }\textbf {\bibinfo {volume} {94}},\ \bibinfo {pages} {032605}
  (\bibinfo {year} {2016})}\BibitemShut {NoStop}%
\bibitem [{\citenamefont {Chandler}\ and\ \citenamefont
  {Garrahan}(2010)}]{chandler2010dynamics}%
  \BibitemOpen
  \bibfield  {author} {\bibinfo {author} {\bibfnamefont {David}\ \bibnamefont
  {Chandler}}\ and\ \bibinfo {author} {\bibfnamefont {Juan~P.}\ \bibnamefont
  {Garrahan}},\ }\bibfield  {title} {\enquote {\bibinfo {title} {Dynamics on
  the way to forming glass: Bubbles in space-time},}\ }\href {\doibase
  10.1146/annurev.physchem.040808.090405} {\bibfield  {journal} {\bibinfo
  {journal} {Annual Review of Physical Chemistry}\ }\textbf {\bibinfo {volume}
  {61}},\ \bibinfo {pages} {191--217} (\bibinfo {year} {2010})}\BibitemShut
  {NoStop}%
\bibitem [{\citenamefont {Lebowitz}\ \emph {et~al.}(1967)\citenamefont
  {Lebowitz}, \citenamefont {Percus},\ and\ \citenamefont
  {Verlet}}]{PhysRev.153.250}%
  \BibitemOpen
  \bibfield  {author} {\bibinfo {author} {\bibfnamefont {J.~L.}\ \bibnamefont
  {Lebowitz}}, \bibinfo {author} {\bibfnamefont {J.~K.}\ \bibnamefont
  {Percus}}, \ and\ \bibinfo {author} {\bibfnamefont {L.}~\bibnamefont
  {Verlet}},\ }\bibfield  {title} {\enquote {\bibinfo {title} {Ensemble
  dependence of fluctuations with application to machine computations},}\
  }\href {\doibase 10.1103/PhysRev.153.250} {\bibfield  {journal} {\bibinfo
  {journal} {Phys. Rev.}\ }\textbf {\bibinfo {volume} {153}},\ \bibinfo {pages}
  {250--254} (\bibinfo {year} {1967})}\BibitemShut {NoStop}%
\bibitem [{\citenamefont {Cheng}\ and\ \citenamefont
  {Ceriotti}(2018)}]{cheng2018computing}%
  \BibitemOpen
  \bibfield  {author} {\bibinfo {author} {\bibfnamefont {Bingqing}\
  \bibnamefont {Cheng}}\ and\ \bibinfo {author} {\bibfnamefont {Michele}\
  \bibnamefont {Ceriotti}},\ }\bibfield  {title} {\enquote {\bibinfo {title}
  {Computing the absolute gibbs free energy in atomistic simulations:
  Applications to defects in solids},}\ }\href
  {https://journals.aps.org/prb/abstract/10.1103/PhysRevB.97.054102} {\bibfield
   {journal} {\bibinfo  {journal} {Physical Review B}\ }\textbf {\bibinfo
  {volume} {97}},\ \bibinfo {pages} {054102} (\bibinfo {year}
  {2018})}\BibitemShut {NoStop}%
\end{thebibliography}%

\end{document}